\documentclass[showpacs,aps,prb]{revtex4}
\usepackage{amsmath,amssymb}
\usepackage{graphicx}
\usepackage{color}

\def\(({\left(}
\def\)){\right)}                       
\def\[[{\left[}
\def\]]{\right]}

\newcommand{\be}{\begin{equation}}
\newcommand{\ee}{\end{equation}}
\newcommand{\bea}{\begin{eqnarray}}
\newcommand{\eea}{\end{eqnarray}}

\begin{document}
\title{Generalization of the cavity method for adiabatic evolution of
  Gibbs states}

\author {Lenka Zdeborov\'a $^1$ and Florent Krzakala $^{2,1}$}
\affiliation{$^1$Theoretical Division and Center for Nonlinear
  Studies, Los Alamos National Laboratory, NM 87545 USA \\$^2$ CNRS
  and ESPCI ParisTech, 10 rue Vauquelin, UMR 7083 Gulliver, Paris
  75005 France}

\begin{abstract}
  Mean field glassy systems have a complicated energy landscape and an
  enormous number of different Gibbs states. In this paper, we
  introduce a generalization of the cavity method in order to describe
  the adiabatic evolution of these glassy Gibbs states as an external
  parameter, such as the temperature, is tuned.  We give a general
  derivation of the method and describe in details the solution of the
  resulting equations for the fully connected $p$-spin model, the
  XOR-SAT problem and the anti-ferromagnetic Potts glass (or
  "coloring" problem). As direct results of the states following method, we present a
  study of very slow Monte-Carlo annealings, the demonstration of the
  presence of temperature chaos in these systems, and the
  identification of a easy/hard transition for simulated annealing in
  constraint optimization problems. We also discuss the relation
  between our approach and the Franz-Parisi potential, as well as with
  the reconstruction problem on trees in computer science.  A mapping
  between the states following method and the physics on the Nishimori line is also
  presented.
\end{abstract}

\date{\today}
\maketitle

\newpage 

\tableofcontents

\newpage 

\section*{Introduction}

Both in classical and quantum thermodynamics, it is often practical to
discuss very slow variations of an external parameter so that the
system remains at equilibrium, and such very slow changes are referred to as adiabatic \cite{BornFock28}. When a macroscopic
system is in a given phase, and if one tunes a parameter, say the
temperature, very slowly then all observables, such as the energy or the magnetization in a magnet, will be given by the equilibrium equation of state.

Such considerations have to be revisited close to a phase transition
where it is impossible to be truly adiabatic, and this is the subject
of modern out-of-equilibrium theories. However, given a system at
equilibrium in a well defined phase, it is always possible to consider
the adiabatic evolution. In the low temperature phases of a
ferromagnet, for instance, the evolution of the magnetization is
different in the two phases (or Gibbs states) corresponding to the
positive or negative magnetization. To describe this theoretically,
one can force the system to be in the Gibbs state of choice (for
instance by adding an external infinitesimal field, or fixing the
boundary conditions) and then study the adiabatic evolution for each
of these phases.

This simplicity, however, breaks when one considers glassy systems
where the energy landscape is very complicated, and especially in the
mean field setting where exponential number of phases (Gibbs states) exists. Adiabatic evolution of phases in mean field glassy systems is, however, a very important problem that has been considered ---via some approximation or in very specific
solvable cases --- in a number of works
\cite{CugliandoloKurchan93,BarratFranz97,LopantinIoffe02,
  MontanariRicci04,KrzakalaMartin02,CaponeCastellani06,RizzoYoshino06,MoraZdeborova07}.
How to deal with this situation in general mean field glassy systems, how to chose a particular phase, and how to follow it adiabatically is the subject of the formalism presented in this work.

Mean-field glassy systems are important in many parts of
modern science. We shall call a system a "mean-field" one
whenever a mean-field treatment is exact for this system: this is the
case for all spin or particle models on fully connected lattices
(such as the Curie-Weiss model of ferromagnets) or on sparse random
lattices that are locally tree-like (such as the Bethe lattice). Over
the last few years, studies of mean-field glassy systems brought
many interesting results in physics as well as in computer
science. Without being exhaustive, we can mention the development of
mean field theories for structural glass formers
\cite{KirkpatrickThirumalai87a,MezardParisi99}, for the jamming
transition and amorphous packing \cite{ParisiZamponi09}, heteropolymer
folding \cite{ShakhnovichGutin89}, or for quantum disordered materials
\cite{IoffeMezard09} on the physics side. On the computer science
side, many results have been obtain using mean field theory on
optimization problems and neural network \cite{MezardParisi87b}, and
more recently random constraint satisfaction problems have witnessed a
burst of new results via the application of the survey propagation
algorithm and related techniques \cite{MezardParisi02,KrzakalaMontanari06}. The theory of
error correcting codes is also closely related to glassy mean field
system \cite{MezardMontanari07}, etc.

A common denominator in all these systems is their complex energy
landscape and a large number of phases (states), whose statistical
features are amenable to an analytical and quantitative description
via the replica and cavity methods
\cite{MezardParisi87b,MezardParisi01}. However, important and deep
questions about the dynamical behavior in these systems remain largely
unsolved, and many of them can be addressed by the knowledge of the
slow dynamics. In order to motivate our approach, let us first discuss
the basic universal features of the thermodynamic behavior of
mean-field glassy models.  As an external parameter, say the
temperature $T$, is tuned, a typical glassy system undergoes the
following changes: At high temperature, the system is in a
paramagnetic/liquid phase. Below the {\it dynamical} glass temperature
$T_d$, this phase shatters into exponentially many Gibbs
states/phases, all well separated by extensive energetic or entropic
barriers, leading to a breaking of ergodicity and to the divergence of
the equilibration time
\cite{CugliandoloKurchan93,BouchaudCugliandolo98,MontanariSemerjian06}. As
the temperature is further lowered, the number of states (relevant for
the Boltzmann measure) may become finite and the structural entropy
(or complexity) vanishes, this defines the {\it static} Kauzmann
transition, $T_K$, arguably similar to the one observed in real glass
formers \cite{Kauzmann48,MezardParisi99}. This scenario is called the
"one-step replica symmetric" (1RSB) picture. In some models
\cite{Gardner85}, the states will divide further into an infinite
hierarchy of sub-states, a phenomenon called "full replica symmetry
breaking" (FRSB) \cite{MezardParisi87b,MezardParisi01}.

\begin{figure}[!ht]
\begin{center}
  \vspace{-0.2cm}
  \includegraphics[width=0.5\linewidth]{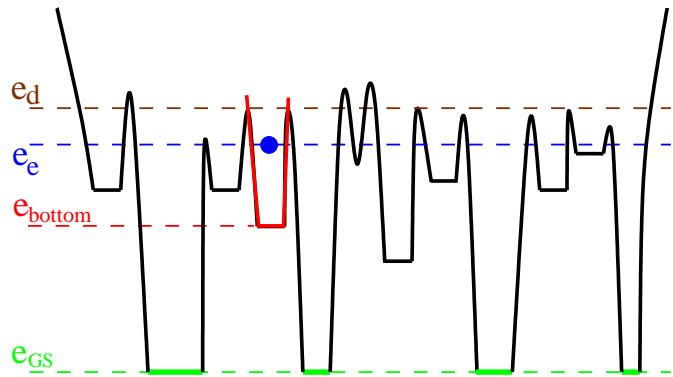}
  \vspace{-0.3cm}
\end{center}
\caption{(color online) A cartoon of the energy landscape in mean
  field glassy systems. The different valleys correspond to different
  Gibbs states and are separated by extensive barriers. For energies
  lower than $e_d$ (the brown line), ergodicity breaks because of
  these barriers.  The ground state energy of the system is $e_{\rm
    GS}$ (in green). The standard cavity and replica method can
  compute how many states of a given size/entropy are present at a
  given energy/temperature $e$/$T$.  The states following method
  we develop in this paper instead pins down one state (in red in the
  figure) that is one of the equilibrium ones at energy $e_e$
  (corresponding to temperature $T_e$, the blue line) and computes
  its properties (entropy, energy) for another temperatures $T_a$: we
  are thus {\it following} a given state as temperature (or
  any other parameter) is changed.  At $T_a=0$, this leads for instance
  to the properties of the bottom of the state as e.g. the limiting
  energy $e_{\rm bottom}$. \label{fig00}}
\end{figure}

The 1RSB picture is well established in many mean field systems, and
the cavity/replica method is able to compute the number, the size or
the energy of the equilibrium Gibbs states. However, with the
exception of few simple models
\cite{CugliandoloKurchan93,BarratFranz97,CaponeCastellani06,KrzakalaMartin02,RizzoYoshino06,MoraZdeborova07},
an analytical description of the dynamics and of the way states are
evolving upon adiabatic change of external parameters is missing. Let
us consider a given setting where the need for adiabatic following is
clear: Imagine an annealing experiment where the temperature $T$ is
changed in time as $T=T_0 - \delta t/N$. Take the thermodynamic limit
$N\!\to\!\infty$ first and then consider a very slow annealing
$\delta\to 0$. As long as we stay in the paramagnetic phase, we expect
that such a slow annealing will equilibrate. The fact that the
equilibration time is finite below $T_d$ can be actually proven
\cite{MontanariSemerjian06,MontanariSemerjian06b} and such annealing
should be thud able to equilibrate down to the dynamical temperature
$T_d$ after which the system get stuck in one of the many equilibrium
Gibbs states.  Computing the energy of the lowest configuration
belonging to this state would thus give the limiting energy for a very
slow annealing (and thus would give a bound to the performance of any
annealing scheme). However, while the standard cavity and the replica
method predict all the properties of an {\it equilibrium} state at a
given temperature $T_e$ (equilibrium temperature), they do not tell
how these properties change {\it for this precise state} when the
temperature changes adiabatically to $T_a \neq T_e$ (actual
temperature). A word of caution: We want to follow the state and stay
in it. Hence by "adiabatic" we mean here slow only {\it linearly} in
the size of the system, corresponding to very long experimental times;
an {\it exponentially} slow annealing always finds the ground state,
but this is of course unfeasibly long.

The extension of the cavity method that we introduced in a recent
Letter \cite{KrzakalaZdeborova09b} precisely answered these questions
by following adiabatically the evolution of any Gibbs state when an
external parameter is changed (for an intuitive and pictorial
description of our goals, see Fig.~\ref{fig00}). This gives detailed
quantitative information about the energy landscape and the long time
dynamics. The aim of this subsequent publication is to derive the
method in general, to discuss in detail the solutions of the resulting
equations, and to discuss relations with some other settings
(reconstruction on trees \cite{MezardMontanari06}, Franz-Parisi
potential \cite{FranzParisi95,FranzParisi97}, Nishimori line
\cite{Nishimori81}). We anticipate that the method will become part of
the standard tool-box for mean field glassy systems and hence a
detailed presentation is appropriate.

The paper is organized as follow: In Sec.~\ref{sec:preli} we give a
brief reminder of the usual cavity method. In the two next section, we
present our formalism for the adiabatic evolution of states from
temperature higher (Sec.~\ref{sec:evolution}) and lower
(Sec.~\ref{sec:general}) than the spin glass static/Kauzmann
transition. We finally solve these equations and present our results
for a fully connected model in Sec.~\ref{pspin_th} and for diluted
models on the Bethe lattice in Sec.~\ref{sec:result-xor}.

\section{Preliminaries}
\label{sec:preli}
In this section, we first review the results of the standard cavity
method that we shall use all along the text. The specific example for
which we shall derive most of the results of this paper is the
$p$-spin model, also called the XOR-SAT problem. However, the
derivation for all other models where the cavity method
\cite{MezardParisi01} can be used goes in the very same lines (and we
will also work later on with the coloring problem). The reader familiar with the cavity method can skip this section and go directly to Sec.~\ref{sec:evolution}.

\subsection{The $p$-spin model and XOR-SAT reminder}
The $p$-spin model is defined by its Hamiltonian \be {\cal H} = -
\sum_a J_a \prod_{i\in \partial a} s_i\, ,\label{Ham_pspin} \ee where
$s_i\in \{-1,+1\}$ are the Ising spins, $a$ are interactions between
$p$-uples of spins, $J_a$ is the strength of the interaction.

In what follows we will focus on two cases of the $p$-spin model:
\begin{itemize}
\item{XOR-SAT (parity check) problem: In this case all the
    interactions $|J_a|=1$. The interactions can be both ferromagnetic
    and anti-ferromagnetic \be Q(J_a)=\rho \delta(J_a+1) + (1-\rho)
    \delta(J_a-1)\, .\label{disorder}\ee In the results section we mostly consider the
    spin glass case $\rho=1/2$. The number of interactions $M$
    (linear equations) is $M=\alpha N$, where $\alpha$ is the
    constraint density. The degree distributions ${\cal Q}(l)$ of variables
    have to be specified here. The number of violated parity checks
    (constraints) is $E=(M+{\cal H})/2$. The values of temperature for
    the $K$-XOR-SAT problems are hence related to those for the $p$-spin
    problem via a multiplicative factor 2, note that here and through the paper $K=p$.}
\item{Fully connected $p$-spin model: The interactions $a$ exist for
    every possible $p$-uple of spins, the mean and variance of $J_a$
    are given by $\langle J_{a} \rangle = J_0 p!/N^{p-1}$ and $
    \langle J_{a}^2 \rangle- \langle J_{a} \rangle^2 = J^2 p!/(2
    N^{p-1})$.}
\end{itemize} 

The XOR-SAT problem was studied and solved in
\cite{RicciWeigt01,MezardRicci03,FranzMezard01,CoccoDubois03}, and its
most important application are the low density parity check error
correcting codes \cite{Gallager62,MacKay99}. The fully connected
$p$-spin model was introduced in \cite{Derrida80,GrossMezard84} and
now stands at the root of the random first order theory of the glass
transition
\cite{KirkpatrickThirumalai87a,KirkpatrickThirumalai87b,KirkpatrickThirumalai89}.

In our examples we will mainly use the ensemble of random regular
graphs, i.e.  ${\cal Q}(l)=\delta(L-l)$ and obtain the fully connected
limit by taking $L \to \infty$. The formulas are, however, written
mostly for a general degree distribution (with a finite second
moment). In the cavity equations we often need the excess degree
distribution, that is the probability distribution of the number of
excess edges given one edge is chosen, that is (denoting ${\overline
  l}$ the average coordination number):
\be q(l) = \frac{(l+1){\cal Q}(l+1)}{\overline l}\, .  \ee

\subsubsection{Liquid phase: Belief propagation equations}

We now summarize in a very brief manner and without extensive
derivations the known cavity equations for the XOR-SAT problem as we
are going to need them for derivation of the states following method. We
are trying the keep the equations in the most general form such that
generalizations to other models are straightforward. The very
principle of the cavity method is that we are working with tree-like
graphs. Random graphs are locally a tree and we thus can work ``as
if'' on a tree (we will eventually have to care of the boundary
conditions and precise relation to random graphs later on). Solving problem on a tree can be done easily with
a recursive procedure that was introduced by Bethe \cite{Bethe35}. We
will, however, use the modern language of computer science, where this
is called the Belief Propagation (BP) equations. 

\begin{figure}[!ht]
\begin{center}
  \vspace{-0.2cm}
  \includegraphics[width=0.15\linewidth]{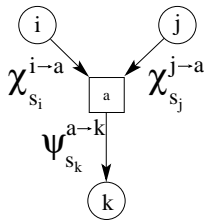}
  \vspace{-0.3cm}
\end{center}
\caption{A sketch of the basic cavity recursion in the factor graph
  representation of XOR-SAT. The square represent the constraints
  involving the product of $p$ spins, while and the circles represent
  the spin variables \label{recursion}. The message passing procedure
  called belief propagation uses messages from constraints to
  variables ($\psi$) and from variables to constraints ($\chi$).}
\end{figure}
For the XOR-SAT problem the BP equations read
\bea  \chi^{i\to a}_{s_i} &=&\frac{1}{Z_\chi^{i\to a}(\{\psi^{b\to i} \},\beta)} \prod_{b\in \partial i \setminus a} \psi_{s_i}^{b\to i}\, , \label{BP_chi} \\
\psi^{b\to i}_{s_i} &=&\frac{1}{Z_\psi^{b\to i}(\{\chi^{j\to b}
  \},\beta)}\sum_{\{s_j\}} e^{\beta J_b \prod_{j\in \partial b} s_j}
\prod_{j\in\partial b \setminus i} \chi^{j\to b}_{s_j}\,
, \label{BP_psi} \eea
where $Z_\psi^{b\to i}$ and $Z_\chi^{i\to a}$ are normalizations
ensuring that $\psi^{b\to i}_{+1}+\psi^{b\to i}_{-1}=1$ and
$\chi^{i\to a}_{+1}+\chi^{i\to a}_{-1}=1$. The quantities $\chi^{i\to
  a}_{s_i}$ (resp. $\psi^{b\to i}_{s_i}$) are interpreted in terms of
messages being send from a variable $i$ to a constraint $a$
(resp. from constraint $b$ to variable $i$) (see Fig.~\ref{recursion}
for a pictorial representation with the so-called "factor graph".).
Message $\chi^{i\to a}_{s_i}$ is a probability that variable $i$ takes
value $s_i$ conditioned on constraint $a$ to be missing from the
graph. Message $\psi^{b\to i}_{s_i}$ is a probability that the
constraint $b$ is satisfied given that variable $i$ takes values
$s_i$.
The recursion could also be written with one single type of message as \be \psi^{a\to i}_{s_i} ={\cal F}(\{\psi^{b\to j}
\},\beta) = \sum_{\{s_j\}}\frac{ e^{\beta J_a s_i\prod_{j\in \partial a\setminus i}
    s_j}}{Z^{a\to i}(\{\psi^{b\to j} \},\beta)} \prod_{j\in\partial a
  \setminus i} \prod_{b\in \partial j\setminus a} \psi_{s_j}^{b\to j}
= \sum_{\{s_j\}} P(\{s_j\}|\{\psi^{b\to j} \},\beta,s_i)
\, . \label{BP_eq}
\ee

Given all messages computed on a given graph, one can compute the Bethe
estimate of the free energy, which is also called the replica
symmetric one (RS) free energy:
\be -\beta F = \sum_a
\log{ Z^{a+\partial a}} - \sum_i (l_i-1) \log{ Z^i }\, , \ee where the
contributions to the free energy are \bea
Z^{a+\partial a}(\{\psi^{b\to i}\},\beta)&=& \sum_{\{s_i\}}  e^{\beta J_a \prod_{i\in \partial a} s_i}  \prod_{i\in \partial a}  \prod_{b\in \partial i-a} \psi_{s_i}^{b\to i}\, , \label{Za}\\
Z^i(\{\psi^{a\to i}\},\beta)&=& \prod_{a\in \partial i}
\psi_{-1}^{a\to i}+ \prod_{a\in \partial i} \psi_{+1}^{a\to i}\,
. \label{Zi} \eea Deriving the free energy with respect to the inverse
temperature we get the energy \be E = \frac{\partial (\beta
  F)}{\partial \beta} = \sum_a E^{a+\partial a}(\{\psi^{b\to
  i}\},\beta) = \sum_a \frac{\sum_{\{s_i\}}J_a \prod_{i\in \partial a}
  s_i \, e^{\beta J_a \prod_{i\in \partial a} s_i}
  \prod_{i\in \partial a} \prod_{b\in \partial i-a} \psi_{s_i}^{b\to
    i}}{\sum_{\{s_i\}} e^{\beta J_a \prod_{i\in \partial a} s_i}
  \prod_{i\in \partial a} \prod_{b\in \partial i-a} \psi_{s_i}^{b\to
    i}}\, .  \ee

All the above equations are written for a given instance (graph, or
given instance of the disorder) of the problem. It is often desirable
to write the BP equations directly in the average form over the graph
and disorder ensemble. This is the replica symmetric cavity equation
\be P(\psi) = \sum_{\{l_i\}} q(\{l_i\}) \int \prod_{i=1}^{K-1}
\prod_{j=1}^{l_i} {\rm d}P(\psi^j) \, \delta[\psi-{\cal
  F}(\{\psi^j\},\beta)]\, , \label{BP_pop_dyn} \ee that can be solved
numerically via the population dynamics technique introduced in
\cite{MezardParisi01} (see also \cite{ZdeborovaKrzakala07} or \cite
{MezardMontanari07} for details).

Whether the BP equations for XOR-SAT are solved on a given random
graph or in the population dynamics they have always the following
fixed point that corresponds to the paramagnetic/liquid phase:
\be \psi^{a\to i}_{+1}=\psi^{a\to i}_{-1}= \chi_{+1}^{i\to a}
=\chi_{-1}^{i\to a} = \frac{1}{2} \, , \label{pspin_mes} \ee %
for all $i$ and $a$. Plugging this solution in the expression for the
free energy we get 
\be -\beta f=-\beta F/N= \frac{\overline l}{K}
\log{(\cosh{\beta})} +\log 2 \, , \label{pspin_rs} \ee and for the
energy we get \be e=E/N=- \frac{\overline l}{K} \tanh{\beta}\,
.  \label{RS_energy} \ee
Hence in this case, the probability that a given constraint is not
satisfied is
 \be \epsilon(\beta) = \frac{1}{1+e^{2\beta}} \,
. \label{eps_rs} \ee

\subsubsection{Glassy solution: One-step replica symmetry breaking}
\label{sec:1RSB}
The replica symmetric liquid solution from the previous section is
asymptotically exact as long as the the point-to-set correlation
length stays finite \cite{MontanariSemerjian06}. This is related to
the reconstruction problem on trees \cite{MezardMontanari06}. When the
point-to-set correlation length diverges, the replica symmetry broken
solution \cite{MezardParisi01} has to be used to describe correctly
the system. 

In the one-step replica symmetry breaking one splits the phase space
into exponentially many Gibbs states, $P^{a\to i}(\psi^{a\to i})$ is
then the probability distribution over states of the cavity message
$\psi^{a\to i}$. We now need to consider all these states and in order
to focus on those with a given free energy $f=-T\log(Z)$, we weight
them according to their Boltzmann weight to a given power $Z^x$, where
$x$ is the so-called Parisi parameter. In the cavity method $x$ is
then used as a Legendre parameter in order to select the states with a
given free energy. With this in mind, the 1RSB self-consistent
recursive equation reads \cite{MezardParisi01}:
\be P^{a\to i}(\psi^{a\to i}) = \frac{1}{{\cal Z}^{a\to i}(\beta)}
\int \prod_{j\in\partial a \setminus i} \prod_{b\in \partial
  j\setminus a} {\rm d}P^{b\to j}(\psi^{b\to j}) \left[Z^{a\to
    i}(\{\psi^{b\to j} \}, \beta) \right]^x \delta[\psi^{a\to i} -
{\cal F}(\{\psi^{b\to j} \},\beta)] \label{1RSB} \ee with ${\cal
  F}(\{\psi^{b\to j} \},\beta)$ and $Z^{a\to i}(\{\psi^{b\to j} \},
\beta)$ being defined in Eq.~(\ref{BP_eq}). The Parisi parameter $x$
is indeed a Legendre parameter conjugated to the internal free energy
of states.The entropy associated with number of states of a given
internal free energy $f$, also called complexity and defined by
$\Sigma(f)=\log{\(({\cal N}_{states}(f)\))}/N$, can be recovered from
the following Legendre transform
\be -\beta x \Phi(\beta,x) = - \beta x f(\beta) + \Sigma(f)\, ,\quad
\quad f(\beta) = \frac{\partial[x \Phi(\beta,x)]}{\partial x} \, . \label{Sigma}
\ee The potential $\Phi(\beta,x)$ is computed from the fixed point of
the 1RSB equations (\ref{1RSB}) as \be \Phi(\beta,x) = \sum_a
\Phi^{a+\partial a} - \sum_i (l_i-1) \Phi^i \, , \ee where \bea
e^{-\beta x\Phi^{a+\partial a}} &=& \int \prod_{i\in\partial a } \prod_{b\in \partial i\setminus a} {\rm d}P^{b\to i}(\psi^{b\to i}) [{Z^{a+\partial a}(\{\psi^{b\to i} \},\beta)}]^x \, ,\\
e^{-\beta x \Phi^i }&=& \int \prod_{a\in\partial i} {\rm d}P^{a\to
  i}(\psi^{a\to i}) [{Z^{i}(\{\psi^{a\to i}\},\beta)}]^x\, .  \eea

The condition for validity of the replica symmetric solution is
recovered by solving the 1RSB equations for $x=1$, that is if at $x=1$
there exist a non-trivial solution of Eq.~(\ref{1RSB}) then RS
solution is not correct and the phase space needs to be divided into
states. This happens at the dynamical temperature $T_d$.

The 1RSB solution is then given by the value of $x$ such that \be x^*
= {\rm argmax}_x[ - \beta f(\beta) + \Sigma(f)|\Sigma(f)\ge 0] \,
. \label{m_phys} \ee Above the Kauzmann temperature $T_K$ one has
$x^*=1$ and $\Sigma(f)> 0$, that is exponentially many states are
relevant to the Boltzmann measure. In this phase the local magnetization (marginal
probabilities) and the thermodynamic potentials, such as the total free
energy, are still given by the replica symmetric solution
Eqs.~(\ref{pspin_mes}--\ref{eps_rs}).

Below the Kauzmann temperature $x^*<1$ and $\Sigma(f)=0$, the
Boltzmann measure is dominated by only a finite number of
states. However, an exponential numbers of sub-dominant
(non-equilibrium) states still exist at any positive temperature.

\subsection{Coloring of graphs, alias the anti-ferromagnet Potts model}
We shall also illustrate some of our results on the anti-ferromagnetic
Potts model on random graphs, mostly known and studied in its zero
temperature version as then it is equivalent to the graph coloring
problem
\cite{MourikSaad02,MuletPagnani02,BraunsteinMulet03,KrzakalaPagnani04,ZdeborovaKrzakala07}. The
Hamiltonian is \be H = \sum_{(ij)\in G} \delta_{s_i,s_j}\,
, \label{Ham_col} \ee where $s_i$ are Potts spins taking one of the
$q$ possible values (colors), $\delta_{i,j}$ is the Kronecker delta
symbol and the sum  all edges of the graph. The phase diagram of
this model at finite temperature is summarized in
\cite{KrzakalaZdeborova07} and all the necessary equations for both
the replica symmetric and glassy solution can be found in
\cite{ZdeborovaKrzakala07}.

\section{Evolution of states above the Kauzmann temperature}
\label{sec:evolution}
In this section we introduce the states following formalism, and derive equations for
the evolution of states that are at equilibrium above the Kauzmann
temperature, $T_e\ge T_K$. In this phase, the paramagnetic replica
symmetric solution (\ref{pspin_mes}--\ref{eps_rs}) correctly describes
all thermodynamic quantities (but misses the ergodicity breaking at
$T_d>T_K$).  In the next section \ref{sec:general} we give a
generalization for equilibrium states below $T_K$ and for meta-stable
states.

In order to get an intuitive idea of what we will do, let us
consider the ferromagnetic Ising model on a random graph. At low
temperature, there are two phases corresponding to the positive and
negative magnetization. In order to study one of these phases, a good
strategy is to first recognize that the random graph is locally a
tree. Then one considers a tree where all spins on the boundaries are
fixed to, say, a value $S=1$, then far away from the boundaries, the system will be in the phase of positive magnetization. By changing the temperature the curve $m^+(T)$ can be computed. We will follow the same strategy, except that choosing the correct boundaries will be slightly more involved.

The main idea behind the equations of state following is that we pick a
configuration at random among the equilibrium ones at a temperature
$T_e$, and then we look at the solution of the belief propagation
equations at a temperature $T_a$ initialized in that configuration. We
will also discuss a special case of factorized replica symmetric
solution where this idea can be actually performed on a single graph.
This is also closely related to the quiet planting discussed in
\cite{KrzakalaZdeborova09,ZdeborovaKrzakala09}.

\subsection{Gedanken experiment on infinite trees}
\label{gedanken}

\begin{figure}[t]
\begin{center}
\includegraphics[width=8.5cm]{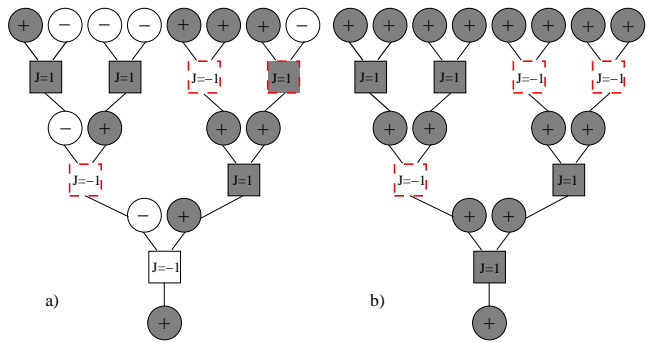}
\end{center}
\caption{(color online) a) Recursive construction of an equilibrium
  configuration at temperature $T_e$ in XOR-SAT. Given the tree, and a
  random choice of interactions (full square $J=1$, empty square
  $J=-1$), one starts from the root, and chooses iteratively the
  configuration of ancestors (full circles $s=1$, empty circle $s=-1$)
  randomly such that it satisfies the constraints with probability
  $1-\epsilon(T_e)$, Eq.~(\ref{eps_rs}), here $\epsilon=3/7$. Violated
  constraints have dashed/red borders.  b) The problem can also be
  Gauge transformed using Eq.~(\ref{Gauge}), see Sec.~\ref{sec:Nish},
  into a fully polarized configuration with all $s=1$ but where the
  $J$'s are chosen from distribution (\ref{Q_N}).  \label{fig0}}
\end{figure}
Let us consider the problem on a large hyper-tree. Let the hyper-tree
have the same distribution of disorder (i.e. the degree distribution,
the distribution of negations, interaction strengths etc.) as the
original problem. Let us consider a measure uniform over all
configurations having energy corresponding to a given temperature $T_e$. To sample
uniformly one configuration from this measure the following steps need
to be done:
\begin{itemize}
   \item[(a)]{Take much larger hyper-tree and start with a random messages on the boundary of the larger hyper-tree and iterate the belief propagation equations at temperature $T_e$ down to the root. This way one created messages taken from the replica symmetric solution on the original hyper-tree.}
   \item[(b)]{Assign a value to the root according to the incoming message. Proceed iteratively up to the leaves of the hyper-tree with the following: Given the value $s_i$ of variable $i$ choose the set of values $\{s_j\}$ according to probability $P(\{s_j\}|\{\psi^{b\to j} \},\beta_e,s_i)$ defined in Eq.~(\ref{BP_eq}), where $a$ is a descendant of $i$ and $b\in \partial j \setminus a$ for each $j\in \partial a \setminus i$.}
\end{itemize} 

Now consider the values of variables from the configuration we picked
on the leaves. This is a boundary condition that defines the
equilibrium Gibbs state at temperature $T_e$ (as long as $T_e\ge
T_K$).  Next consider the belief propagation equations at temperature
$T_a \neq T_e$, initialize the messages on the leaves of the
hyper-tree in the configuration we picked (i.e. $\psi_{s}=1$ and $\psi_{r}=0$ for all $r\neq s$ if we picked value $s$) and iterate down to the
root. The result of these iteration does describe properties (free
energy, energy, size, overlap) of the Gibbs state at the temperature
$T_a \neq T_e$.

In case the original temperature was above the dynamical glass temperature, $T_e>T_d$, the solution of the belief propagation at $T_a$ will not be different from the result of the pure BP at $T_a$. That is because all the equilibrium configurations above $T_d$ lie in the same paramagnetic state. 

When the original temperature is below the dynamical glass temperature, $T_K \le T_e \le T_d$, the situation is much more interesting. Then the equilibrium configuration we picked lies in one of the exponentially many equilibrium Gibbs states and the belief propagation equations at a different temperature do describe adiabatic evolution of that Gibbs state.  

In the next subsections we shall translate the above reasoning into
the cavity equations, and describe the population dynamics technique
used to solve them.

\subsection{The simplest case: Factorized RS solution.}
The simplest form of the equations for adiabatic evolution of states
can be written when the replica symmetric solution is {\it
  factorized}, i.e. when the values of the messages are the same. This
is the case in the XOR-SAT problem where there is a BP fixed point in
which for all $i$ and $a$ the message $\psi^{a\to
  i}=1/2$. Furthermore, this fixed point gives an asymptotically exact
results above the Kauzmann temperature $T_K$.

When the RS solution is factorized the step (a) in the construction of
the equilibrium configuration can be skipped and the probabilities
$P(\{s_j\}|\{\psi^{b\to j} \},\beta_e,s_i)$ depend only on the values of
variables and the inverse temperature $\beta_e$. In the XOR-SAT in
particular we have from (\ref{BP_eq}) \be P(\{s_j\}|\{\psi^{b\to j}
\},\beta_e,s_i)= \frac{e^{\beta_e J_a s_i \prod_{j\in \partial a \setminus
      i} s_j }}{2^{K-1} \cosh{(\beta_e J_a)}} \, .
\label{sj_values}
\ee Meaning that a clause is unsatisfied with probability
$\epsilon(\beta_e)$ given by (\ref{eps_rs}). These probabilities are
used according to step (b) to choose an equilibrium configuration on
the hyper-tree. Then belief propagation equations at a temperature
$T_a$ are initialized on the leaves in that configuration and
iterated. As usual for belief propagation equations a probability
distribution of the values of messages can be written. This time one
has to distinguish between messages sent to the variables which were
assigned value $+1$ in the equilibrium configuration, and those that
were assigned $-1$. Given the probabilities to choose values of
variables (\ref{sj_values}), the two probability distributions have to
satisfy the following self-consistent equation \be P_s(\psi) = \sum_J
Q(J) \sum_{\{l_i\}} q(\{l_i\}) \sum_{\{s_i\}} \frac{ e^{J\beta_e s
    \prod_{i} s_i} }{2^{K-1}\cosh{\beta_e J}} \int \prod_{i=1}^{K-1}
\prod_{j=1}^{l_i} {\rm d}P_{s_i}(\psi^j) \, \delta[\psi-{\cal
  F}(\{\psi^j\},\beta_a)]\, , \label{pop_pl} \ee where $q(\{l_i\})$ is
the excess degree distribution, and ${\cal F}(\{\psi^j\},\beta_a)$ is
defined in (\ref{BP_eq}). Note the use of inverse temperature
$\beta_a$ in the BP equations represented by the delta function.
Given a Gibbs state that is one of the equilibrium ones at temperature
$T_e$, Eq.~(\ref{pop_pl}) encode its properties when the temperature
is changed to $T_a$.

The learned reader will have recognize that, when $T_e=T_a$, this is
nothing but the 1RSB equation \cite{MezardMontanari06} (at Parisi
parameter $x=1$). This is actually quite normal, since when the two
temperatures are equal, we are just describing the properties of a
typical state, which is what the 1RSB method does. Similar equations
when the temperatures are equals were thus considered in many works
\cite{ZdeborovaKrzakala07,Semerjian07,Zdeborova08,KrzakalaZdeborova09}.

To solve Eq.~(\ref{pop_pl}) with the population dynamics we represent
$P_s(\psi)$ by an array of values for each value of $s$. To update one
element in the array $P_s$ we first pick degrees $l_i$ from the excess
degree distribution, then based on value of $s$ and
Eq.~(\ref{sj_values}) we pick the values $\{s_i\}$. After that, for
each $i$ we pick $l_i$ random elements in the array $P_{s_i}$ and
based on (\ref{BP_eq}) we compute a new value of $\psi$ and replace
one element in the array $P_s$. We repeat many times until (based on
computation of some average quantities) the convergence is reached. It
is also important to note that the initial state of the populations
corresponding to the boundary conditions on the hyper-tree is \be
P_s(\psi) = \delta\left[{\psi_s \choose \psi_{-s}} - {1 \choose 0}
\right]\, .  \ee

The internal Bethe free energy of the state is \bea -\beta_a
f(\beta_a) &=& \alpha\sum_J Q(J) \sum_{\{l\}} q(\{l\}) \sum_{\{s_i\}}
\frac{ e^{J\beta_e \prod_{i} s_i}}{2^{K}\cosh{\beta_e J}}
\int\prod_{i=1}^K \prod_{j=1}^{l_i} \left[ {\rm d}\psi^{j}
  P_{s_i}^{j}(\psi^{j}) \right] \log{Z^{a+\partial
    a}(\{\psi^{j}\},\beta_a)}\nonumber\\ &-& \sum_{l} {\cal Q}(l)
\frac{l-1}{2} \sum_{s_i} \int \prod_{i=1}^l \left[ {\rm d}\psi^{i}
  P_{s_i}^{i}(\psi^{i}) \right] \log{Z^{i}(\{\psi^{i}\},\beta_a)}\,
. \label{free_pl} \eea The value can be computed using a population
dynamics procedure based on the fixed point for Eq.~(\ref{pop_pl}).

\subsection{The case of a general (non-factorized) RS solution}
In a general case when the replica solution is not factorized, e.g. in
the canonical case of the random K-SAT problem, the situation is a bit
more complex. In the gendanken experiment of section \ref{gedanken}, the uniform boundary conditions have been created with (and thus depends on) the replica symmetric marginals $\overline{\psi}$. The procedure described in the gedanken experiment translates to a more general form of equations, that are exact on a tree. The equivalent of
eq.~(\ref{pop_pl}) then reads
\bea
\overline \psi_s \overline P_s(\psi|\overline \psi) {\cal P}_{\rm
  RS}(\overline \psi) &=& \sum_J Q(J) \sum_{\{l\}} q(\{l\}) \int
\prod_{i=1}^{K-1} \prod_{j=1}^{l_i} \left[ {\rm d} \overline \psi^j
  {\cal P}_{\rm RS}(\overline \psi^j) \right] \, \delta\left[\overline
  \psi - {\cal F}(\{\overline \psi^j\},\beta_e)\right] \nonumber \\ &&
\sum_{\{s_i\}} e^{\beta_e C(J,s,\{s_i\})} \frac{\prod_{i=1}^{K-1}
  \prod_{j=1}^{l_i} \overline \psi^j_{s_i}}{Z(\{\overline
  \psi^{j}\},\beta_e)} \int \prod_{i=1}^{K-1} \prod_{j=1}^{l_i}
\left[{\rm d}\psi^j \overline P_{s_i}^{j}(\psi^j|\overline
  \psi^j)\right] \, \delta\left[\psi - {\cal
    F}(\{\psi^{j}\},\beta_a)\right]\, , \label{final_m1} \eea 
where $C(J,s,\{s_i\})$ is an arbitrary interaction between spins $s, \{s_i\}$ of "strength" $J$, in case of XOR-SAT we had $C(J,s,\{s_i\})=J\beta_e s \prod_{i} s_i$. This
equation is maybe easier to understand from the population dynamics
procedure used to solve it. Now we have $|s|+1$ different arrays to
represent the messages. In the first array we initially put an
equilibrated replica symmetric population (values obtained by solving
the simple RS equations by population dynamics). In the array
corresponding to value $s$ we initially put a message completely
polarized in the direction $s$.

Updating one element have to be done in all the $|s|+1$ arrays
simultaneously. One first chooses the degrees $l_i$, then one chooses
the corresponding number of random elements in the population. One
uses the first array to compute the new value corresponding to the
first array in the new element. To compute a new value corresponding
to array $s$ one uses the values on the first array to draw a
configuration of values $\{s_i\}$ using probabilities \be
P(\{s_i\}|\{\overline \psi^j\},\beta_e,s) = e^{\beta_e  C(J,s,\{s_i\})}
\frac{\prod_{i=1}^{K-1} \prod_{j=1}^{l_i} \overline
  \psi^j_{s_i}}{Z(\{\overline \psi^{j}\},\beta_e)} \, .  \ee Finally
using elements in arrays corresponding to values $s_i$ one computes a
new value. This done for every value of $s$ finalizes one step. All is
repeated until convergence is reached. The expression for the free
energy is analogous to (\ref{free_pl}) using the same generalization
as for (\ref{final_m1}).

We note that when $\beta_e=\beta_a$, the above equations are actually
already known, and are exactly equivalent to the 1RSB equations at
$x=1$. Again, this is just because in that case, we are just
describing the properties of a typical state. With equal temperatures
the above form of the 1RSB equations at $x=1$ appeared in
\cite{MontanariRicci08,Zdeborova08,ZdeborovaKrzakala09} (to which we
refer the reader interested to see how the present derivation
generalizes) and similar equations appeared in the context of the
analysis of an idealized BP decimation algorithms in
\cite{MontanariRicci07,RicciSemerjian09}.

\subsection{The relation to the problem of reconstruction on trees}
\label{sec:reconstruction}
In the special case when $T_e=T_a$ the above equations are thus
equivalent to the 1RSB equations when the Parisi parameter $x=1$, and
are closely related to the problem of reconstruction of trees, an
important setting in computer science and information theory, as was
first realized by M\'ezard and Montanari \cite{MezardMontanari06}. 

In the reconstruction on trees, a single (configuration) is spread
from the root of the tree to its leaves with some given rules and
noise level, and the task is then to reconstruct (infer) the value of
the root based on the configuration on the leaves. In particular in
a model with a factorized replica symmetric solution, constructing an
equilibrium configuration on the tree has a simple local
interpretation, as e.g. in \cite{MezardMontanari06}. The noise level
corresponds to the equilibrium temperature $T_e$.
This spreading construction is {\it precisely} the one we have used in
Fig.~\ref{fig0} for the XOR-SAT problem: starting from a value of the
spin chosen randomly, we have chosen iteratively the configuration of
the other variables randomly such that it violates the constraints
with probability $\epsilon$, corresponding to $T_e$. 

In the reconstructing on trees one thus applies BP starting form the leaves, using the values on the
boundaries as starting conditions, to generate the marginal
distributions of the variables within the tree. This is precisely what
we have done, the only difference is that in the reconstruction
formalism, one {\it knows} the value of $\epsilon$ that has been used
in order to construct the configuration on the tree, and therefore, one
used the same value in the BP equation in the recovery process. 

The states following problem is thus a generalization of the reconstruction on trees, where one first generates the configuration with a value $\epsilon_e$
(corresponding to a temperature $T_e$) but then apply the BP equation
with a different value $\epsilon_a$ (corresponding to a temperature
$T_a$).

\subsubsection{Reconstruction in a noisy channel without knowledge of
  the noise}
The method of states following can thus be viewed as a variant of
reconstruction on trees. In this interpretation the noise of the
channel is described by the inverse temperature $\beta_e$. If this
value is unknown, the task of reconstructing the values will be done
with an priori different temperature $\beta_a$: The behavior of
equations (\ref{pop_pl}) thus describe the reconstruction problem
where the noise level of the channel is not known.

Two interesting remarks can be done at this point.  First, it follow
from the maximum likelihood principle that the best chance to
reconstruct corresponds to $\beta_e=\beta_a$, and in fact, this give a
direct way to recover the noise value by maximizing the free energy:
On a tree, both the noise value and the marginal distribution can thus be
recovered in the reconstruction process. A second point is that, as we
will see from the behavior of the states following method,
in some cases although reconstruction is possible at
$\beta_a=\beta_e$ it might {\it not} be possible at some
$\beta_a>\beta_e$ (that is when trying to reconstruct by assuming the
number of mistake {\it smaller} than the actual value), which is
rather counterintuitive.

\subsubsection{A new bound for noisy reconstruction}
A last point we shall mention is that out method provides a simple way
to have new bounds on noisy reconstruction. Consider indeed a
problem where we have generated the boundaries with a noise level
$\beta_e$. We now use a very simple algorithm: we simply do the BP
recursion with $\beta_a \to \infty$, that is, assuming that no mistakes were done in the process. In that case, a simple bound of the
reconstruction threshold can be obtained by considering a probability
when the boundary condition directly imply the correct value of the root
\cite{Semerjian07,Sly08}. A similar procedure for $T_e=0$ was called
{\it naive reconstruction} in \cite{Semerjian07,Zdeborova08}. 

In the case $\beta_a \to \infty$ the equations simplify and can be cast in a set of coupled equations with two variables variables -- one being the probability that the value of the root is implied in the actual value, second being the probability that the value is implied wrongly. As long as the first probability is larger than the second, which is always the case in the cases we studies, this leads to new bounds on the noisy reconstruction problem. Some of these values are given in Sec.~\ref{sec:WP} for the XOR-SAT problem. In fact, it is simply the generalization of the {\it naive reconstruction} bound to the case of noisy channels \cite{Semerjian07,Sly08}. 

\subsection{Quiet planting: How to simulate impossible to simulate
  models?}
\label{sec:planting}
The construction we have described in this section is related to the notion of quiet
planting, which turns out to be a powerful way to perform
simulations for the mean field models, that would not be possible
otherwise.

Let us first stress that the thought experiment of choosing an
equilibrium configuration, that lead us to the derivation on above
equations, is feasible {\it only} on trees. On a random graph we would
encounter problems as soon as the long cycles would start appearing
when proceeding from a node playing the role of the root. Indeed,
choosing an equilibrium configuration on a given random graph below
the dynamical temperature $T_d$ requires as far as we know an
exponential time.

This difficulty, however, can be bypassed in the special cases where
the replica symmetric solution is factorized and in this case the
adiabatic evolution of states is realizable also on a single graph
instances and this has interesting algorithmic applications: it is
possible to create a graph {\it and} an equilibrated configuration at
the same time. The point is that when the RS solution is factorized it
is possible to create a typical random graph (from the ensemble under
consideration) and a configuration that is an equilibrium
configuration at temperature $T_e\ge T_K$ on that graph. This concept
was called "quiet planting" and was discussed by the authors in
\cite{KrzakalaZdeborova09,ZdeborovaKrzakala09}.

Let us first define quiet planting in the XOR-SAT problem and then
justify the above claimed properties. Planting an equilibrium
configuration in XOR-SAT at a given temperature (or equivalently at a
given energy) works as follows: First choose a random configuration of
spins $\{\sigma_i\}$, then choose a random instance from the ensemble
under consideration restricted to the fact that $(1-\epsilon) M$
constraints are satisfied by the chosen configuration of spins and
$\epsilon M$ are not satisfied. Thus, given the configuration, choose
at random $\epsilon M$ ($(1-\epsilon) M$ resp.) constraints out of all
the possible satisfied (unsatisfied resp.) constraints. The fraction $\epsilon$ is a function of the temperature $\beta_e$, and it is given by Eq.~(\ref{eps_rs}). This way for one given clause, out of the $2^{K-1}$ configurations that
do not satisfy that clause each will happen with probability $\epsilon
/ 2^{K-1}$, each satisfying configurations will appear with
probability $(1-\epsilon)/2^{K-1}$. If we condition on the value of
one variable contained in the clause we obtain probabilities
(\ref{sj_values}). Thus if we look on a finite neighborhood of a
variable in a very large planted hyper-graph we will obtain a
hyper-tree statistically identical to the one described in section
\ref{gedanken}. Consequently, Eq.~(\ref{pop_pl}) and (\ref{free_pl})
are the cavity equations describing the properties of the planted
graph. As typical properties of the graph follow from the solution of
Eq.~(\ref{pop_pl}), the planted graph and the planted configuration
will have the same typical properties as a random graph and an
equilibrium configuration. Hence, justification of the name "quiet"
planting, i.e. planting that does not change properties of the
ensemble.

Note here that the above argument was possible only because the
probabilities (\ref{sj_values}) were independent of the values of
messages $\{\psi^{b\to j}\}$. On the other hand, whenever this is the
case, i.e. whenever the replica solution is factorized, the above
argument is valid. In a general factorized case, i.e. for non-symmetric interactions or when disorder in the interactions is present, the planting procedure have to be slightly generalized. The marginal probabilities are used to plant a configuration with a proper number of each value (proper magnetization). Based on the RS solution one has to compute probabilities that a given type of constraint has given set of
values on its neighboring variables and plant the constraints
accordingly. An example of this general procedure at zero temperature
was described in detail in \cite{ZdeborovaKrzakala09}.

It shall be noted at this point that the equivalence between the
planted and purely random ensemble has been proven rigorously in
\cite{AchlioptasCoja-Oghlan08} in the zero temperature case in the
region of parameters where the second moment of the number of
solutions is smaller than some constant times the square of the first
moment. E.g. in the coloring problem for 3 colors the above holds till
average degree $c_q(3)=3.83$, for 4 colors until $c_q(4)=7.81$, to be
compared with the Kauzmann transition, also called the condensation
transition, $c_c(3)=4$ and $c_c(4)=8.46$. In the factorized models the annealed free energy,
$\log{\langle Z \rangle}$, starts to differ from the quenched one
${\langle \log Z \rangle}$ at the Kauzmann transition  \cite{ZdeborovaKrzakala07}, and then the
equivalence between the two ensembles breaks down.

Even before the proofs of \cite{AchlioptasCoja-Oghlan08} the
equivalence between the planted and random ensemble for XOR-SAT for
connectivities below the condensation transition was proven in
\cite{MontanariSemerjian06}, appendix A. In this special case the
equality of the annealed and quenched free energies is directly linked
to the absence of hyper-loops in the graph
\cite{MontanariSemerjian06}.  Authors of \cite{MontanariSemerjian06}
used the equivalence between the planted and random ensembles and the
fact that the planted configuration is one of the equilibrium
configurations as a handy tool to "equilibrate" their Monte-Carlo
simulations even at temperature where the usual equilibration is
impossible in feasible times. 

The fact of generating {\it for free} an equilibrium configuration
together with a typical realization of the disorder for all
temperature $T>T_K$ is extremely useful, and allows to perform
simulation that would be impossible otherwise: indeed for all the range
of temperatures $T_K\le T\le T_d$, it is unfeasible to find an
equilibrium configuration as soon as $N$ is not ridiculously
small. With the new method, this limitation disappears! The present
authors have already used this in
\cite{KrzakalaZdeborova09,ZdeborovaKrzakala09}, where one benefited
from the fact that Monte Carlo, belief propagation and other dynamical
procedures can be initialized in a truly equilibrium
configuration. Later on, in section \ref{sec:result-xor}, we will use this trick of quiet planting to confirm numerically, through
Monte-Carlo simulations, the results of the states following method.

\subsection{Reformulation using mapping on the Nishimori line}
\label{sec:Nish}
A last, and maybe the most striking, relation to previous
works arise when one considers Gauge transformations. Let us
consider, again, the $p$-spin model. The equations for adiabatic evolution of states can
be further simplified by exploiting a Gauge invariance. Indeed for any spin
$i$, the transformation
\be s_i \to -s_i \quad {\rm and} \quad 
J_{a} \to - J_{a} 
\quad {\forall} a \in \partial i 
\label{Gauge}
\ee 
keeps the Hamiltonian Eq.~(\ref{Ham_pspin}) invariant. As shown in
Fig.~{\ref{fig0}}, this allows to transform the equilibrium spin
configuration on any graph into a uniform one (all $s=1$), the disorder
distribution then changes from (\ref{disorder}) with $\rho=1/2$ to \be
Q_{\rm NL}(J)=\epsilon(T_e) \delta (J+1) + [1-\epsilon(T_e)] \delta (J-1)\,
, \label{Q_N} \ee 
where $\epsilon(T_e)$ is given by (\ref{eps_rs}).  Since
all $s=1$, there is no need to distinguish between the $+1$ and the
$-1$ sites. Eq.~(\ref{pop_pl}) reduces to the usual replica symmetric
cavity equation for a problem with mixed
ferromagnetic/anti-ferromagnetic interactions at temperature $T_a$
initialized in the uniformly positive state \be P_s(\psi) = \sum
Q_{\rm NL}(J) \sum_{\{l_i\}} q(\{l_i\}) \int \prod_{i=1}^{K-1}
\prod_{j=1}^{l_i} {\rm d}P(\psi^j) \, \delta[\psi-{\cal
  F}(\{\psi^j\},\beta_a)]\, , \quad \quad P_{\rm init}(\psi) =
\delta\left[ {\psi_1 \choose \psi_0} - {1\choose 0} \right] \,
. \label{pop_pl_N} \ee

The distribution of interactions (\ref{Q_N}) is given by the
Nishimori-like \cite{Nishimori81,NishimoriBook01} relation between
temperature $T_e$ and density of anti-ferromagnetic couplings
$\epsilon$, and arises because at $T_a=T_e$ the overlap with the
equilibrium configuration, playing a role of magnetization $m$ in the
Gauge transformed model, is equal to the overlap $q$ between two
typical configurations from the state, a well known properties on the
so-called Nishimori line (that is the line defined by the Nishimori
relation in the temperature/ferromagnetic bias plane).

The Gauge invariance have thus transformed the task of following an
equilibrium state in a glassy model into describing the evolution of
the ferromagnetic state in a ferromagnetically biased model with the
standard cavity approach. As we will derive in Sec.~\ref{RS_pfollow}
adiabatic evolution of states in the fully connected $p$-spin model for $T_e\ge
T_K$ is thus equivalent to solving the $p$-spin model with an
additional effective ferromagnetic coupling (\ref{J_eff}), and one can
thus readily take the solution of the $p$-spin in the literature,
e.g. \cite{NishimoriWong99,NishimoriBook01}, to obtain properties of
equilibrium states.

Similar mapping exist for all mean field models where the replica
symmetric solution is factorized (see for instance
\cite{NishimoriStephen83} for glassy Potts models), however, the
resulting model is not always very natural nor already known. For the
$p$-spin model the evolution of a glassy state being equivalent to
melting of the ferromagnetic state on the Nishimori line has deep
consequences for the physics of glasses, as will be discussed
elsewhere \cite{KrzakalaZdeborova10}.

\section{Evolution of states: General cavity equations for any
  temperature}
\label{sec:general}
In this section we introduce a method of states following that is
suitable at any temperature, and where replica symmetry breaking is
thus taken into account. The method is set for any general 1RSB
states, at any value of the Parisi parameter $x$ and any temperature,
as long as the corresponding states are stable towards further steps
of replica symmetry breaking. Stability of the following equations
towards RSB for different values of $\beta_a$ is interesting and will
be discussed later.

\subsection{Adiabatic evolution of 1RSB states}
In order to understand the general equations for the adiabatic evolution of states, let
us first briefly recall how are derived the cavity 1RSB equations that describe the equilibrium states. We work at inverse temperature $\beta_e$ where many states exist, each of them has a corresponding BP fixed point, i.e. a message $\psi$ on each link. As described in Sec.~\ref{sec:1RSB}, the 1RSB
method uses the distribution of messages $P(\psi,\beta_e)$
over all states with a given free energy $f$. In order to select the free energy, we consider the BP recursion in every possible
state, but we reweight each state according to the
Boltzmann weight $Z^x(\psi,\beta_e)$. This leads to Eq.~(\ref{1RSB}). 

One intuitive way to understand Eq.~(\ref{1RSB}) is to think about the problem on a tree and consider many possible
boundary conditions $P_{\rm init}(\psi,\beta_e)$. In order to select boundary conditions that lead to the state of free energy $f$ we reweight $P(\psi,\beta_e)$ at each steps with the Boltzmann weight $Z^x(\psi,\beta_e)$. Eventually, for different $x$ such fixed point will describe states with different free energy $f$. For more details on this derivation see \cite{KrzakalaMontanari06,MontanariRicci08,Zdeborova08}. 

In order to write the equations for the adiabatic evolution of the
1RSB states, we first need to describe the state via Eq.~(\ref{1RSB}),
and second we use another distribution $\tilde P(\psi,\tilde \psi)$
that describes the same state at a new temperature $\beta_a$. The
equilibrium states at temperature $\beta_e$ arise if one uses the
reweigthing $Z^x(\psi,\beta_e)$. The probability distribution $\tilde
P(\psi,\tilde \psi)$ thus needs to be reweighted with the same factor
$Z^x(\psi,\beta_e)$ in the state following method. Thus, the
generalization of the 1RSB equations to the state following is a
recursion on both $P(\psi,\tilde \psi)$ and $\tilde P(\psi,\tilde
\psi)$ as follows 
\bea \tilde P^{a\to i}(\psi^{a\to i},\tilde
\psi^{a\to i}) = \frac{1}{{\cal Z}^{a\to i}(\beta_e,\beta_a)} && \int
\prod_{j\in\partial a \setminus i} \prod_{b\in \partial j\setminus a}
{\rm d}\tilde P^{b\to j}(\psi^{b\to j},\tilde \psi^{b\to j})\,
\left[Z^{a\to i}(\{\psi^{b\to j} \}, \beta_e) \right]^x \nonumber \\
&&\delta[\psi^{a\to i} - {\cal F}(\{\psi^{b\to j} \},\beta_e)] \,
\delta[\tilde \psi^{a\to i} - {\cal F}(\{\tilde \psi^{b\to j}
\},\beta_a)] \, .
\label{1RSB_pl_2}
\eea The distribution $\tilde P$ is initialized as \be \tilde P^{a\to
  i}(\psi^{a\to i},\tilde \psi^{a\to i}) = P^{a\to i}(\psi^{a\to i})
\, \delta(\psi^{a\to i}-\tilde \psi^{a\to i})\, , \ee where the
$P^{a\to i}(\psi^{a\to i})$ is the solution of the usual 1RSB
equations (\ref{1RSB}) describing the equilibrium state at an inverse
temperature $\beta_e$. Equation (\ref{1RSB_pl_2}) then describes
adiabatic evolution of this state at an inverse temperature
$\beta_a$. Note that the reweighting factor $Z^x$ comes from the
messages $\psi$ as the inverse temperature $\beta_e$ -- again, this is
the key element assuring that we are looking into the same state at a
different temperature.

The internal free energy of the state at temperature $\beta_a$ is given in terms of node and link contributions, as usual in the 1RSB cavity method:
\be
       -\beta_a F(\beta_a) = - \beta_a \sum_a  F^{a+\partial a}(\beta_a) + \beta_a \sum_i (l_i-1) F^i(\beta_a)\, ,
\ee
where 
\bea
   - \beta_a  F^{a+\partial a}(\beta_a)  &=& \frac{\int  \prod_{i\in\partial a} \prod_{b\in \partial i\setminus a} {\rm d}\tilde P^{b\to i}(\psi^{b\to i},\tilde \psi^{b\to i})\left[ \log{Z^{a+\partial a}(\{\tilde \psi^{b\to i} \},\beta_a) } \right]\left[Z^{a+\partial a}(\{\psi^{b\to i} \}, \beta_e) \right]^x}{\int  \prod_{i\in\partial a} \prod_{b\in \partial i\setminus a} {\rm d}\tilde P^{b\to i}(\psi^{b\to i},\tilde \psi^{b\to i}) \left[Z^{a+\partial a}(\{\psi^{b\to i} \}, \beta_e) \right]^x} \, , \label{free_a}\\
   - \beta_a   F^i(\beta_a) &=& \frac{\int  \prod_{a\in \partial i} {\rm d}\tilde P^{a\to i}(\psi^{a\to i},\tilde \psi^{a\to i})  \left[ \log{Z^{i}(\{\tilde \psi^{a\to i} \},\beta_a)}\right] \left[Z^{i}(\{\psi^{a\to i} \}, \beta_e) \right]^x}{\int  \prod_{a\in \partial i} {\rm d}\tilde P^{a\to i}(\psi^{a\to i},\tilde \psi^{a\to i}) \left[Z^{i}(\{\psi^{a\to i} \}, \beta_e) \right]^x} \, . \label{free_i}
\eea
And for the corresponding energy we have
\be
   E(\beta_a)  =  \sum_{a} \frac{\int  \prod_{i\in\partial a} \prod_{b\in \partial i\setminus a} {\rm d}\tilde P^{b\to i}(\psi^{b\to i},\tilde \psi^{b\to i}) E^{a+\partial a}(\{\tilde \psi^{b\to i} \},\beta_a)  \left[Z^{a+\partial a}(\{\psi^{b\to i} \}, \beta_e) \right]^x}{\int  \prod_{i\in\partial a} \prod_{b\in \partial i\setminus a} {\rm d}\tilde P^{b\to i}(\psi^{b\to i},\tilde \psi^{b\to i}) \left[Z^{a+\partial a}(\{\psi^{b\to i} \}, \beta_e) \right]^x} \, . \label{energy_gen}
\ee

Eqs.~(\ref{1RSB_pl_2}-\ref{energy_gen}) are written for a given instance of the problem. It is instructive to describe how to solve them on average over the graph ensemble using the population dynamics method \cite{MezardParisi01}. We need to keep a population (representing the average over graph) of couples of messages (one for $\psi$, the other for $\tilde \psi$). Then the population is iterated in the exact same way as in the usual case \cite{MezardParisi01}, the whole population of couples is reweighted using the reweighting factor  $[Z^{a+\partial a}(\{\psi^{b\to i} \}, \beta_e)]^x$ computed from the elements $\psi$ at inverse temperature $\beta_e$. 

\subsection{When states themselves divide into states}
\label{sec:stab}

So far we supposed that the state we are following does not exhibit an instability towards replica symmetry breaking. This assumption may break when temperature $T_a$ is low enough. To check for the local stability we can use a variant of any known method, see e.g. \cite{MontanariParisi04} or appendix C of \cite{Zdeborova08}.
One of the methods simplest to implement in the population dynamics is the monitoring of deviation of two replicas. For that we first need to find an equilibrated population at a temperature $T_a$, then we copy this population and introduce a small noise. Further the two copies (replicas) are updated with the same random choices and the deviation of the two is monitored. If the deviation is going to zero the state is locally stable, if the deviation is growing the state is not stable towards replica symmetry breaking, that is the state has the tendency to divide into many smaller states. This second case can be treated in the following way.  

If the state to be followed is not stable towards replica symmetry breaking then applying the 1RSB method within this state shall lead better result about its behavior. The following equations together with (\ref{1RSB},\ref{1RSB_pl_2}) describes the method
\be
      P_2^{a\to i}(\tilde P^{a\to i}) =  \frac{1}{{\cal Z}_2^{a\to i}} \int  \prod_{j\in\partial a \setminus i} \prod_{b\in \partial j\setminus a} {\rm d} P_2^{b\to j}(\tilde P^{b\to j}) \left[{\cal Z}^{a\to i}(\beta_e,\beta_a) \right]^{x_2} \delta\left[\tilde P^{a\to i} - {\cal F}_2(\{\tilde P^{b\to j} \},\beta_e,\beta_a)\right] \, ,
\ee
where the functional ${\cal F}_2(\{\tilde P^{b\to j} \},\beta_e,\beta_a)$ is define by Eq.~(\ref{1RSB_pl_2}). Said in words on every edge next to the population corresponding to Eq.~(\ref{1RSB}) one would have to keep a population of populations, each corresponding to a sub-state. Each of these populations would be reweighted using the reweighting from Eq.~(\ref{1RSB}). A second reweighting with Parisi parameter $x_2$ would have to be done on the level of populations. On top of all that in the non-factorized cases a population of these object would have to be kept to account for the average over the graph. Numerical resolution of such equations becomes involved and we let their implementation for the diluted models for future works.

\section{First application:  Adiabatic evolution of states in the fully connected
  $p$-spin model}
\label{pspin_th}
Now that we have presented the method for adiabatic evolution of states, let us
show how does the solution of the equations behave and what can be
learned about the physics of the $p$-spin problem. We will also
describe here the connection of the states following method to the Franz-Parisi
potential \cite{FranzParisi95,FranzParisi97} and with the physics on
the Nishimori line.

One of the simplest cases to which Eq.~(\ref{pop_pl}) can be applied is the
fully connected $p$-spin model. The static replica solution of the model is in
\cite{GrossMezard84}.  In appendix \ref{pspin_rem} we show how to
re-derive the replica symmetric equations for the fully connected
$p$-spin model as a limit of infinite connectivity of the cavity
(belief propagation) equation (\ref{pop_pl}). Here we only summarize
the equations needed to explain how to obtain the solution for state
following.

\subsection{The equilibrium solution of the fully connected $p$-spin model}
As discussed in details in appendix \ref{pspin_rem}, BP simplifies in
the fully connected model where the amplitudes of all interactions are
small, and become
 \be m^{i\to a} = \tanh{\left(
    \beta \sum_{b\in\partial i \setminus a} J_b \prod_{j\in \partial
      b\setminus i} m^{j\to b} \right)} \, .
\label{cavity}
\ee
where $m$ the local cavity magnetization. The replica symmetric
solution can then be written in terms of the distribution of such
cavity magnetization that are Gaussian because of the central limit
theorem \cite{MezardParisi85b,MezardParisi85c}: we thus need only the
average magnetization $m=\langle m^{i\to a} \rangle$ and the average
overlap between configuration $q=\langle (m^{i\to a})^2 \rangle$. The
recursion reads:
\bea
m&=&   \int_{-\infty}^{\infty} {\cal D}y  \tanh{\left( \beta J y \sqrt{p q^{p-1}/2}+ \beta J_0 p m^{p-1} \right)} \, ,\label{eq_m}\\
q&=& \int_{-\infty}^{\infty} {\cal D}y \tanh^2{\left( \beta J y
    \sqrt{p q^{p-1}/2}+ \beta J_0 p m^{p-1} \right)}\, ,\label{eq_q}
\eea where we call ${\cal D}y={\rm d}y \, e^{-\frac{y^2}{2}}
/\sqrt{2\pi}$ the Gaussian integration. The free energy is a function
of the fixed point of the above equations: \be -\beta f =
\frac{1}{4}\beta^2 J^2 (p-1) q^p - \beta J_0 (p-1) m^p +
\frac{1}{4}\beta^2 J^2 - \frac{1}{4} \beta^2 J^2 p \, q^{p-1} + \int
{\cal D}y \log{2\cosh{\left( \beta J y \sqrt{p \, q^{p-1}/2}+ \beta
      J_0 p m^{p-1}\right)}}\, ,\label{free_en} \ee and for the
replica symmetric energy density we have \be e = \frac{\partial (\beta
  f)}{\partial \beta} = -J_0 m^p -\frac{1}{2} \beta J^2 (1-q^p)\,
. \label{p_energy} \ee

The 1RSB solution with the value of Parisi parameter $x$ is obtained
in a similar way, and the corresponding fixed point equations are 
\bea
m &=& \int {\cal D}u  \left[  \frac{\int {\cal D}v \, \cosh^x{(\beta G)} \, \tanh{(\beta G)} }{\int {\cal D}v  \, \cosh^x{(\beta G)}}\right] \, , \label{1RSB_m}  \\
q_1 &=& \int {\cal D}u  \left[  \frac{\int {\cal D}v  \, \cosh^x{(\beta G)}\,  \tanh^2{(\beta G)} }{\int {\cal D}v  \, \cosh^x{(\beta G)}} \right]\, , \label{1RSB_q1} \\
q_0 &=& \int {\cal D}u \left[ \frac{\int {\cal D}v \, \cosh^x{(\beta
      G)} \, \tanh{(\beta G)} }{\int {\cal D}v \, \cosh^x{(\beta G)}}
\right]^2 \, , \label{1RSB_q0} \eea 
where 
\be 
G=J u\sqrt{\frac{p}{2}q_0^{p-1}}  + Jv \sqrt{\frac{p}{2}q_1^{p-1}-\frac{p}{2}q_0^{p-1}}  +J_0 p m^{p-1}
\ee is a sum of two Gaussian
variables, and ${\cal D}$ is the Gaussian integral. The parameter
$q_1$ is the average self-overlap and $q_0$ the average overlap
between states. The 1RSB Parisi (replicated) free energy reads \bea -
\beta x \Phi(\beta,x) &=& - x \beta J_0 p m^p + \frac{1}{4} (1-x)x
(p-1) \beta^2 J^2 q_1^p + \frac{1}{4} x^2 (p-1)\beta^2 J^2 q_0^p +
\frac{1}{4}\beta^2 J^2 x - \frac{1}{4}\beta^2 J^2 x p q_1^{p-1} \nonumber \\ &+&
x\log{2} + \int {\cal D}u \log{\int {\cal D}v \, \cosh^x{\beta G}} \,
,\label{1RSB_free} \eea the free energy is derived as
$f(\beta)=\partial \Phi(\beta,x)/\partial x$.

\subsection{Equations for adiabatic evolution of states for $T_e\ge T_K$}
\label{RS_pfollow}
Let us give a derivation of state following equations for the fully connected $p$-spin
model using the equivalence with the planted ensemble. We think about the fully connected model as the large connectivity version of the diluted model and used the planting procedure described in sec.~\ref{sec:planting}. We first plant
an equilibrium configuration at inverse temperature $\beta_e$, then
initialize the belief propagation equations in this solution and
iterate to a fixed point at another inverse temperature
$\beta_a$. When the temperature of planting is larger than the
Kauzmann temperature, $\beta_e<\beta_K$, then the planting can be done
in a very natural way. One first takes the replica symmetric energy at
$\beta_e$ and computes the corresponding fraction $\epsilon$ of
interactions that are not satisfied at that energy. Then when
constructing the planted graph one first chooses a random
configuration, the sign of interactions is then chosen in
such a way that fraction $\epsilon$ of them is unsatisfied and
$1-\epsilon$ satisfied.

The value of $\epsilon$ in the fully connected $p$-spin model is
computed as follows. Let us assume $J_0=0$, as this is really the case
we are interested in, recall that we rescale the interactions in the fully connected model as $\langle J_a^2 \rangle= J^2 p!/(2N^{p-1})$, hence $J_a=\pm J\sqrt{p!/(2N^{p-1})}$, there is total of $N^p/p!$ interactions. The energy we want to achieve is given by
(\ref{p_energy}), hence $\epsilon$ has to satisfy \be \epsilon =
\frac{1}{2} - \beta_e \frac{J(1-q^p)}{2} \sqrt{\frac{p!}{2N^{p-1}}}\,
.  \ee Moreover as we consider only $\beta_e<\beta_K$, in the $p$-spin
model this means that $q=0$.

Now let us keep in mind the above planting, moreover consider that
spin $i$ was planted $+1$ (without loss of generality) and look back
at equation (\ref{cavity}), considering $J_0=0$. The terms in the sum
in the argument of the $\tanh{}$ are independent (by the assumption of
replica symmetry within the planted state) and their statistics is
thus ruled by the central limit theorem. Thus our aim is to compute
the mean and variance of the argument. If the interactions would not
be correlated with the planted configuration the mean would be zero
(remind we have $J_0=0$). If we satisfy every interaction with probability $1-\epsilon$, there will be $1-2\epsilon$ more satisfied interactions that unsatisfied ones. These $1-2\epsilon=2 \beta_e J \sqrt{p!/8}/ N^{(p-1)/2}$ interactions are biasing spin $i$
in the direction $+1$. The mean of the argument of the $\tanh{}$ is
thus \be \mu= \beta_a \frac{N^{p-1}}{(p-1)!} \frac{2\beta_e J
  \sqrt{p!/8}}{N^{(p-1)/2}} \sqrt{\frac{J^2 p!}{2 N^{p-1}}} m^{p-1} =
\beta_a \, \beta_e J \sqrt{p!/8} J \frac{\sqrt{2 p!}}{(p-1)!} m^{p-1}
= \beta_a \left[ \beta_e J^2 /2 \right] p \, m^{p-1}\, .  \ee 
The planting only influences the directions of the interactions, thus in
the variance computation nothing changes and we have again \be \sigma
= \beta_a^2 J^2 p\, q^{p-1}/2\, .  \ee Thus parameters $m=\langle
m^{i\to a} \rangle$, and $q=\langle (m^{i\to a})^2 \rangle$ are ruled
again by equations (\ref{eq_m}-\ref{eq_q}) with inverse temperature
$\beta_a$ and effective coupling \be J^{\rm eff}_0=\beta_e J^2/2\,
. \label{J_eff} \ee Parameter $m$ now measures how far from the
equilibrium planted configuration is a typical configuration at
$\beta_a$.  Also the free energy equation (\ref{free_en}) applies to
this case with inverse temperature $\beta_a$ and effective $J_0^{\rm
  eff}$ given by (\ref{J_eff}). The free energy here is, however, free
energy of the planted state and thus the complexity, defined by (\ref{Sigma}) with $x=1$, can be computed
considering the difference \be \Sigma = -\beta_a f(J_0=0) + \beta_a
f(J_0^{\rm eff})\, . \label{comp} \ee

Consequently all the physics of adiabatic evolution of states above
the Kauzmann temperature for the $J_0=0$ model can be induced from the
known phase diagram of the $J_0\neq 0$ model and Eq.~(\ref{J_eff}) is
the Nishimori line condition \cite{Nishimori81,NishimoriBook01} for
the fully connected $p$-spin model with a nonzero $J_0$: This
illustrates the general equivalence we have discussed in
Sec.~\ref{sec:Nish} between the states following method and the original
model on the Nishimori line.

The dynamical temperature can be interpreted as the spinodal point of
the planted state, thus at $T_d$ we start to have a nontrivial
solution if $\beta_e=\beta_a$. And at $T_K$ the complexity
(\ref{comp}) becomes negative at $\beta_e=\beta_a$. Iterating
Eqs.~(\ref{eq_m}-\ref{eq_q}) we indeed obtain values of the two
critical temperatures as known from the 1RSB solution of the $p$-spin
model. For $p=3$ we have $T_K=0.6513$ and $T_d=0.6815$.

In the case of fully connected $p$-spin model the state following when
they start to be unstable (divide into sub-states), described in generality in
Sec.~\ref{sec:stab}, can be done easily (at least on the 1RSB level)
by using the mapping on a model with effective ferromagnetic
coupling. Again, for the model with $J_0=0$ the 1RSB solution inside a
state is equivalent to the standard 1RSB solution in a model with $J^{\rm eff}_0=\beta_e J^2/2$, Eqs.~(\ref{1RSB_m}-\ref{1RSB_q0}).

\subsection{Equations for adiabatic evolution of states for $T_e< T_K$}
\label{sec_pspin_1RSB}
We now want to follow states below the Kauzmann temperature, or
metastable states above $T_K$ (i.e. at a Parisi parameter
$x\neq 1$). As far as we know there is no "planting" interpretation
for this case, the mapping into ferromagnetically biased model on the Nishimori line does not work either in this case. The underlying equilibrium measure at $T_e<T_K$  become more complicated, the derivation then must follow by rewriting equations (\ref{1RSB},\ref{1RSB_pl_2}) in the large connectivity limit.

For simplification we note that in the $p$-spin model with $J_0=0$ we have $m=q_0=0$ thus the only non-trivial parameter describing the 1RSB state we aim to follow is $q_1$, given by the equation (\ref{1RSB_q1}), this summarizes Eq.~(\ref{1RSB}).  

To rewrite Eq.~(\ref{1RSB_pl_2}) we need to introduce overlap $\tilde q_1$ and a correlation between the two populations 
\bea 
      \tilde q_1 &=& \int {\cal D}Q(P)  \left[  \frac{\int {\rm d}\tilde P(m^{i\to a},\tilde m^{i\to a})  \, Z(\{m^{i\to a}\},\beta_e)^x \,  (\tilde m^{i\to a})^2 }{\int {\rm d}P(m^{i\to a})  \, Z(\{m^{i\to a}\},\beta_e)^x} \right] 
\equiv \langle \langle (\tilde m^{i\to a})^2 \rangle_{\tilde P, P} \rangle_Q\, , \\
     c &=& \int {\cal D}Q(P)    \left[  \frac{\int {\rm d}\tilde P(m^{i\to a},\tilde m^{i\to a}) \, Z(\{m^{i\to a}\},\beta_e)^x \,  m^{i\to a} \tilde m^{i\to a} }{\int {\rm d}P(m^{i\to a})  \, Z(\{m^{i\to a}\},\beta_e)^x }\right] 
\equiv \langle \langle m^{i\to a} \tilde m^{i\to a} \rangle_{\tilde P,P} \rangle_Q \, .
\eea 
In appendix \ref{app:1RSB} we remind the derivation of the standard 1RSB equations for the fully connected $p$-spin model. What we nee here goes in a very similar manner. We define  
\be X= \sum_{b\in\partial i \setminus a} J_b
\prod_{j\in \partial b\setminus i} m^{j\to b}\, , \quad \quad  \tilde X= \sum_{b\in\partial i \setminus a} J_b
\prod_{j\in \partial b\setminus i} \tilde m^{j\to b}\, . \ee
and obtain similarly as in appendix \ref{app:1RSB}
\bea
     \sigma^2_{1}&=&\langle \langle X^2 \rangle_{P,\tilde P} \rangle_Q =   J^2 \frac{p}{2} q_1^{p-1}\, , \\
     \tilde \sigma^2_{1} &=& \langle \langle \tilde X^2 \rangle_{P,\tilde P} \rangle_Q = J^2 \frac{p}{2} \tilde q_1^{p-1}\, ,\\
     \rho &=& \frac{\langle \langle X \tilde X \rangle_{P,\tilde P} \rangle_Q}{ \sigma_{1} \tilde \sigma_{1} }=  \left( \frac{c}{\sqrt{q_1\tilde q_1}} \right)^{p-1}\, .
\eea
The final self-consistent equations for $\tilde q_1$ and $c$ are then 
\bea
  \tilde q_1 &=&  \frac{\int {\cal D}\{u,v\}  \, \cosh^x{(\beta_e v)}\,  \tanh^2{(\beta_a u)} }{\int {\cal D}v  \, \cosh^x{(\beta_e v)}}\, ,  \\
  c  &=&   \frac{\int {\cal D}\{u,v\} \, \cosh^x{(\beta_e v)}\,  \tanh{(\beta_e v)} \tanh{(\beta_a u)} }{\int {\cal D}v  \, \cosh^x{(\beta_e v)}} \, ,
\eea
where 
\bea
   {\cal D}\{u,v\}  &=&  \frac{1}{2\pi \sigma_1 \tilde \sigma_1 \sqrt{1-\rho^2}} \exp{\left[-\frac{1}{2(1-\rho^2)}\left( \frac{u^2}{\tilde \sigma^2_1} +  \frac{v^2}{\sigma^2_1}  -  \frac{2 \rho u\,  v}{\sigma_1\tilde \sigma_1} \right)\right]} {\rm d}u  \,  {\rm d}v \, , \\
   {\cal D}v &=& \frac{1}{\sigma_1 \sqrt{2\pi}}  \exp{\left(-\frac{v^2}{2\sigma_1^2}\right)} {\rm d}v  \, .
\eea

The free energy of the followed state can be obtained by plugging
expressions (\ref{z_site}) and (\ref{z_link}) into
(\ref{free_a}-\ref{free_i}) and getting \be -\beta_a f =
\frac{1}{4}\beta_a^2 J^2 (p-1) \left(\tilde q_1^p - 2x
  \frac{\beta_e}{\beta_a} c^p\right ) + \frac{1}{4}\beta_a^2 J^2 -
\frac{1}{4}\beta_a^2 J^2 p \tilde q_1^{p-1} + \frac{\int {\cal
    D}\{u,v\} \, \cosh^x{(\beta_e v)}\, \log{[2 \cosh{(\beta_a u)}
    ]}}{\int {\cal D}v \, \cosh^x{(\beta_e v)}}\, .  \ee The energy is
then obtained by deriving $e=\partial (\beta_a f)/\partial \beta_a$. Again, these equation are similar to standard replica equation with a kind ferromagnetic bias, but do not have as simple form as was given by the mapping on Nishimori line for $T_e>T_K$.

\subsection{What happens when one follows states: Turning cartoons
  into data}
\label{sec:plot-pspin}
So far we were describing ideas and the formalism for the method of
adiabatic following of Gibbs states. In the remaining subsections we
describe and discuss results which can be obtained about the energy
landscape and the structure of states for the fully connected $p$-spin
model, based on the previously derived equations.

One obvious application of the states following method is to compute
how does the energy of equilibrium states evolve with
temperature. Such energy-temperature diagrams (volume or entropy is
sometimes plotted on the $y$-axes, or density is plotted as a function
of the pressure) appear in many works about glassy systems
\cite{BouchaudCugliandolo98}, for recent examples see
\cite{KrzakalaKurchan07,MariKrzakala09}. Except for a few very
simplistic models such as the spherical $p$-spin models
\cite{CugliandoloKurchan93,BarratFranz97,CaponeCastellani06}, the
random energy or random entropy model \cite{KrzakalaMartin02} or the
random subcube model \cite{MoraZdeborova07} (where the dynamics is
exactly solvable), all these diagrams were drawn as qualitative
schemes, or as results of Monte-Carlo simulations. Moreover, in the
field of glassy dynamics, the energy landscape is often cartooned by
drawing many valleys of different sizes and depths. The states
following method allows to draw the above mentioned figures with {\it
  actual quantitative} data for any model solvable via the cavity or
replica method.

\begin{figure}[!ht]
  \includegraphics[width=0.495\linewidth]{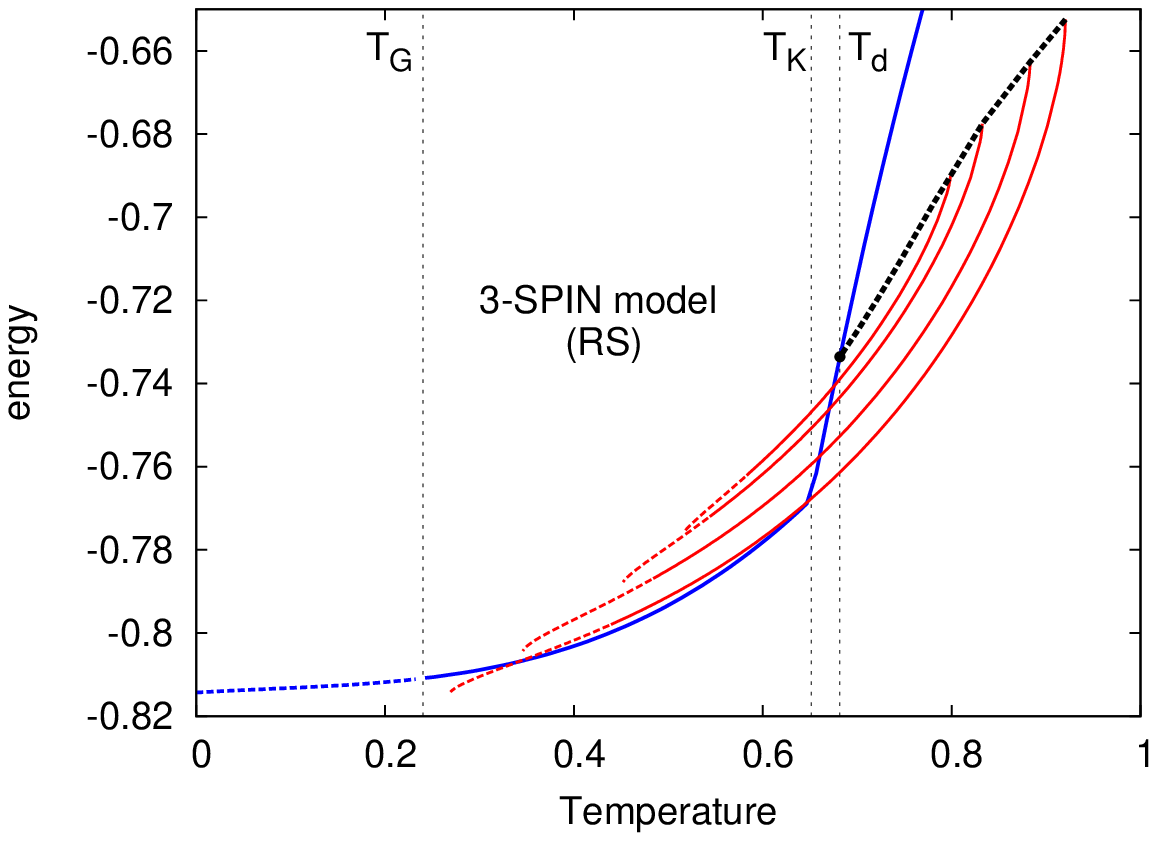}
  \includegraphics[width=0.495\linewidth]{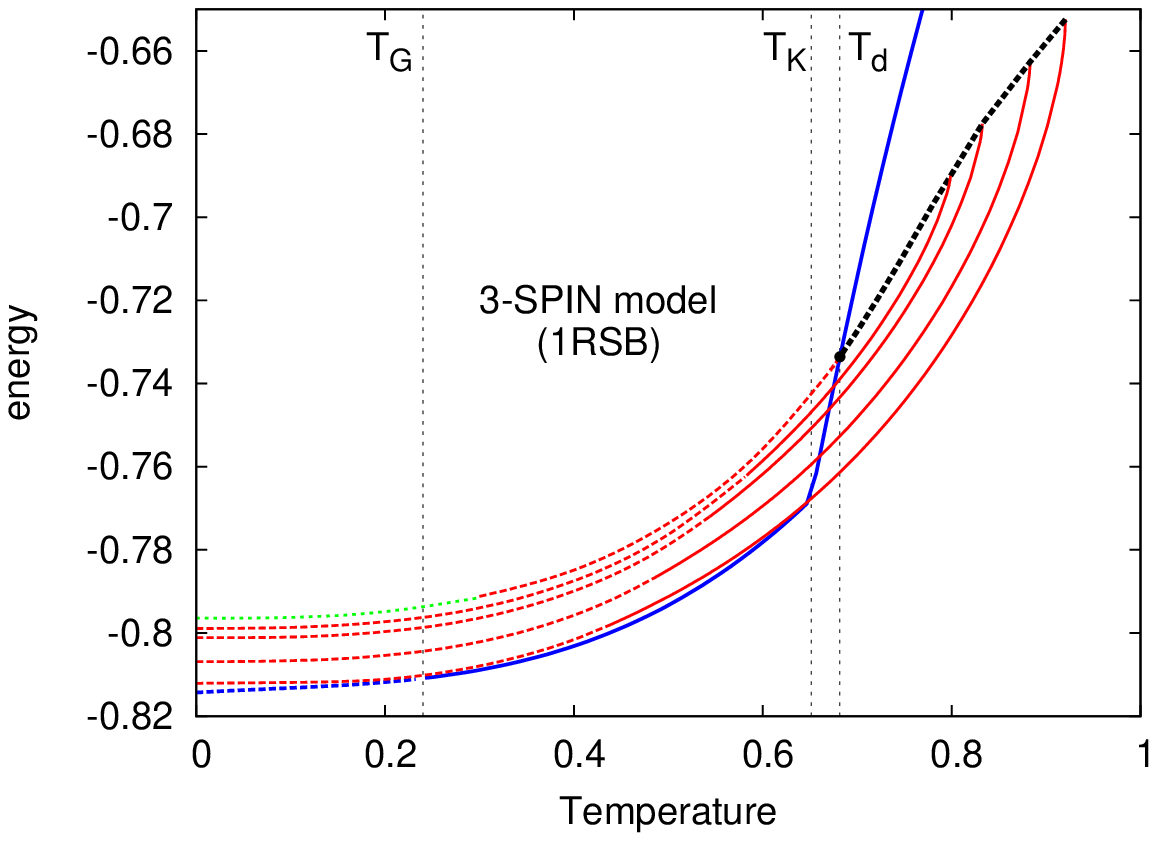}
  \caption{\label{fig_follow} (color online): Adiabatic evolution of states in the
    fully connected $3$-spin model, where $T_d=0.6815\ldots$ and
    $T_K=0.6513\ldots$. The continuous blue curve shows the equilibrium
    energy $e_e(T)$ of the model versus the temperature (with the
    Gardner transition towards full RSB phase at low temperature). We have
    applied the states following method and the red curves (five
    roughly parallel curves) mark the adiabatic evolution of states
    from equilibrium with $T_e=0.6815,0.675,0.67,0.66,0.6513$ for
    temperatures $T_a \neq T_e$. Left: This is the RS result where we
    follow states using the RS ansatz, Eqs.~(\ref{eq_m}-\ref{eq_q}) and (\ref{p_energy}). Upon warming, the states exist
    until meeting a spinodal point at much larger temperature. Upon
    cooling the states can be followed until they become unstable
    against 1RSB (dashed), eventually a non-physical spinodal point
    makes the RS solution vanish. Note that the equilibrium state at $T_e=T_d$
    vanishes as soon as $T_a<T_e$ and is thus not even seen on this
    picture.  Right: We follow states using the
    1RSB ansatz (the dashed part of the red curves), Eqs.~(\ref{1RSB_m}-\ref{1RSB_free}). The 1RSB solution is actually also
    unstable to further steps of RSB and (most probably) the full RSB
    should be used instead. The 1RSB is therefore a (good) lower bound
    to the correct result. Note also that the non-physical spinodal
    points for low temperatures do not appear and that the 1RSB
    approach has mostly cured the problem, with the exception of
    the states corresponding to $T_e \approx T_d$. The green dotted
    line is an example of a region where even the 1RSB equations do not
    have a physical solution. The green dotted line is a lower bound
    computed by the construction suggested in Sec.~\ref{sec_approx}.}
\end{figure}

The left part of Fig.~\ref{fig_follow} shows how does the energy of states depend on
the temperature $T_a$. The blue line is the
equilibrium energy of the system at a given temperature. This curve is
divided into four parts, the part above the dynamical temperature
$T_d$ represents a high temperature liquid phase. Between the
dynamical $T_d$ and Kauzmann $T_K$ temperature is the dynamical glass
phase, where the free energy or energy are still given by the liquid
result, but the equilibrium state is divided into exponentially many
Gibbs states. Below the Kauzmann temperature $T_K$ the static line is
obtained by solving the 1RSB equations (\ref{1RSB_m}-\ref{1RSB_free}), at this point its derivative
changes discontinuously. This 1RSB solution becomes unstable below
the Gardner temperature $T_G$ \cite{Gardner85,MontanariParisi04} below which the line is dashed, as it is
no longer exact, the correct FRSB energy would be higher.

Each of the red lines (five roughly parallel lines crossing the
figure) is obtained by following the evolution of one of the
exponentially many states equilibrium at some $T_K<T_e<T_d$. When the
state is heated the energy grows up to a spinodal point where the
state disappears, i.e. when the only solution of equations
(\ref{eq_m}-\ref{eq_q}) with $J_0$ given by (\ref{J_eff}) has $m=0$. As $T_e$ approaches $T_d$ the
spinodal point is at lower and lower temperatures, states very near to
$T_d$ are lost almost immediately when heated. This is an
interesting result as in the spherical models, the state at $T_d$
could be heated to much larger temperature: that is an
unphysical property of the spherical model that disappears in the Ising
model we have considered here.

When the state is cooled down, the energy is decreasing, but slower than
the equilibrium energy. As soon as the temperature changes the state
goes out of equilibrium: this corresponds to the notion of glassy
states trapping the dynamics up in the energy landscape. In
Fig.~\ref{fig_follow} left we plotted the energy of states supposing
they are stable against replica symmetry breaking. We found, however,
that this was not always the case and thus denoted the unstable, thus
unphysical, part of the curves by dashing. Indeed, the dashed
part of the lowest state curve even crosses the equilibrium line, which is
unphysical, a clear sign that the replica symmetry is broken. The left ends of
the red lines (five roughly parallel lines) correspond to another
spinodal point, in the sense that the non-trivial, $m\neq 0$, solution
of the RS state following equations (\ref{eq_m}-\ref{eq_q}) ceases to exist. This spinodal point does, however, not have a physical interpretation, as the states are unstable towards RSB in that region. We will see in Sec.~\ref{sec:ferro} that this unphysical spinodal is related to a known problem in the spin-glass with ferromagnetic bias.

The right hand side of Fig.~\ref{fig_follow} depicts the same quantities as the
left hand side, the difference is that for the adiabatic evolution of states
we used the 1RSB equations (\ref{1RSB_m}-\ref{1RSB_free}). The
part where this changes the result is distinguished by dashing. We checked that even the
1RSB description of the states is not stable towards more step of RSB,
so that the exact description of the dashed parts requires a full RSB solution.
The 1RSB result, however, gives a much better ---and physical--- approximation of the
correct behavior. Still, for the upper states a non-physical spinodal
point remains; this can be seen on the highest red curve
when its dashed part finishes and turns into dotted green, we will
describe in Sec.~\ref{sec_approx} how the green dotted line was obtained. 
Obtaining the FRSB is only a technical problem of solving the corresponding equations as the mapping to the partly ferromagnetic model (\ref{J_eff}) is valid on any level of RSB.

Note that our results correspond well to the solution of the
dynamical equations in the spherical approximation, where indeed the
transition towards more steps of replica symmetry breaking was observed
\cite{BarratFranz97,CaponeCastellani06}: we expect actually this
behavior to be quite universal and to be observed in any
spin glass model with an 1RSB equilibrium solution.

One comment is in order here. The reader familiar with the
phenomenology of the $p$-spin model will certainly find many similarities between our results and the one advocated in \cite{MontanariRicci03}. In that
work, the authors considered like us the states that are at higher free
energies than the equilibrium ones for a given temperature $T$, and
found that at high energies these states are always unstable towards FRSB,
just as we see in Fig.~\ref{sec_approx}. There is, however, a major
difference between our works: in \cite{MontanariRicci03} the authors
were looking at the {\it typical} excited states at a given free
energy $f$ and temperature $T$, that is, at the most numerous ones. In
our present work we instead concentrate on following the states that were typical at a given temperature. The point is
that as soon as the temperature is changed, these states become out of
equilibrium and not typical. If one wants to focus on the states that are the most important ones for the free energy landscape, it is {\it necessary} to
consider their basins of attraction, as we do, and this is why we have
developed the following state method in the first place. We will come
back on this point when we will discuss the iso-complexity problem in
Sec.~\ref{sec:WP}.

\begin{figure}[!ht]
  \includegraphics[width=0.48\linewidth]{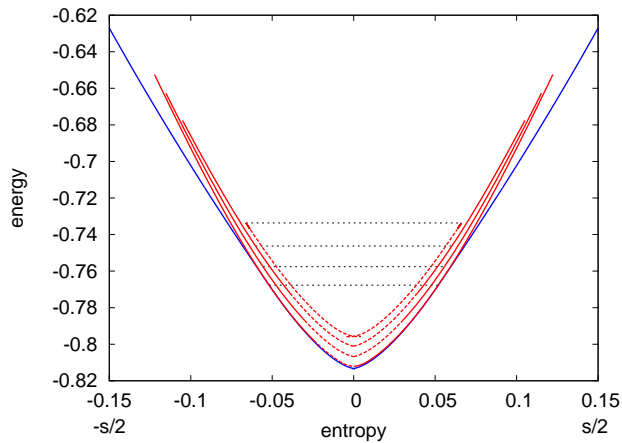}
  \caption{\label{fig_follow2} (color online): A direct quantitative
    look at the shape of states in the energy landscape of the fully
    connected $3$-spin model. The energy of four different states is
    plotted as a function of their entropy; we plot $s/2$ and $-s/2$
    such that the width corresponds to the entropy. The shape of the
    curves show how the valleys looks in the energy landscape.  The
    most outer curve (blue) is the total equilibrium energy
    versus entropy. The four inside red curves correspond to different
    states that are at equilibrium at temperatures
    $T_e=0.6815,0.67,0.66,0.6513$, and their equilibrium energy $e_e$
    are marked by horizontal black dotted lines. The highest bottom of these
    states marks a lower bound on the best possible limiting energy for a slow annealing.}
\end{figure}

Fig.~\ref{fig_follow2} presents the same data as Fig.~\ref{fig_follow}
in a different perspective. It is a sort of direct look at the shape of states in the energy
landscape. The $y$-axis is still the energy, the $x$-axis depicts the
size (entropy) of the state. More precisely we plot a line at $-S/2$
and $S/2$. The blue (most outer) line is the entropy of the equilibrium
(static) solution. The four red lines correspond to different Gibbs
states. The horizontal dashed lines depict energy at which these
states are the equilibrium Gibbs states. Note that according to the
laws of thermodynamics the derivative of the energy at the minima have
to be zero, whereas figure \ref{fig_follow2} shows a slight non-physical cusp. This
is dues to the 1RSB approximation that underestimates the entropy, the
FRSB solution would have the correct derivative.

\subsection{Below the Kauzmann transition: Level crossings and
  temperature chaos}
\label{sec:chaos}

\begin{figure}[!ht]
  \includegraphics[width=0.495\linewidth]{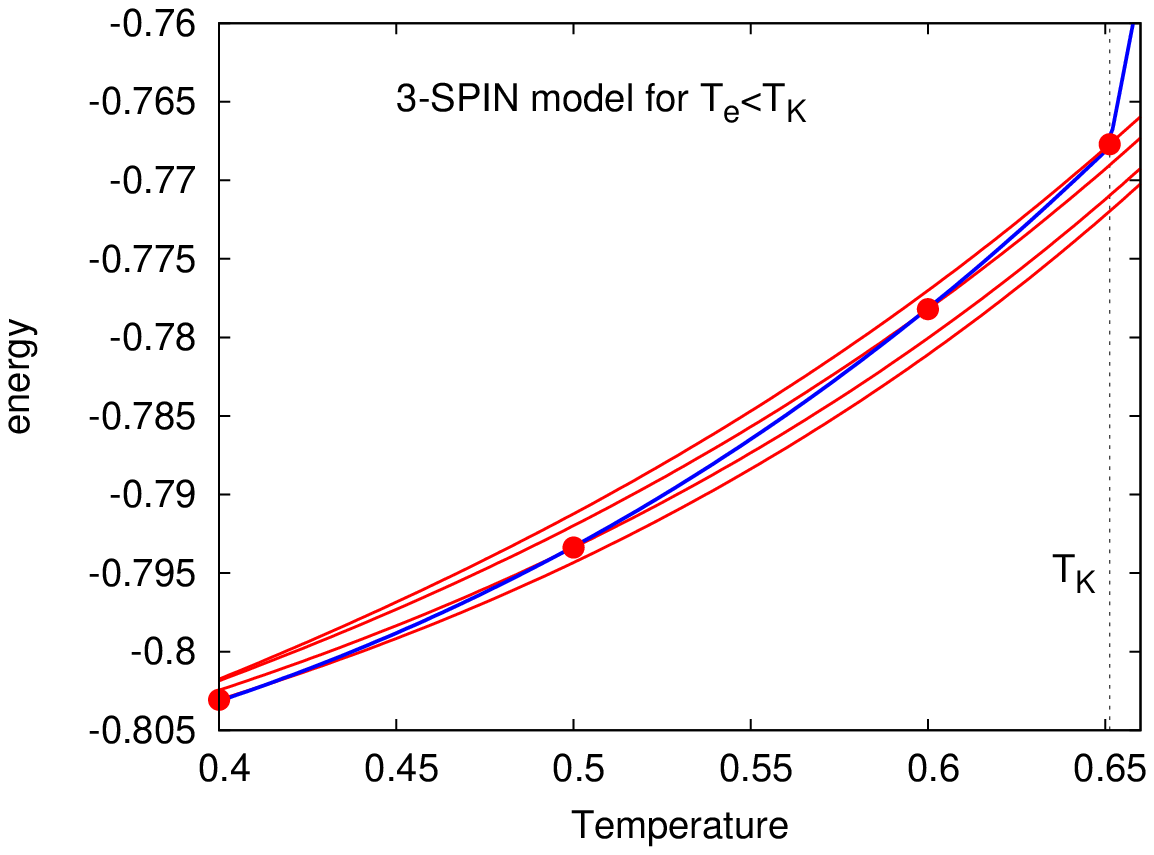}
  \includegraphics[width=0.495\linewidth]{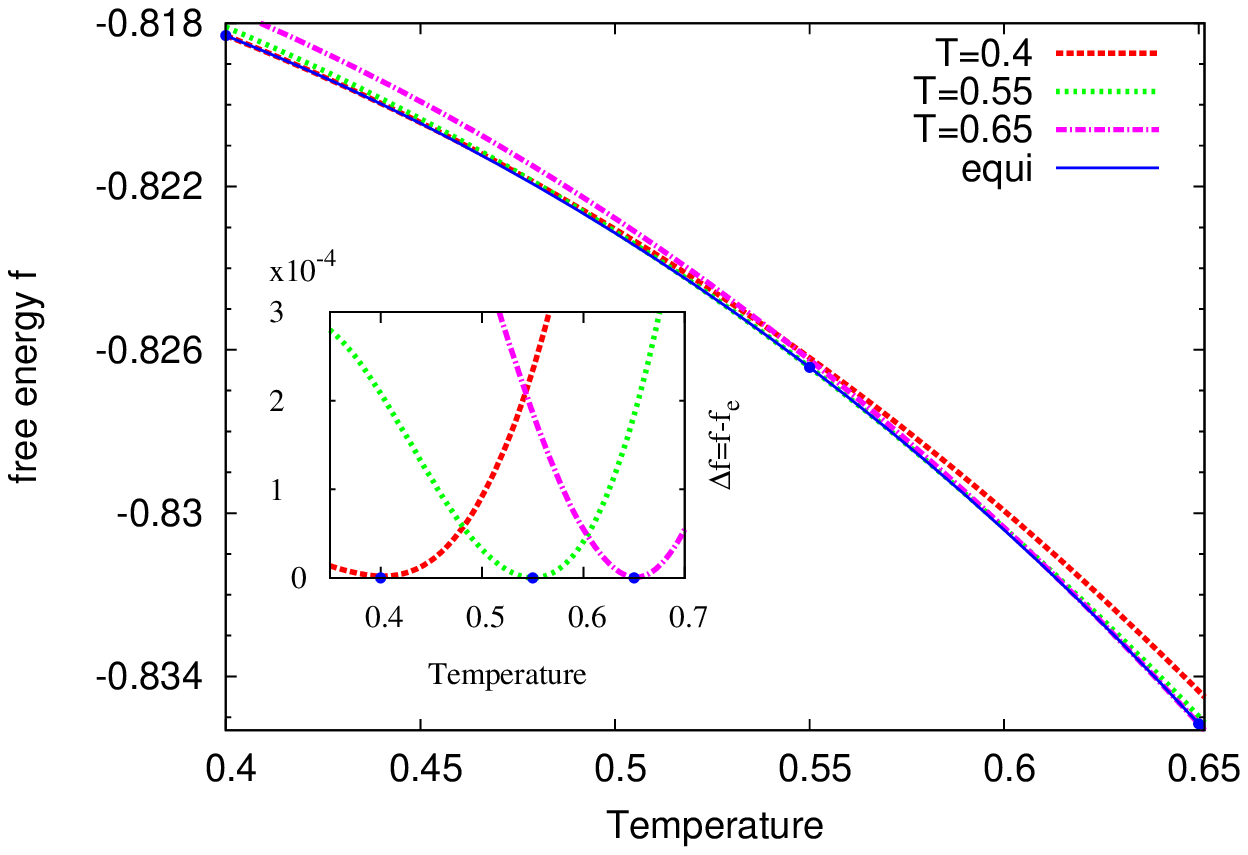}
  \caption{\label{fig_chaos} (color online): Explicit demonstration of
    the presence of temperature chaos below the Kauzmann temperature
    in the fully connected $3$-spin model. Left: the energy as a
    function of temperature for states below the Kauzmann transition: the
    four curves correspond to temperatures $T_e=0.4, 0.5, 0.6$ and
    $0.6513$. Right: The crossing of free energies of different states
    at a function of temperature. The lower envelope (blue) is the
    equilibrium free energy: it results from the crossings of many
    states; here we show three such states that are the equilibrium
    ones at temperatures $T_e=0.4, 0.55, 0.65$. Inset: to make the
    crossing more evident, we plot the difference $\delta F$ between
    the free energy of these three different states and the
    equilibrium free energy.}
\end{figure}

We now turn to the description of static spin glass phase,
$T<T_K$. Fig.~\ref{fig_chaos} uses equations derived in Sec.~\ref{sec_pspin_1RSB}
and depicts the evolution of states that are at equilibrium below
$T_K$. Fig.~\ref{fig_chaos} left gives the energy as a function of
temperature for three states that are at equilibrium (marked by red points)
at some temperature $T_e<T_K$. In Fig.~\ref{fig_chaos} right, we plot
the free energy as a function of temperature of these states, the lower envelope of the
free energies of all the states is then the equilibrium free energy.
In order to enhance the differences, in the inset, we subtracted the equilibrium free energy from the free energies of the three states.

These plots clearly show that, although a finite number of states
dominates the partition sum (which is the very definition of the glass
phase below $T_K$), these states become out-of equilibrium as soon as
the temperature is slightly changed. Even though for all $<T_K$ the
partition sum is dominated by a finite number of state: {\it these
  states change entirely when the temperature is slightly
  modified}. This is the phenomenon of temperature chaos that appears
due to free energy levels crossing.

Temperature chaos has been discussed extensively in spin glasses, see
for instance
\cite{FisherHuse86,BrayMoore87,BanavarBray87,Kondor89,FranzNifle95},
it is crucial in the interpretation of memory and rejuvenation
experiments \cite{SasakiNemoto00,JonssonYoshino02}. It existence was
subject of debates, as its absence was advocated in many papers
\cite{MuletPagnani01,Rizzo01,BilloireMarinari00}, as well as its
presence
\cite{AspelmeierBray02,RizzoCrisanti02,SasakiMartin02b,KatzgraberKrzakala07}.
Our results allow to finally clearly demonstrate that temperature
chaos is present in the Ising fully-connected $p$-spins, and that it
arises through many level crossings, as advocated in
\cite{KrzakalaMartin02,RizzoYoshino06}.

\subsection{The phase diagram of evolving states and the mapping to a
  ferromagnetic $p$-spin model}
\label{sec:ferro}

\begin{figure}[!ht]
  \includegraphics[width=0.495\linewidth]{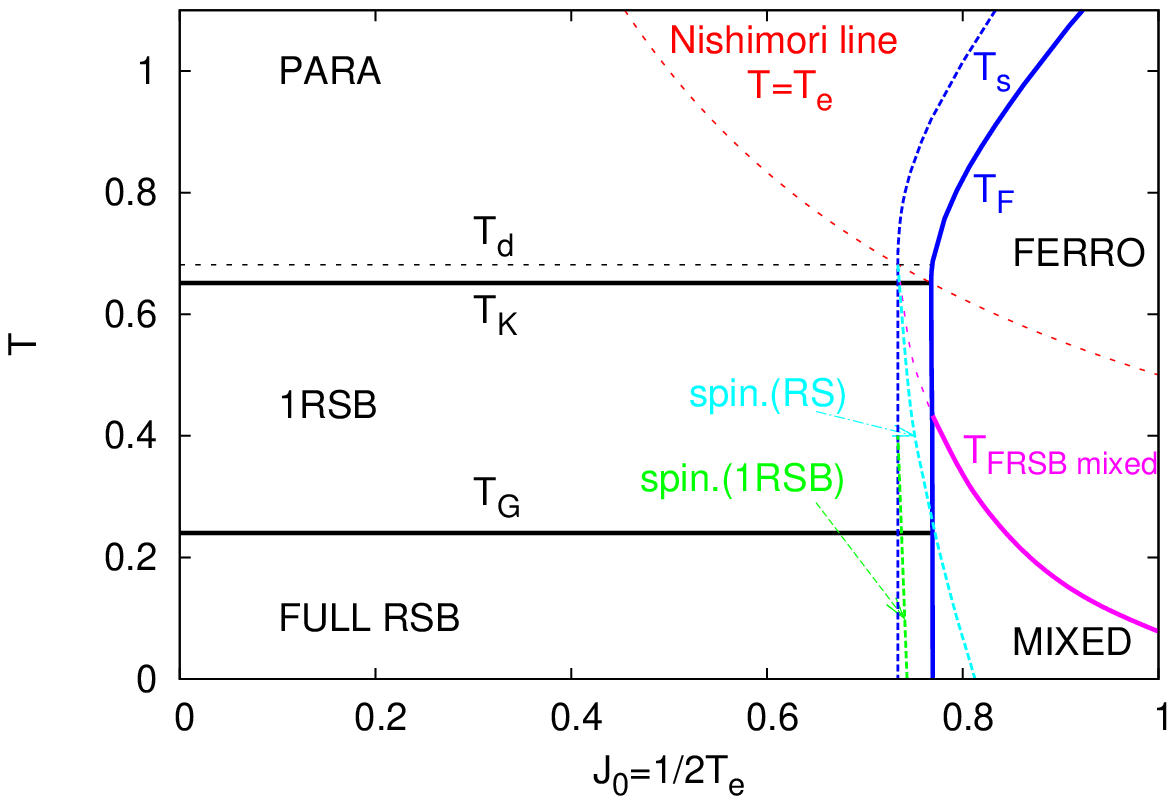}
  \includegraphics[width=0.495\linewidth]{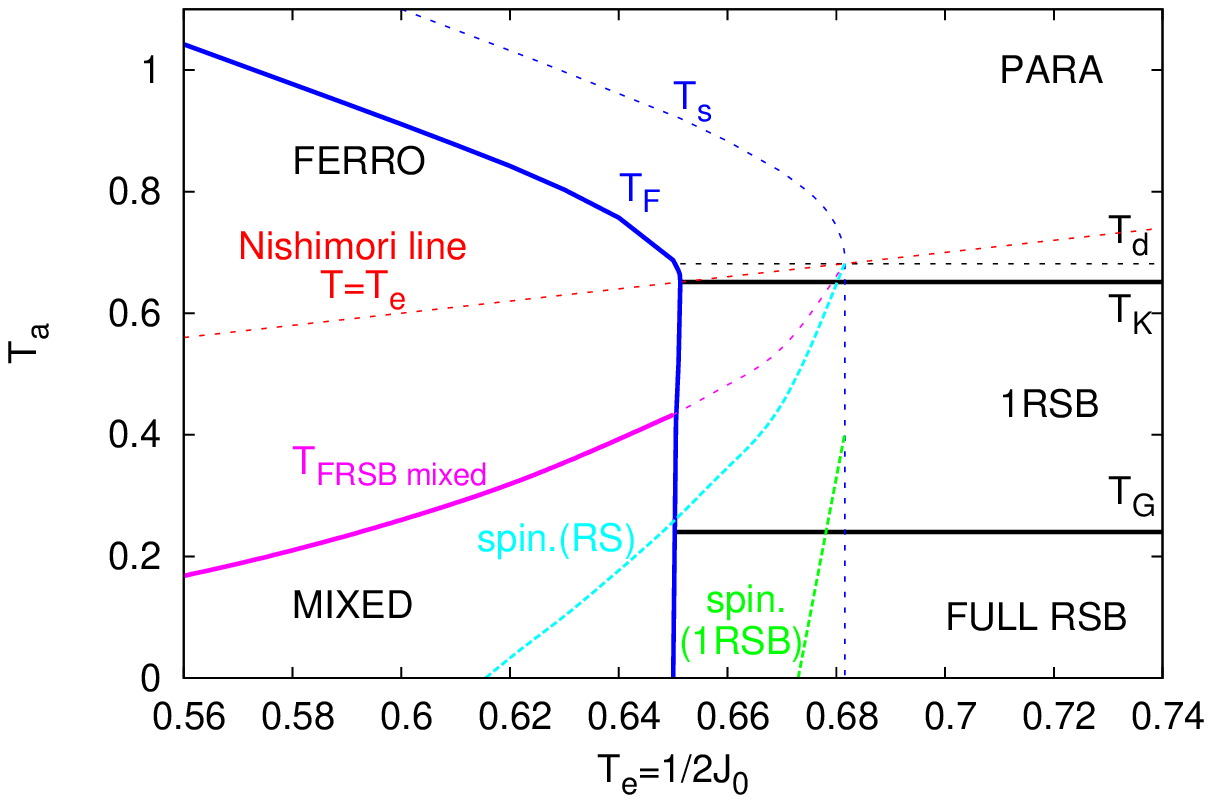}
  \caption{\label{fig_J0_phase} (color online): The phase diagram of
    the fully connected 3-spin model with a ferromagnetic bias (left),
    or equivalently, the phase diagram in function of the two
    temperatures $T_a$ and $T_e$ in the state following formalism.
    True phase transition are marked by thick lines, while spinodals
    and the Nishimori line are draw thin and dashed.  The two phase
    diagrams are given to highlight how the physics of states
    following can be understood intuitively from the ferromagnetically
    biased model. Note that the line denoting the transition from spin
    glass to ferromagnet (computed at the 1RSB level) is almost but
    not completely straight.}
\end{figure}

As we derived in Secs.~\ref{sec:Nish} and \ref{RS_pfollow}, all the physics of the adiabatic evolution of equilibrium states above the Kauzmann
temperature for the $J_0=0$ $p$-spin model can be induced from the known, see
e.g. \cite{NishimoriWong99,NishimoriBook01}, phase diagram of the
$J_0\neq 0$ $p$-spin model. We will now discuss the phase diagram in the
context of results given in Figs.~\ref{fig_follow} and
\ref{fig_follow2}. Fig.~\ref{fig_J0_phase} shows two drawings of the
phase diagram, the $y$-axes in both is the actual temperature. The
$x$-axes on the left is the standard ferromagnetic bias $J_0=\beta_e
J^2/2$, on the right the $x$-axes is the equilibrium temperature
$T_e=J^2/(2J_0)$ (in both figures we took $J=1$). The red (dashed) line in both parts is $T_a=T_e = J^2/(2J_0)$, on the left this is the Nishimori line, on the right this is the
equilibrium line. The task is to follow states that are the
equilibrium ones on this line for $T_e\ge T_K$. The horizontal black lines depict the location of the dynamical, Kauzmann and Gardner temperatures
for $T_a$.

The phase diagram of the $p$-spin model with a ferromagnetic bias
$J_0$, Fig.~\ref{fig_J0_phase} left hand side, see also
\cite{NishimoriWong99,NishimoriBook01} has five thermodynamic phases
separated by thick lines in the figure: The paramagnetic phase at high
temperature (PARA). The 1RSB spin glass phase for low enough bias
$J_0$ and $T<T_K$ that becomes a full RSB phase for $T<T_G$. The
ferromagnetic phase (FERRO) for $J_0$ large enough and $T<T_F$ that
becomes a mixed phase with both ferromagnetic and full RSB order
(MIXED) for $T<T_{FRSB mixed}$. At low enough $J_0$ the system is a
spin glass with an ergodicity breaking transition at $T_d$ and then
the Kauzmann and Gardner phase transitions at $T_K$ and $T_G$. At
larger $J_0$, the system undergoes a first-order ferromagnetic
transition at $T_F$. The ferromagnetic state is thermodynamically
stable starting from the spinodal temperature $T_s>T_F$, at $T_F$ it
becomes thermodynamically dominant. Below the pink line, $T_{\rm
  FRSB\, mixed}$, the replica symmetric solution describing the
ferromagnetic state ceases to be stable toward RSB and the system
transits into the FRSB mixed phase. The mixed phase have some
unphysical spinodals: below the light-blue line there is no
non-trivial ferromagnetic RS solution. This is cured by the 1RSB
approach, but only down to the green line, below which there is no
non-trivial ferromagnetic 1RSB solution. The correct spinodal should
be a vertical line in the FRSB computation. The Nishimori line is the
red dashed curve: notice how it crosses the ferromagnetic transition
exactly when $T_F=T_K$ and the spinodal at $T_s=T_d$.

The right hand side of Fig.~\ref{fig_J0_phase} depicts the same diagram as a function of the equilibrium temperature $T_e$. Following a state at equilibrium at temperature $T_e$ to temperature $T_a$ is equivalent to looking to the ferromagnetic state on the Nishimori line at $T_e$, and then to move vertically to other temperatures $T_a$. The vertical strip of temperatures $T_K\le T_e \le T_d$ is hence particularly relevant for state following. The blue spinodal ferromagnetic line $T_s$ thus corresponds to the high temperature spinodal line in state following (see  Fig.~\ref{fig_follow}). The pink line $T_{\rm FRSB\, mixed}$ to the point where the state divides into many sub-states and develops a (presumably) full replica symmetry breaking. However, we are unable to follow the state with the RS formalism below the light blue line that correspond to an unphysical spinodal. This is cured in part by the 1RSB formalism, but the same problems arises in this case below the green line, so that a FRSB solution is eventually needed to describe the adiabatic evolution of states at $T_e=T_d$.

For $T_e<T_K$ ($J_0>J^2/(2T_K)$) the phase diagram shows a
ferromagnetic phase which would correspond to following a state that almost surely does not exist for a typical instance of the problem. To follow states equilibrium below $T_K$ the mapping to a model with a ferromagnetic bias breaks.

Two comments are in order about the spinodal and the ferromagnetic
transition.  Let us first discuss the spinodal, which we believe to be
vertical bellow $T_d$: the FRSB ferromagnetic solution in the mixed
phase must exist up to the vertical blue line. However, different
levels of replica symmetry breaking have different unphysical spinodal
lines beyond which no-nontrivial ferromagnetic solution exist at the
RSB level (the blue spinodal for RS and green for 1RSB are
depicted). Such behavior is not unheard of: the very same phenomena
takes place in the study of the Sherrington Kirkpatrick model in
external magnetic field \cite{SherringtonKirkpatrick75} (that is, for
$p=2$). There also the boundary of the mixed is a vertical line in the
FRSB solution, but differs at all finite levels of RSB
\cite{GabayToulouse81,MezardParisi87b}. Similar features were observed
in the dilute mean field spin glasses as well
\cite{KwonThouless88,CastellaniKrzakala05}. In the state following
method the lack of a ferromagnetic solution near to the true FRSB
spinodal translates into difficulty of obtaining a sensible 1RSB upper
bound on the low temperature adiabatic evolution of states with $T_e
\approx T_d$.

Let us now consider the ferromagnetic transition line $T_F$ bellow
the Nishimori line. Nishimori proved \cite{NishimoriBook01} that the line was either vertical, or bending towards the ferromagnetic phase, this is also apparent from the states-following interpretation. Although it
might not be completely visible from Fig.~\ref{fig_J0_phase} (left), the
analysis of the 1RSB equations shows that the line is
bending slightly towards lager $J_0$ as $T$ is lowered, although the effect is
very small (to the best of our knowledge, this was an unknown feature
of this phase diagram). Interestingly, this has a clear interpretation
in the states-following formalism. At zero temperature,
the energy of the ferromagnetic state at $J_0=1/(2T_K)$ is equal to the
bottom energy of the equilibrium state at $T_K$. As discussed in Sec.~\ref{sec:chaos} chaos and level
crossings make these states to have a larger energy than the true
equilibrium one at $T<T_K$; as a consequence, the ferromagnetic state at $T=0$
and $J_0=1/(2T_K)$ {\it must} have bottom-energy larger than the ground state energy of the system, so that the ferromagnetic transition can only happen for larger values of
$J_0$. Interestingly chaos disappears in the large $p$ limit (as well
as in the spherical approximation) and this is why this line is
strictly straight in the phase diagram of the random energy model
\cite{Derrida80}, and in the spherical $p$-spin model 
\cite{GillinSherrington01}\footnote{Note that there have been a
  considerable amount of efforts to discover this effect in finite-dimensional
  spin glass, see for instance \cite{HasenbuschPelissetto08}, and it
  is therefore interesting to observe it in mean field models as
  well.}.

\subsection{Relation to the Franz-Parisi potential}
The idea of exploring one of the many phases in glassy mean field
systems is a very natural one, and is therefore not new. Our
states following approach is actually related to the one pioneered by Franz and Parisi
years
ago \cite{FranzParisi92,FranzParisi97,BarratFranz97,FranzParisi98},
which is now commonly referred to as the {\it Franz-Parisi potential}. Their idea was to study glassy systems in presence of an
attractive coupling among two real replicas, one of which being at
equilibrium. Looking to the free energy of the copy when its overlap
with an equilibrium configuration is tuned allows to compute the local
free energy potential around this equilibrium point.

What the states following method is actually doing is to focus directly on the minimum
of the Parisi-Franz potential, thus bypassing the need of an
attractive couplings and making the formalism much simpler and
applicable easily to the models on sparse graphs. The Franz-Parisi potential
can, however, be obtained within the states following method if we fix the overlap
between the state at temperature $T_e$ and $T_a$. The purpose of the
present paragraph is to explain how to do this in the
$p$-spin model. The reason is two-folds: (a) we want to make the
correspondence with the Franz-Parisi formalism and (b) looking at
these free energies turns out to be extremely instructive to
understand the unphysical spinodal points and related issues discussed
in the previous section.

Conveniently enough, in our mapping to a model with an effective
ferromagnetic coupling $J_0$, the "magnetization" parameter $m$ (\ref{eq_m}) is nothing else than the overlap of the configuration under study and the planted
one. This demonstrates the usefulness of the above mapping: The free energy at fixed magnetization of a $p$-spin model with a ferromagnetic bias $J_0$ at the  temperature $T$ is equal to the Franz-Parisi potential for the spin glass problem with temperature $T_e=J^2/(2J_0)$ and $T_a=T$.

Fixing the magnetization, i.e. ensuring $\sum_i s_i/N=m$, is done by
introducing a Lagrange multiplier $h$ and writing the partition
function of the system with a fixed magnetization as $Z_m=Z_h
e^{-N\beta h m}$, where $Z_h$ is a partition function of the model
with an external magnetic field, i.e., with Hamiltonian $H_h = H -
h\sum_i s_i$. Once we compute the free energy of the model with
external magnetic field $f(h)$ the free energy of the system with
magnetization fixed to $m$ is recovered via $f(m)=f(h)+hm$. To fix a
value $m=m^*$ we need to ensure that $\partial f(h) / \partial h
|_{h^*} = - m^*$; the $f(m)$ is thus a Legendre transform of $f(h)$.

This allows us to derive easily the Franz-Parisi potential
in the $p$-spin model. Actually, it also allows to obtain
instantaneously all the (not straightforward) Franz-Parisi computations in the
spherical and mixed spherical $p$-spin models just by looking to the
equilibrium free energy of the model with a ferromagnetic bias. The reader is
invited for instance to compare the free energy in
\cite{GillinSherrington01} with the Franz-Parisi potential in
\cite{BarratFranz97,FranzParisi98}.

\subsubsection{Franz-Parisi potential at the replica symmetric level}
At the RS level, the equations for the $p$-spin model with an external field $h$ become:
\bea
m&=&   \int_{-\infty}^{\infty} {\cal D}y  \tanh{\left( \beta_a J y \sqrt{p q^{p-1}/2}+ \beta_a J_0 p m^{p-1} + \beta_a h\right)} \, ,\label{eq_m_for_fixed_m}\\
q&=& \int_{-\infty}^{\infty} {\cal D}y \tanh^2{\left( \beta_a J y \sqrt{p
      q^{p-1}/2}+ \beta_a J_0 p m^{p-1} + \beta_a h \right)}\,
,\label{eq_q_for_fixed_m} 
\eea
with ${\cal D}y={\rm d}y \, e^{-\frac{y^2}{2}} /\sqrt{2\pi}$. The free energy is given by (\ref{free_en}) with $\beta_a h$ added in the argument of the $\cosh{}$.

\begin{figure}[!ht]
  \includegraphics[width=0.48\linewidth]{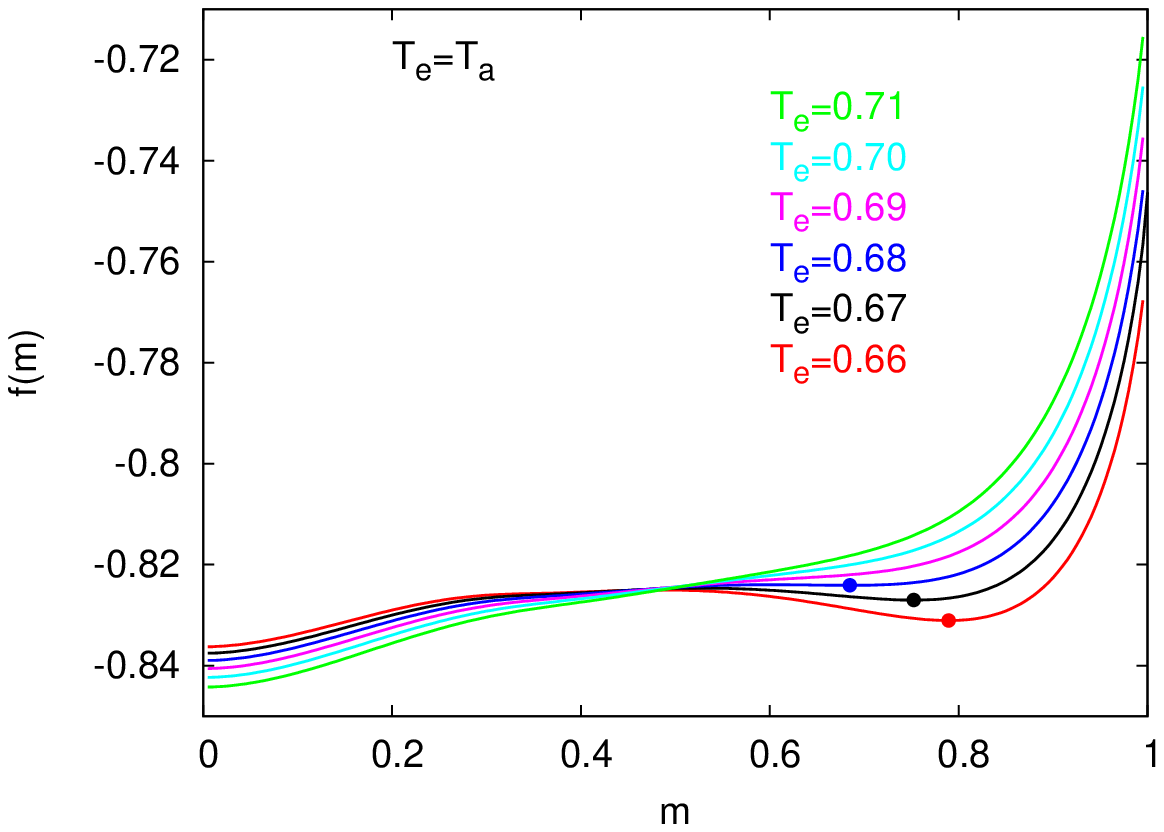}
  \includegraphics[width=0.48\linewidth]{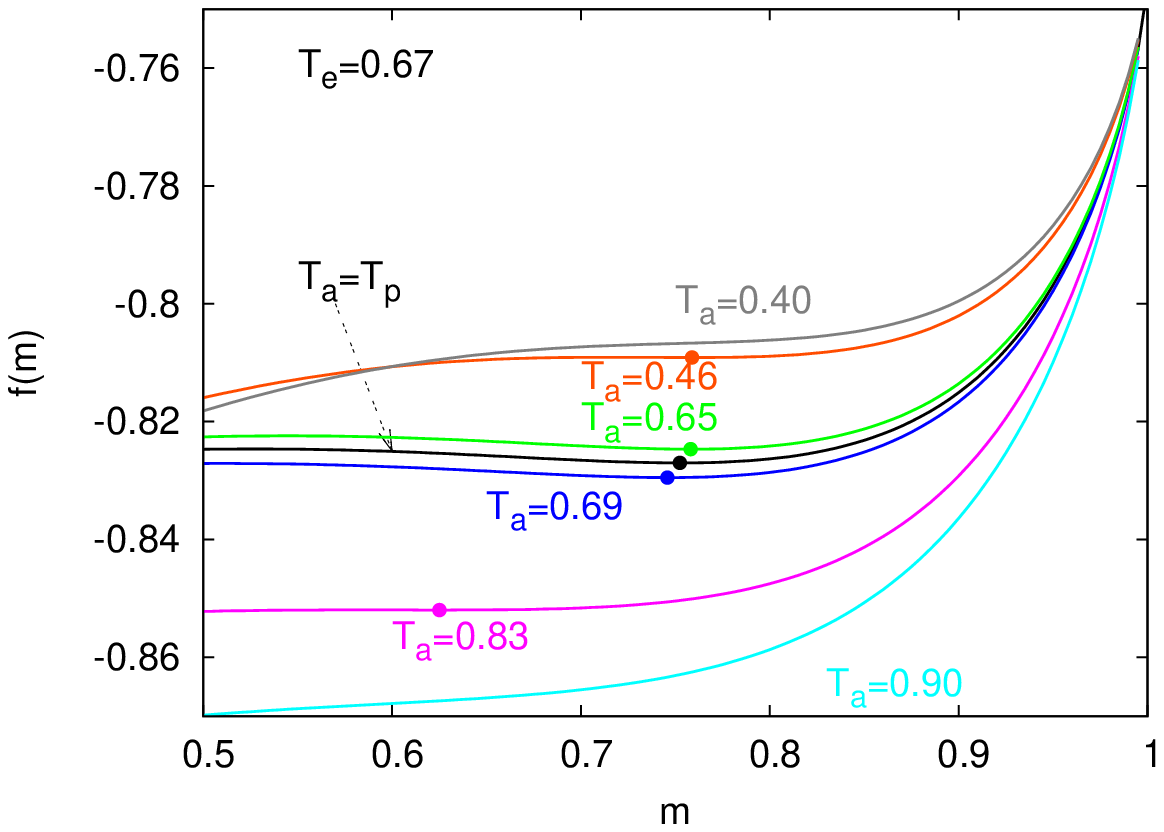}
  \caption{\label{fig_pspin} (color online) The replica symmetric
    free energy of a state at temperature $T_a$ given the distance $m$
    from an equilibrium configuration at $T_e$ in the $3-$spin fully
    connected model. The points indicate the
    minima corresponding to what the state following method finds. Left: In this case $T_a=T_e$; a minimum starts to form
    around the equilibrium configuration for $T_e<T_d=0.6815$ while in
    the liquid phase $T_e>T_d$ no such minimum exist.  Right: We now
    use $T_a\neq T_e$ (here $T_e=0.67$) and we observe how the curves
    are evolving with $T_a$. Upon warming, the minimum happens at
    lower free energies and $m$ is decaying, indicating that the
    state gets larger until finally a spinodal point is reached and no
    more minimum exist; this is the moment where the state melts into the liquid.  Upon cooling, we thus expect that both the free energy and
    $m$ increase. This is indeed the case, but for low temperature
    (here $T_a=0.46$) we start to observe a non monotonous behavior
    for $m$. Worse, if the temperature is again lowered (here
    $T_a=0.4$) the minimum disappears. These are non-physical features
    and are clear signs that replica symmetry must be broken for low
    temperatures.}
\end{figure}

Note that when the free energy is non-convex (as it is in the present
case), one has to be extremely careful in solving these
equations. Indeed if one simply chooses $h$ once and for all and
simply performs a recursion of
Eqs.~(\ref{eq_m_for_fixed_m}-\ref{eq_q_for_fixed_m}) some values of
$m$ will never be obtained. A good method is to first choose the
desired value $m^*$, and then to fix the magnetic field $h$ {\it at
  each iteration} such that Eq.~(\ref{eq_m_for_fixed_m}) is satisfied.
This can be easily generalized in the 1RSB equation. In
Fig.~\ref{fig_pspin}, we show the results of this procedure.

The left side of Fig.~\ref{fig_pspin} shows the free energy
of configuration at a distance $m$ from the planted one in the ferromagnetically biased model; this is the
Franz-Parisi potential. One sees that for $T_e>T_d=0.6815$ there is no
minimum except the trivial one at $m=0$; for $T_e<T_d=0.6815$, however, a second minimum appears, with a finite value of the overlap: this is precisely
the one found in the states following approach, which is only performing a gradient
descent in this free energy starting from the point $m=1$, thus
directly focusing on the non trivial-minima of the free energy potential.

The right hand side of Fig.~\ref{fig_pspin} shows the free energy
potential of configuration at a distance $m$ from the planted one when
the temperature is different from the planted temperature. We have
used $T_e=0.67$ and we can see how the free energy of the state
changes with temperature. This is actually very instructive.  When
rising the temperature the bottom free energy of the state decreases
(as a free energy should with temperature because of the positivity of
entropy) while the overlap $m$ at the minimum get smaller: this is the
sign that the state become larger. At even larger temperature, a spinodal point is met and the minimum (as well as the state) stop to exist.

When decreasing the temperature, we expect that both the free energy
and $m$ increase, as the state gets smaller and deeper in the free
energy landscape. This is indeed the case initially, but for low
temperature (here $T_a=0.46$) we start to observe a non-monotonous
behavior for $m$. If
the temperature is further lowered (here $T_a=0.4$) the minimum
disappears, as we have seen in the previous chapter. These are
non-physical features that show that the replica symmetric assumption
is incorrect and that we need to break the replica symmetry.

\subsubsection{Franz-Parisi potential at the replica symmetry broken level}
\label{sec_approx}

\begin{figure}[!ht]
  \includegraphics[width=0.48\linewidth]{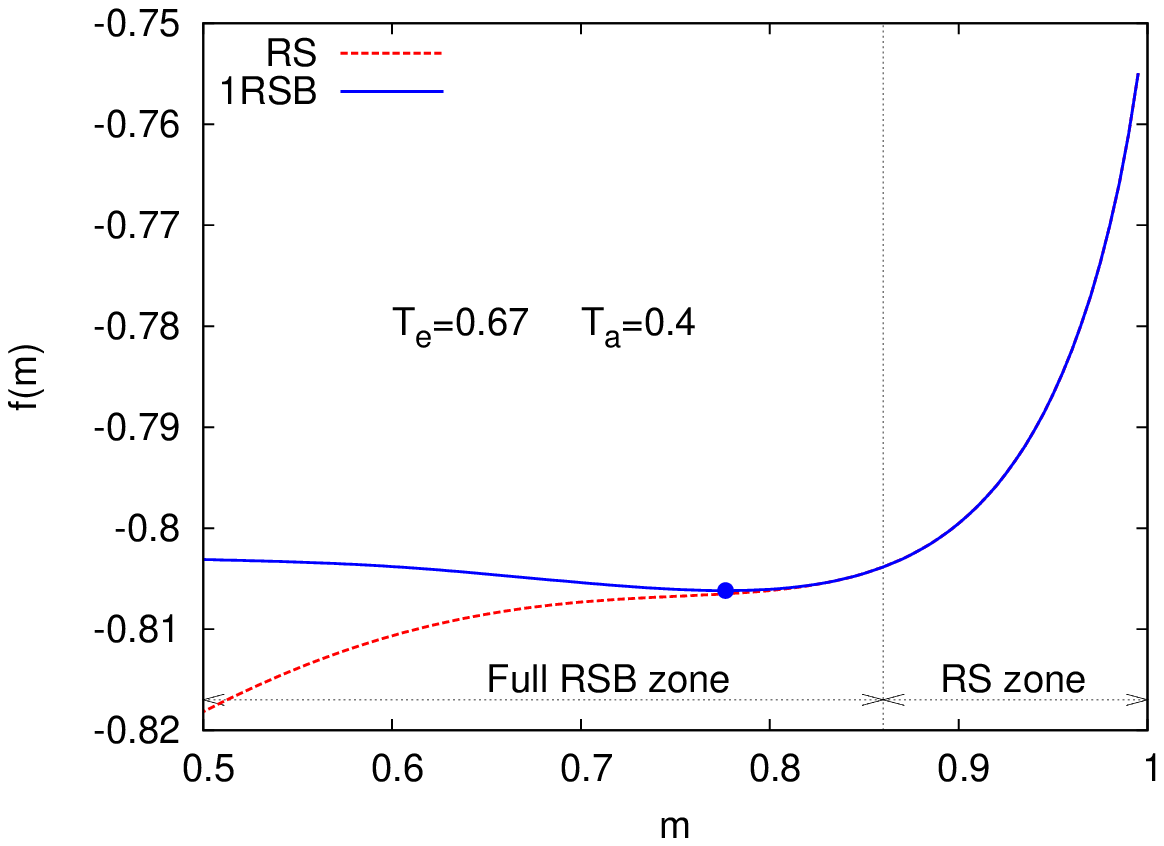}
  \includegraphics[width=0.48\linewidth]{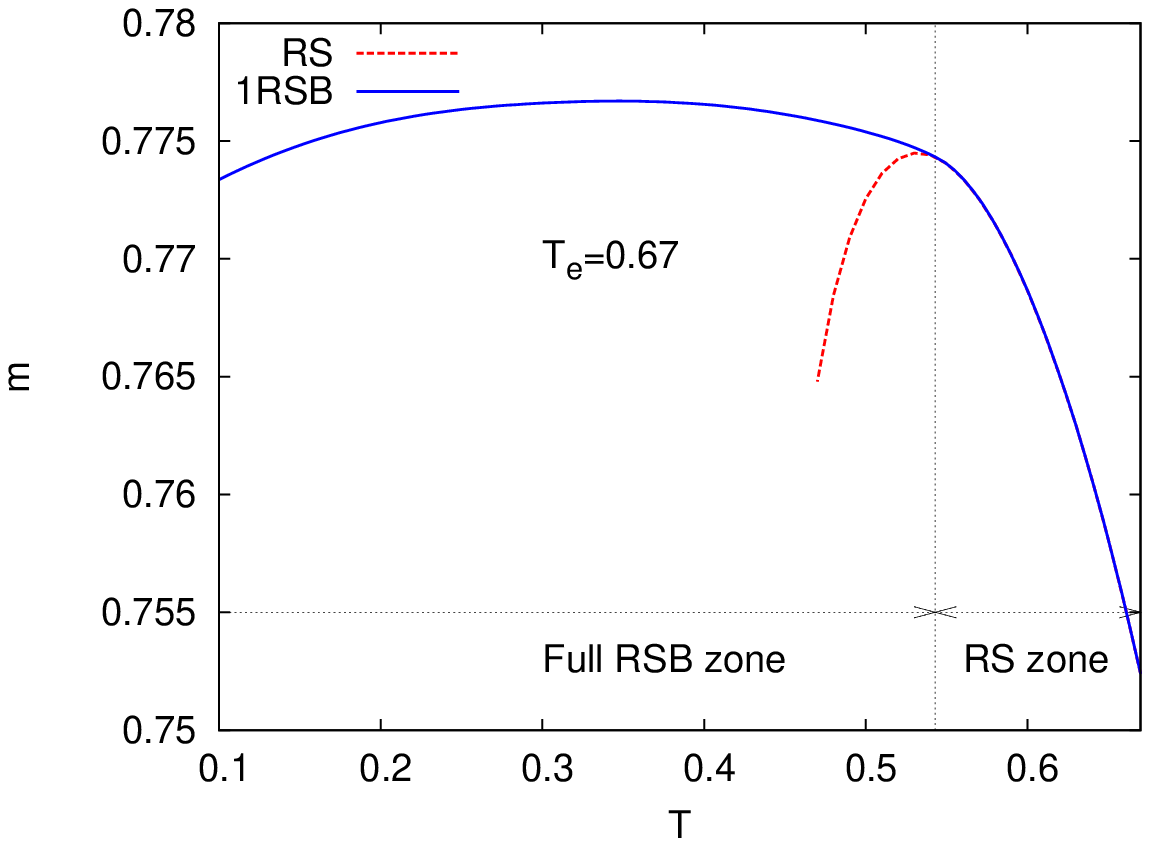}
  \caption{\label{fig_pspin_1rsb} (color online) Left: Comparison between the RS Franz-Parisi potential lower curve (red) and the 1RSB one upper curve (blue). The point marks the minima on the 1RSB curve. Right: Magnetization of the minima as a function of temperature $T_a$. Again the lower (red) curve is the RS result, the upper (blue) curve is the 1RSB result. The increasing or non-existing part of the curve in unphysical, the size of the unphysical region is much smaller in the 1RSB result.}
\end{figure}

To obtain the 1RSB approximation of the Franz-Parisi potential we need to fix the parameter $m$ in the 1RSB equations using again an auxiliary magnetic field $h$. The result for $T_a=0.4$ and $T_e=0.67$ is shown in Fig.~\ref{fig_pspin_1rsb} left. The lower line (red) is the replica symmetric result, the upper line (blue) is the 1RSB result. The two curves differ in the RS unstable zone on the left side of the plot. Unlike the unstable RS result, the 1RSB Franz-Parisi potential has a secondary physical minimum at about $m=0.776$. Moreover, as general in replica theory, the 1RSB free energy is always larger than the RS one.  

Right part of Fig.~\ref{fig_pspin_1rsb} shows the dependence of the magnetization at the minimum as a function of temperature $T_a$. 
Following states becomes studying ferromagnetically biased model, in ferromagnets magnetization usually grows as the temperature decreases. Hence, the part where $m(T_a)$ increases (or does not exist) is unphysical and will decrease in the FRSB solution. We can see that the physical region extends into lower temperatures $T_a$ for the 1RSB result. We also observed that the 1RSB magnetization is systematically larger than the RS one. 

The above findings suggest a method how to obtain a lower bound on the energy of the state even at temperatures where the 1RSB solution does not exist (the 1RSB Franz-Parisi potential does not develop the secondary minima). At such temperature $T_a$ the magnetization at the real minima (which we would observe in the FRSB result) have to be larger than the maxima of magnetization $m_m$ in the 1RSB result over all $T_a$. As the FRSB free energy is larger that the 1RSB one the free energy at that minima have to be larger than the 1RSB free energy at $m_m$. Using this receipt we can thus obtain a lower bound on the free energy (and energy) which is probably not far from the true result. This is how we obtained the green dotted part in Fig.~\ref{fig_follow}. 

Note, however, that the above described construction of the lower
bound is not very elegant and requires the calculation of the full
Franz-Parisi potential $f(m)$. It is interesting to see if better
approximation to the FRSB result can be obtained using different
techniques.

\section{Second application: Energy landscape in constraint
  satisfaction problems and diluted  models}
\label{sec:result-xor}
We have discussed at length the various aspects of adiabatic evolution
of Gibbs states for the simple case of fully connected $p$-spin model
because many of those aspects repeat for the computationally more
involved models on sparse random graph like XOR-SAT or graph
coloring. In this section we present results for those two models.

\subsection{Following states in diluted spin models}

\begin{figure}[!ht]
\includegraphics[width=0.495\linewidth]{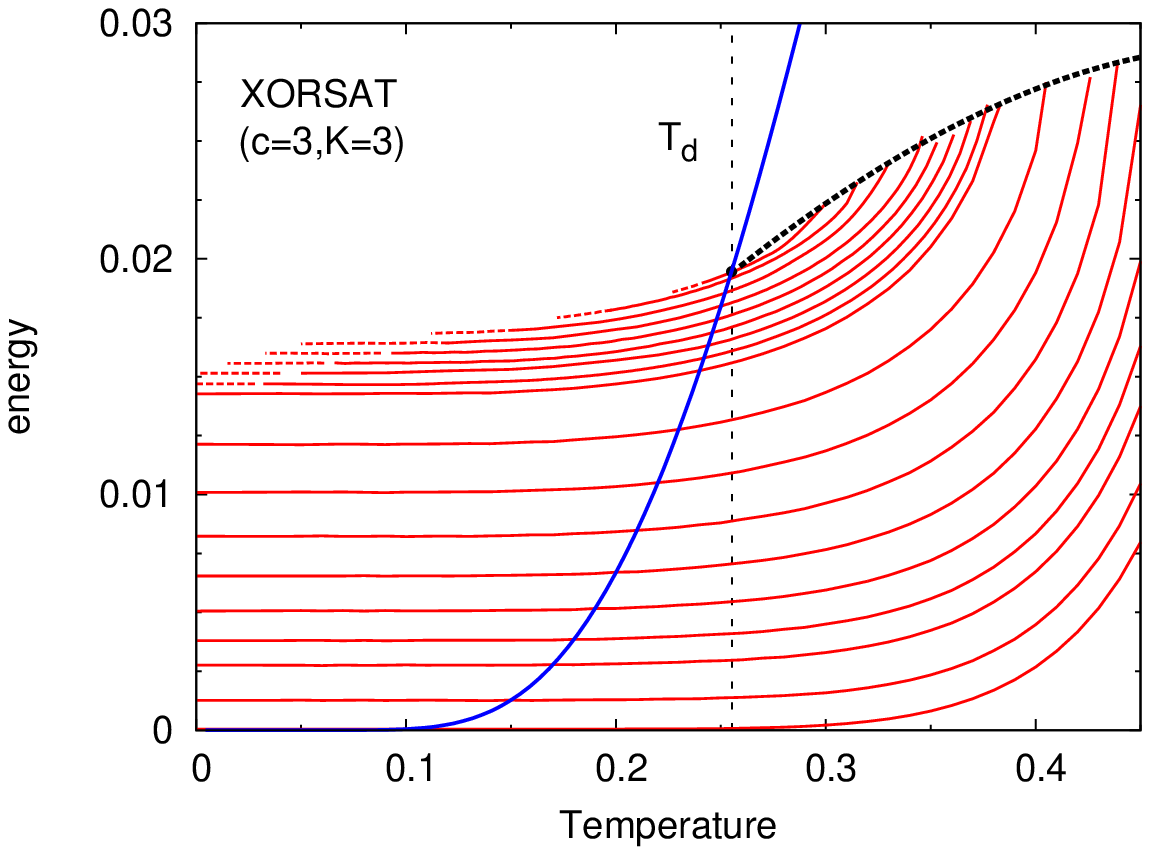}
\includegraphics[width=0.495\linewidth]{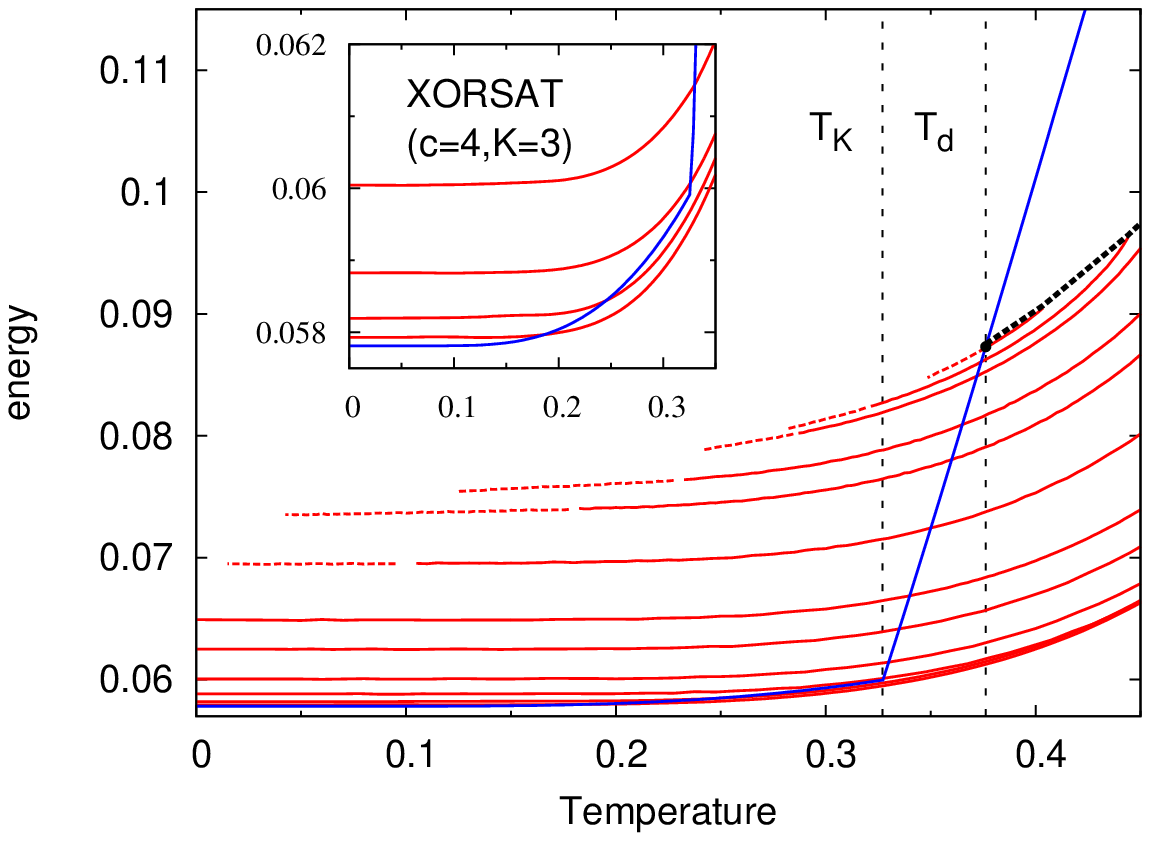}
\caption{(color online) Representative behavior of states in diluted
  mean field systems. Left: The XOR-SAT problem with $K=3, L=3$. Right:
  the XOR-SAT problem with $K=3, L=4$.  The value of energy here is the
  number of violated constraints per variable. The blue line crossing
  the diagram is the equilibrium energy computed from the standard
  cavity method, Eqs.~(\ref{1RSB}--\ref{m_phys}). The vertical lines
  denote the dynamical and Kauzmann temperatures. The red lines depict
  adiabatic evolution of states that are the equilibrium one at the
  temperature $T_e$ where the red curve crosses the blue one. The state
  evolution curves are obtained by solving equations
  (\ref{pop_pl}-\ref{free_pl}) when $T_e\ge T_K$, and
  Eqs.~(\ref{1RSB_pl_2}-\ref{energy_gen}) when $T_e<T_K$ (in the inset
  of the right hand side). The dashed part of the red curves depicts
  the region of temperatures where the state is no longer stable
  towards replica symmetry breaking and splits into many
  sub-states. The ends of the red curves at nonzero temperature
  correspond to the non-physical spinodal points beyond which
  Eqs.~(\ref{pop_pl}-\ref{free_pl}) have only the trivial liquid
  solution. \label{fig1}}
\end{figure}

In Fig.~\ref{fig1} we plot the energy of states versus temperature for
the 3-XOR-SAT problem with degree of variables $L=3$ (left), $L=4$
(right). The behavior is extremely similar to the one of the fully
connected model and the very same feature are observed: the spinodal
upon heating, the transition towards symmetry breaking upon cooling,
and the unphysical spinodal. Note that these plots have been obtain
with the RS procedure and a first remark is that the RS computation
gives a more complete results than in the fully-connected case. Since
equilibrium states {\it do not} develop the Gardner instability, it is not surprising that the instability towards RSB is less strong in the
out-of-equilibrium states as well. However, we observed once again
that the states at equilibrium close to $T_d$ undergo a
FRSB transition and decompose into many marginally stable sub-states:
this seems to be an universal features of spin-glasses.

\begin{figure}[!ht]
  \includegraphics[width=0.49\linewidth]{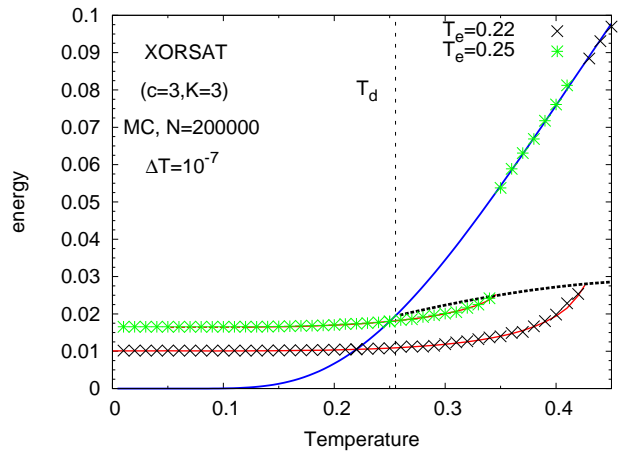} 
  \caption{\label{fig:simu} (color online) Following states in the
    XORSAT problem with $c=3$ and $K=3$ with Monte-Carlo simulations. On
    this picture, we have reproduced the data of Fig.~\ref{fig1}
    with Monte-Carlo simulation (black and green crosses). We
    have prepared a large XORSAT system at equilibrium for $T=0.25$
    and $T=0.22$ and performed slow cooling and heatings with a
    Monte-Carlo procedure: when the dynamics is slow enough (we have
    changed the temperature only by a factor $\Delta T=10^{-8}$ at
    each Monte-Carlo steps) the energy in the simulation follows
    perfectly the prediction of the states following formalism. We
    have thus succeed in predicting the adiabatic evolution of the
    dynamics starting from equilibrium.}
\end{figure}

In order to check these results, we have performed the following
numerical simulation. We have first prepared two large XOR-SAT system
with $c=3$ and $K=3$ and $N=200000$ spins at equilibrium for
temperature $T=0.22$ and $T=0.25$. Of course, since these temperatures
are below the dynamic transition, this would be an impossible task if
we could not use the planting trick described in
Sec.~\ref{sec:planting} that allows to prepare at virtually no
computational cost a random instance {\it together} with an equilibrated
configuration. Then, we have used a metropolis Monte-Carlo algorithm
initialized in the planted configuration to follow the state upon slow
cooling and slow heating. As shown in Fig.~\ref{fig:simu}, the results
of the simulation agree perfectly with the theoretical predictions,
including the location of the high temperature spinodals.

As already pointed out computing the limiting energy of adiabatic
simulated annealing corresponds to the zero temperature $T_a=0$ energy of a
typical state with $T_e=T_d$, and this requires to consider replica
symmetry breaking within the states. At least the 1RSB computation
plus the analysis suggested in Sec.~\ref{sec_approx} is needed to
compute lower bounds on the energy achieved by the infinitely slow
annealing.  This is numerically involved and we will thus address it
in subsequent works. Another way of accessing this energy-value would
be to solve the dynamical equations
\cite{SompolinskyZippelius81,SompolinskyZippelius82,CugliandoloKurchan94,BouchaudCugliandolo98},
which at current time seems to be even much harder task. Despite these
limitations, a very useful new insight about the energy landscape and
limitations of simulated annealing and other stochastic local search
algorithms can be obtained from the results that we already have from the states following method, as we
explain in the next section.

\begin{figure}[!ht]
  \includegraphics[width=0.49\linewidth]{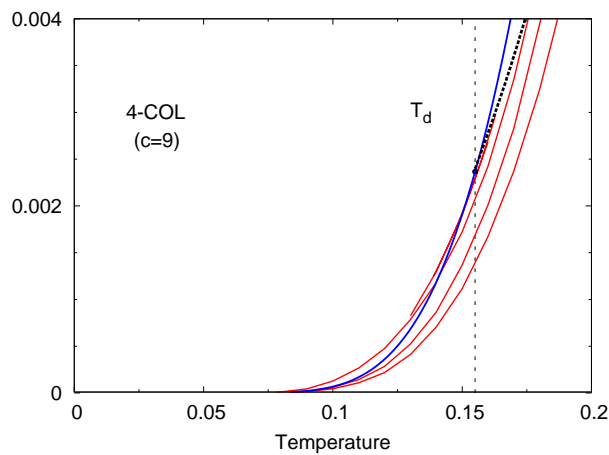} 
  \caption{\label{fig_col} (color online): State following for the 4-coloring of 9-regular random graphs. Equilibrium in blue, several states in red. Unlike in the XOR-SAT problem, here the states descent very fast to zero energies.}
\end{figure}

\subsection{Canyon versus valleys}
Let us turn our attention to Fig.~\ref{fig_col}. It depicts in the
same manner as before the evolution of states for the 4-coloring of 9-regular
random graphs. Unlike in the examples in Fig.~\ref{fig1} we see that
all depicted states fall very fast down to zero energy. Again due
to RSB instabilities, so far, we are not able to show explicitly that
the state with $T_e=T_d$ goes down to zero energy. But in any case we
see that the asymptotic behavior at zero temperature is rather
different.

\begin{figure}[!ht]
  \hspace{-0.35cm}
  \includegraphics[width=0.48\linewidth]{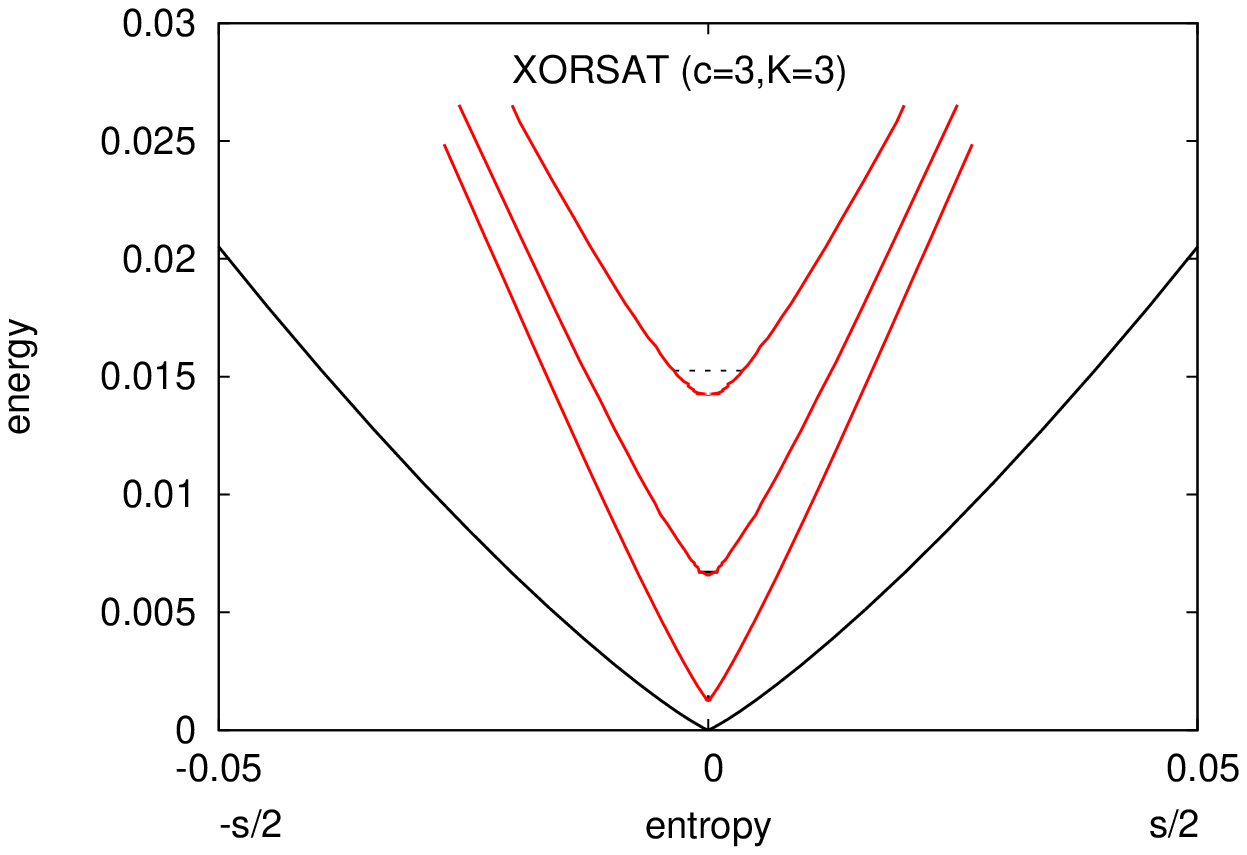}
  \includegraphics[width=0.48\linewidth]{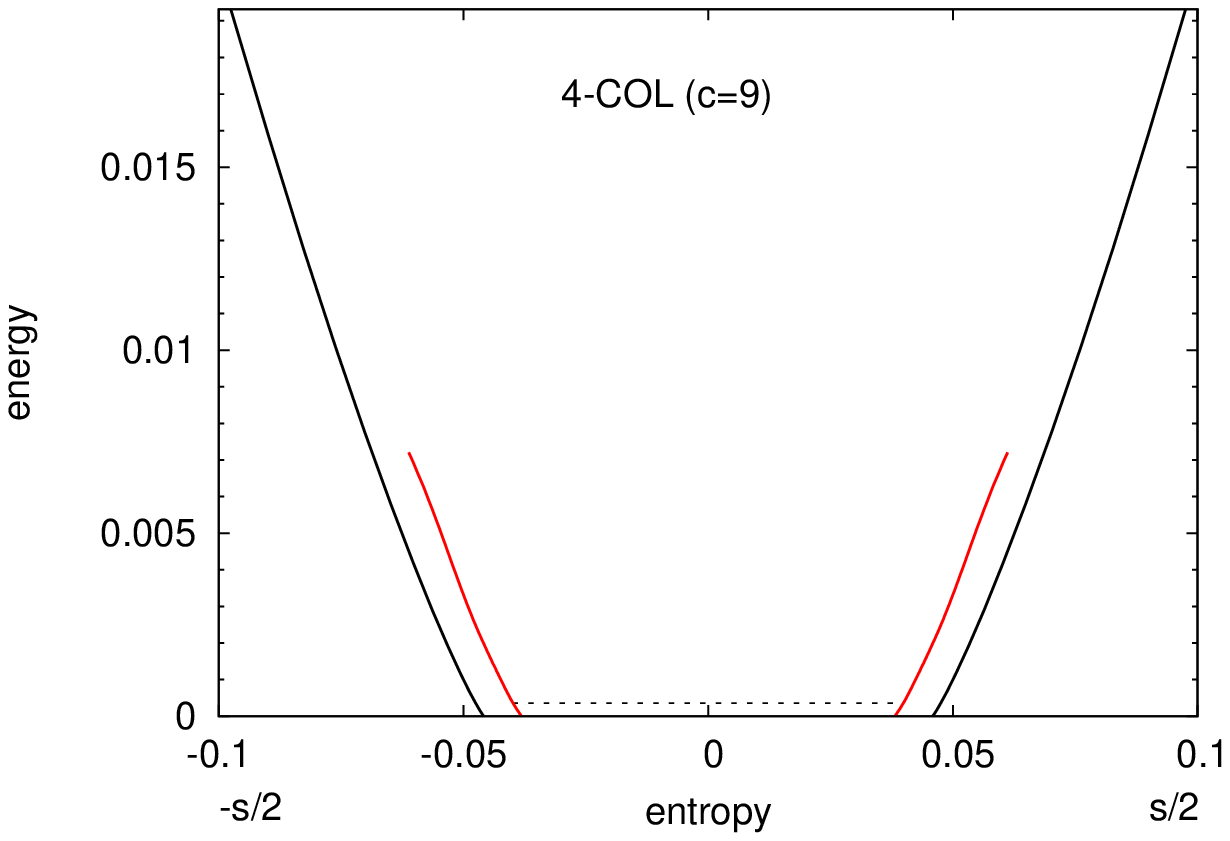}
  \caption{\label{fig_states} (color online): Data from
    Fig.~\ref{fig1} and \ref{fig_col} plotted in order to visualize
    the energy landscape. The energy $e$ is plotted against the
    entropy $s=\beta (e-f)$ for different equilibrium states. We
    plotted $s(e)/2$ and $-s(e)/2$ such that the width corresponds to
    the logarithm of the number of configurations at energy $e$ for
    the Gibbs state. Left: XOR-SAT with $c=3$, $K=3$.  Right:
    4-coloring of random graphs with $c=9$. The black curve
    corresponds to the equilibrium total entropy. The red curves are
    different equilibrium states, corresponding to $T_e=0.15,0.2,0.24$
    (left), and $T_e=0.12$ on (right), the energies corresponding to
    $T_e$ are depicted by horizontal black dashed lines. The left side
    states, with their finite energy bottoms, remind us of the valley
    in Fig.~\ref{fig00}, while the right side reminds of deep canyons
    that all reach the ground state energy level. Note that our distinction between canyons and valleys is purely energetic, both canyons and valleys can have both zero or positive bottom entropy.}
\end{figure}

In order to be precise, we now distinguish between two types of
states:
\begin{itemize}
     \item{Canyons are states with bottoms at the ground state energy.}
     \item{Valleys are states with bottoms strictly above the ground state energy.}
\end{itemize}
By definition there is at least on canyon-state in every system. The
definition is rather intuitive when looking to the carton of the energy
landscape in Fig.~\ref{fig00}. The difference between canyons
and valleys is accentuated in Fig.~\ref{fig_states} where data from
Figs.~\ref{fig1} and \ref{fig_col} are plotted in order to visualize
the shape of the states. The energy is plotted against the entropy
$s=\beta_a (e-f)$ for different equilibrium states. We plotted
$s(e)/2$ and $-s(e)/2$ such that the width corresponds to the
logarithm of the number of configurations at energy $e$ for the Gibbs
state. The left hand size is for the $L=3$ 3-XOR-SAT, right hand side
for 4-coloring of 9-regular random graphs. The blue (the most outer)
curve corresponds to the equilibrium energy and entropy. The different
red curves are shapes of states equilibrium at energy $e_e$ depicted
by the horizontal black dashed lines. The bottoms of depicted states
on the left are at positive energy hence these states are valleys,
whereas the bottom of the state depicted on the right is at zero
energy, this is hence a canyon.

Based on the distinction between canyon-states and valley-states, we
now describe two distinct types of energy landscape, depending on the
basin of attraction of states at the dynamical transition $T_e=T_d$:
\begin{itemize}
\item{In the canyons-dominated landscape, a typical equilibrium state at
    $T_e=T_d$ is a canyon.}
   \item{In the valleys-dominated landscape those states are valleys.}
\end{itemize}
In the previous examples, the cases of the XORSAT problem we showed
have valleys dominated landscape and the coloring example is canyons dominated. This
has a deep algorithmic consequence: an adiabatically slow simulated
annealing {\it is able} to find the ground state in the
canyons-dominated landscape, but {\it this is not the case} for the
valleys-dominated landscape. In constraint satisfaction problems where
one can change continuously the connectivity, we thus expect that
there will be a {\it sharp transition} $c_{cv}$ from the
canyons-dominated landscape to the valleys-dominated landscape as the
density of constraints is increased. This phase transition must happen
between the clustering and satisfiability threshold, $c_d\le c_{cv}
\le c_s$. We will now argue that the canyons/valleys transition
$c_{cv}$ is upper bounded by the rigidity transition introduced in
\cite{ZdeborovaKrzakala07,Semerjian07}.

\subsection{The warning propagation limit and the bottoms of states}
\label{sec:WP}

It is possible to derive analytical equations for the energy of
the bottoms of equilibrium states, i.e. in the zero temperature limit,
$\beta_a \to \infty$. This is simply the limit of Eq.~(\ref{pop_pl})
that takes a simpler closed form when $T_a=0$. Here we present these equations and their derivation for the XOR-SAT problem. We also derived corresponding equations for the graph coloring.

We consider for simplicity that degree of every
variable in the XOR-SAT problem is fixed to $L$ and all interactions are antiferromagnetic $J_a=-1$. For our purpose it is convenient to rewrite
Eq.~(\ref{pop_pl}) in terms of probability $\epsilon$ that a
constraint is violated in the planted configuration,
Eq.~(\ref{eps_rs}).  \be P_s(\psi) = \frac{1}{2^{K-2}} \sum_{\{s_i\}}
\left[ (1-\epsilon) \delta_{1,s+\sum_i s_i} + \epsilon
  \delta_{0,s+\sum_i s_i} \right] \int \prod_{i=1}^{K-1}
\prod_{j=1}^{L-1} {\rm d}P_{s_i}(\psi^j) \delta(\psi-{\cal
  F}(\{\psi^j\}))\, ,\label{eq_wp} \ee where the sums in the Kronecker deltas are modulo 2, and ${\cal F}(\{\psi^j\})$ is defined by the BP equation
(\ref{BP_eq}) at zero temperature $\beta_a \to \infty$. In this limit
we can write the warning propagation version of Eq.~(\ref{eq_wp})
which can be solved without the use of population dynamics.

Let us introduce the following probabilities
\begin{itemize}
   \item $\mu$ is a probability that a given constraint is forcing variable into a value in which it was planted (warning from $a$ to $i$)
   \item $\eta$ is a probability that a given constraint is forcing variable into a value in which it was not planted  (warning from $a$ to $i$)
   \item $\tilde \mu$ is a probability that a variable is being forced into a value into which it was planted  (warning from $i$ to $a$)
   \item $\tilde \eta$ is a probability that a variable is being forced into a value into which it was not planted  (warning from $i$ to $a$)
\end{itemize}
The following equations are then linking the above probabilities
\bea
   \tilde \mu &=& \sum_{s=0}^{\frac{L-2}{2}} \sum_{r=1}^{L-1-2s} \frac{(L-1)!}{s! (r+s)! (L-1-r-2s)!} \, \mu^{s+r} \eta^s (1-\mu-\eta)^{(L-1-r-2s)} \, , \label{B_first}\\
\tilde \eta &=& \sum_{s=0}^{\frac{L-2}{2}} \sum_{r=1}^{L-1-2s} \frac{(L-1)!}{s! (r+s)! (L-1-r-2s)!}\,  \mu^s \eta^{s+r} (1-\mu-\eta)^{(L-1-r-2s)} \, ,
\eea
and 
\bea
    \mu &=& (1-\epsilon) \sum_{r=0}^{\frac{K-1}{2}}\frac{(K-1)!}{(2r)!(K-1-2r)!} \, \tilde\mu^{K-1-2r} \tilde \eta^{2r} +  \epsilon \sum_{r=0}^{\frac{K-2}{2}}\frac{(K-1)!}{(2r+1)!(K-2-2r)!} \, \tilde\mu^{K-2-2r} \tilde \eta^{2r+1}\, , \\
   \eta &=&(1- \epsilon) \sum_{r=0}^{\frac{K-2}{2}}\frac{(K-1)!}{(2r+1)!(K-2-2r)!} \, \tilde\mu^{K-2-2r} \tilde \eta^{2r+1} + \epsilon \sum_{r=0}^{\frac{K-1}{2}}\frac{(K-1)!}{(2r)!(K-1-2r)!} \, \tilde\mu^{K-1-2r} \tilde \eta^{2r}\, . \label{B_last}
\eea
Sanity check is that the above equations give $\mu=(1-(1-\mu)^{L-1})^{K-1}$ and $\eta=0$ when $\epsilon=0$, this is the equation for appearance of the hard fields derived in \cite{Semerjian07}.

The corresponding energy is expressed as
\be
      E= \sum_{r=1}^{L/2} r P_{i+\partial i}(r) -L (1-1/K) P_a\, ,\label{B_energy}
\ee
where $P_{i+\partial i}(r)$ is the probability that $r$ contradictions happened when a variable $i$ and all its neighbors are added. And $P_a$ is the probability that a contradiction happened when the constraint $a$ was added. We have
\bea
   P_a &=& (1- \epsilon) \sum_{r=0}^{\frac{K-1}{2}}\frac{K!}{(2r+1)!(K-1-2r)!} \, \tilde\mu^{K-1-2r} \tilde \eta^{2r+1} + \epsilon \sum_{r=0}^{\frac{K}{2}}\frac{K!}{(2r)!(K-2r)!} \, \tilde\mu^{K-2r} \tilde \eta^{2r}\, , \\
P_{i+\partial i}(r)&=& \sum_{s=1}^{L-2r} \frac{L!}{r! (r+s)! (L-s-2r)!}\,  \mu^{r+s} \eta^{r} (1-\mu-\eta)^{(L-s-2r)}+ \nonumber \\ &&\sum_{s=1}^{L-2r} \frac{L!}{r! (r+s)! (L-s-2r)!}\,  \mu^{r} \eta^{r+s} (1-\mu-\eta)^{(L-s-2r)}+ \frac{L!}{r! r! (L-2r)!}\,  \mu^{r} \eta^{r} (1-\mu-\eta)^{(L-2r)} \, .
\eea

\begin{figure}[!ht]
  \includegraphics[width=0.495\linewidth]{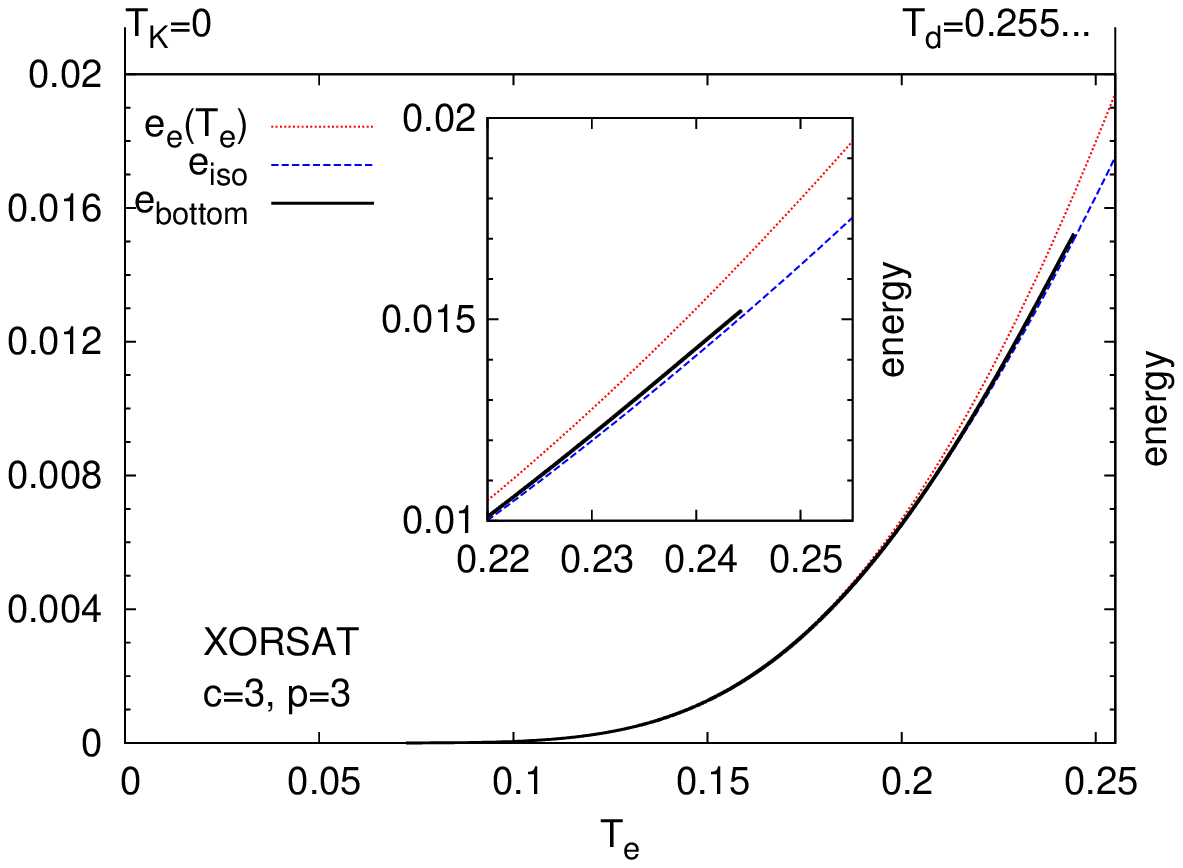}
  \includegraphics[width=0.495\linewidth]{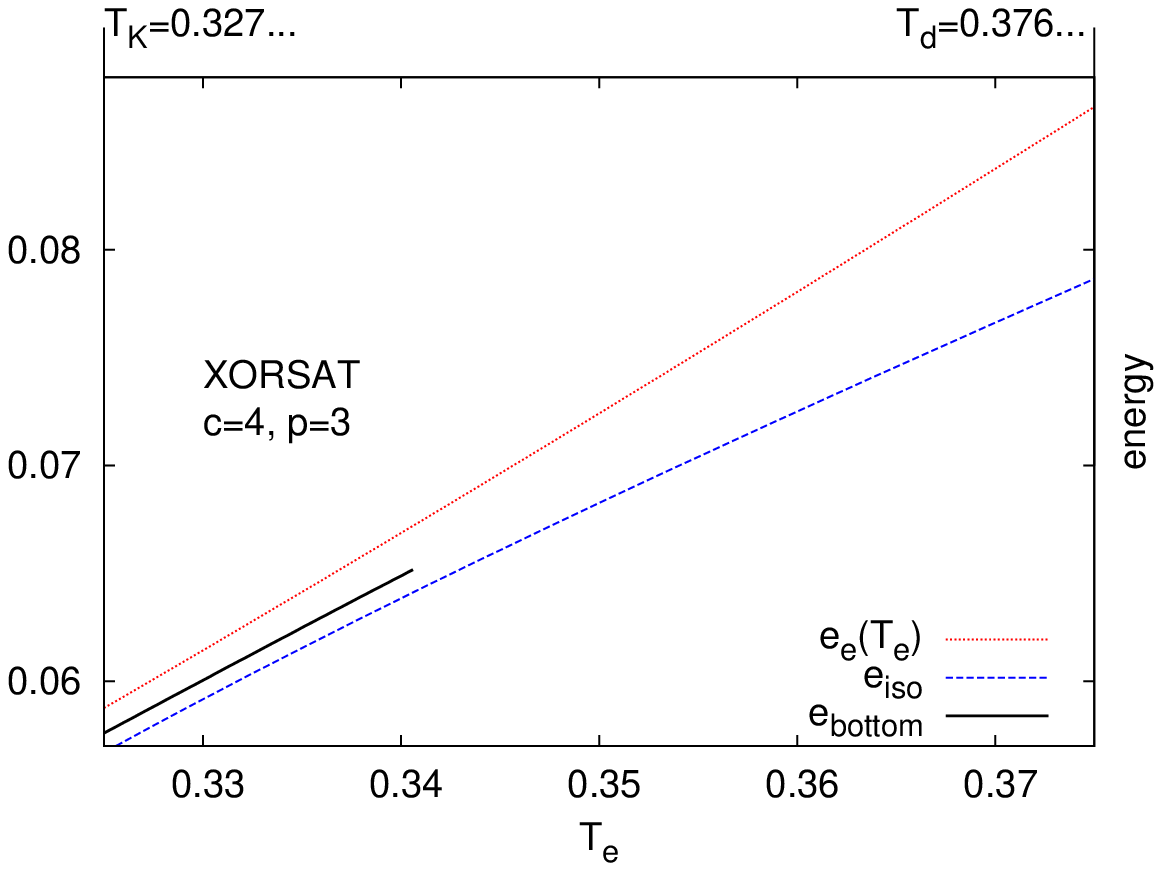}
  \caption{\label{fig3} (color online) 
Comparison between the exact
  adiabatic evolution of states and the iso-complexity lower
  bound. The black line is the energy of the bottoms of states that
  were the equilibrium ones at temperature $T_K \le T_e \le T_d$ in
  XOR-SAT for $c=3$, $p=3$ (left, the inset is a zoom) and for $c=4$,
  $p=3$ (right). The red (upper-most) line is the equilibrium energy
  at $T_e$. The blue line is the iso-complexity lower bound from
  \cite{MontanariRicci04}.}
\end{figure}

The solution of these equations is depicted
in Fig.~\ref{fig3} by a black line. Because of the relaxation within
the state, the bottom is significantly lower than the equilibrium
energy (red line). 

Let us compare the state following result with an interesting heuristic idea to
estimates the bottoms of states that was developed in
\cite{MontanariRicci04} (see also \cite{BarratFranz97}). It uses an approach called iso-complexity.
Instead of exactly following the states, the authors proposed instead
to count the number of states at a given temperature $T_e$, and then
to consider the energies at $T<T_e$ for which the number of state is
equal to the one at $T_e$. Iso-complexity leads, however, only to a
lower bound, because ending up at lower energies would be exponentially unprobable. We see in Fig.~\ref{fig3} that indeed
the true bottom is always at larger energy than given by the iso-complexity
computation of \cite{MontanariRicci04}.

Note also that above certain temperature $T_e$ the equations
do not have any non-trivial solution, this corresponds again to the to
the non-physical spinodal point is observed in Fig.~\ref{fig1}. The
physical reason for this is that the states are unstable against RSB
and that we should have used the RSB formalism. The noise level $\epsilon$ corresponding to this spinodal point is summarized in Table~\ref{tab:reconstruction}.

\begin{table}[!ht]
  \begin{center}
    \begin{tabular}{|c|l|l|c|c|c|}
     \noalign{\smallskip}     \hline
      $K$ & $L$ & $\epsilon$ & $\eta$ & $\mu$  \\ \hline 
       3 & 3 & 0.01665(1) & 0.088265 & 0.571804   \\ \hline 
       3 & 4 & 0.05184(1)  & 0.094346 & 0.886841  \\ \hline 
       3 & 5 & 0.09558(1)  & 0.128800 & 0.723037  \\ \hline 
       4 & 3 & 0.00656(1)  & 0.011682 & 0.907334  \\ \hline
       4 & 4 & 0.03179(1)  & 0.059680 & 0.933018  \\ \hline 
       4 & 5 & 0.06673(1)  & 0.087771 & 0.794195  \\ \hline 
       4 & 6 & 0.08815(1)  & 0.123228 & 0.857183  \\ \hline 
       5 & 3 & 0.00349(1)  & 0.006069 & 0.936580  \\ \hline 
       5 & 4 & 0.02298(1)  & 0.043819 & 0.952231  \\ \hline 
       5 & 5 & 0.05259(1)  & 0.068310 & 0.830129  \\ \hline 
      \noalign{\smallskip}
    \end{tabular}
    \caption{Largest values of $\epsilon$ with a non trivial solution
      of Eqs.~(\ref{B_first},\ref{B_last}) in the XOR-SAT problem. This gives a lower bounds
      to the largest possible values of the noise in the
      noisy reconstruction on trees. \label{tab:reconstruction}
}
\end{center}
\end{table}

\subsection{Where the really hard problem really are?}
We shall now argue that the canyons/valleys transition $c_{cv}$ is
upper bounded by the rigidity transition
\cite{ZdeborovaKrzakala07,Semerjian07}. In the limit $T_e\to 0$ ($\epsilon \to 0$), the equations for the bottom-energy of the states (\ref{B_first}-\ref{B_last}) reduce to the
equations for frozen variables in the equilibrium zero temperature
states from \cite{ZdeborovaKrzakala07,Semerjian07}. 

In the same limit
the bottom of a low $T_e$ state (\ref{B_energy}) is positive if
there are frozen variables in the ground state $T_e=0$. Also in
order to have a non-trivial (i.e. $E>0$) solution at finite $T_e$, we need a non-trivial solution at $T_e=0$: hence if there are no frozen
variables in the ground state then Eqs.~(\ref{B_first}-\ref{B_last})
have only the trivial paramagnetic solution for low $T_e$ which means
that either the bottoms of low $T_e$ states are at zero energy {\it
  or} that we encountered again the previously discussed instability
that prevents us to follow some states down to zero temperature using
only the replica symmetric approach.

On the other hand, when there are frozen variables in the equilibrium
state at zero temperature the landscape is always valleys-dominated
and simulated annealing (and presumably any simple algorithm with
local moves) will not be able to find the ground state: these
corresponds to truly difficult problems.

The instability towards RSB unfortunately prevents us from showing
explicitly that the canyons-valleys transition is strictly larger than
the dynamical transition (in the model such as K-SAT or graph coloring
where the later does not coincide with the rigidity
transition). However, it is reasonable to expect this is the case and
indeed the behavior of simulated annealing observed in simulation
confirms this \cite{MourikSaad02}. The existence of a phase with
canyon-dominated landscape thus explains the unreasonable efficiency
of some stochastic local search algorithms
\cite{ArdeliusAurell06,KrzakalaKurchan07}. The really hard problem
requires the landscape not only to be glassy {\it but also} not to
have states going down to zero energy with a large basins of
attraction, i.e. to have a valleys dominated energy landscape.

Our analysis also provides an insight about the types of solutions
that are achieved by simulated annealing or stochastic local
search. Those solutions are clearly not equilibrium ones and instead
belong to the bottoms of states that undergo full-step replica
symmetry breaking at low temperatures. The existence of the above
discussed spinodal line means that there is no low temperature belief
propagation fixed point associated to these states. This was indeed
observed numerically in previous works
\cite{ZdeborovaKrzakala07,DallAstaRamezanpour08,LiMaZhou08} -- when BP
is initialized in a solution found by some heuristic algorithm it
always converges back to the replica symmetric fixed point. The
procedure called whitening
\cite{Parisi03,ManevaMossel05,BraunsteinZecchina04,SeitzAlava05,ZdeborovaKrzakala07}
never finds a non-trivial fixed point either when initialized in
solutions found by survey propagation or other heuristics.

Our results explicitly explain why solutions found by polynomial
heuristics have quite different properties from the equilibrium
solutions that are usually described by the cavity method. This shows
how futile are the attempts to study clustering, BP fixed points, and
other equilibrium predictions starting from solutions obtained by
heuristics solvers! Instead exhaustive search, planting techniques or
other provably equilibrium procedures have to be used if one wants to
consider equilibrium configurations.

\begin{figure}[!ht]
  \includegraphics[width=0.495\linewidth]{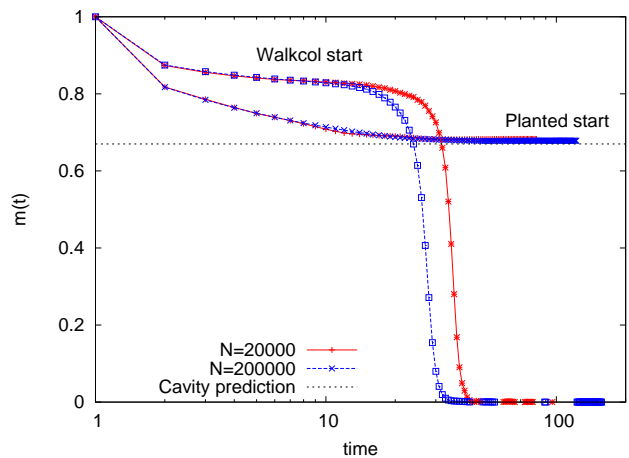}
  \caption{\label{whitening} (color online) Iteration of Belief
    Propagation in the 4-coloring problem of graph with average
    Poissonian degree $c=8.4$.  The average magnetization of BP
    messages is plotted versus the number of iterations. When
    initialized in the equilibrium solution (obtained by planting), BP
    converge to the non trivial magnetization, in agreement with
    cavity prediction. On the very same graph, when initialized in a
    solution obtain by Walk-COL, it however converges to the trivial
    fixed point. This shows solutions found by heuristic solver are
    very different from the equilibrium ones, and instead belong to
    clusters that do not have an associated BP fixed point, as
    expected from the picture obtained by the following state method.}
\end{figure}

In order to illustrate this we created an instance of the coloring
problem using the quiet planting procedure and ran belief propagation
initialized both in the planted configuration, and then in a solution
found by the Walk-Col algorithm introduced in
\cite{ZdeborovaKrzakala07}. Fig.~\ref{whitening} shows how the
magnetization evolves with number of iterations. BP initialized in the
planted configuration converges to a non-trivial fixed point at a
value of magnetization that describe the "width" of the corresponding
equilibrium states, which is perfectly in agreement with the cavity
prediction \cite{ZdeborovaKrzakala07}. However, BP initialized in the
Walk-Col solution converges to a trivial fixed point after an
intermediate plateau corresponding to flattening of the underlying
potential. The potential does however not have a minimum hence BP ends
up in the trivial fixed point \footnote{Interestingly the
  magnetization of the plateau is higher than the equilibrium
  magnetization. Suggesting that clusters found by heuristic solves
  are smaller than the equilibrium ones, this is somewhat
  counterintuitive and requires further investigation.}. This plateau
corresponds to the deep minima found is experiments with whitening
procedure, when a number of changes is plotted as a function of the
number of interactions \cite{Parisi03,ManevaMossel05}. The same
behavior is observed in the entropy at a certain distance from a
solution investigated in \cite{DallAstaRamezanpour08,Zhou09}, which is
the Franz-Parisi potential at zero temperature.

This show unambiguously that many types of solutions exist in these
problems, and that one should not confuse the {\it equilibrium}
thermodynamic solutions of the standard cavity approach, with the {\it
  out-of-equilibrium} solutions, that should be studied with the
formalism we have introduced here.

\section{Conclusions and discussion}
We have described how to follow adiabatically Gibbs states in glassy
mean field models, answered some long-standing questions about the
glassy energy landscape, and we have computed for a first time the
residual energy after an adiabatically slow annealing from
equilibrium. We have described the behavior of out-of-equilibrium
states, and demonstrated the presence of temperature chaos in these
mean-field models. We have found new features of the energy landscape,
and identified a new transition from a canyons-dominated landscape to
a valleys-dominated one that allows to quantitatively understand why
ground state configurations are sometimes easy to find despite the
presence of a glass transition. We have also shown that these
out-of-equilibrium ground-states have different properties than the
equilibrium ones, thus explaining many apparent discrepancies between
theory and simulations in the literature. Finally, we have also checked
some of these results using Monte-Carlo simulations.

On the methodological side of our work, the states following method we
have developed has interesting connections to the reconstruction on
trees and it is also closely related to the Franz-Parisi potential. In
some models it can be re-interpreted via the planting of an
equilibrium configuration, which is in particular useful for speeding
up simulations and we shall pursue on this aspect in forecoming
works. A curious and interesting connection between the properties of
the glassy systems and ferromagnets on the Nishimori line is found and
its consequences for the physics of the glass transition will be also
explored in subsequent works.

The method of states following has, however, one drawback that arises
due to the instability of states towards full-step replica symmetry
breaking at lower temperatures. Usually in such a situation replica
symmetric or 1RSB approach can be used as a sensible and very accurate
approximation. However, we found ourself here in the rather particular
situation where there is {\it no} non-paramagnetic RS or 1RSB solution
for states with $T_e$ close to $T_d$ at low $T_a$ (as illustrated in
Fig.~\ref{fig_follow}), in a region where we anticipate FRSB to be the
correct solution. Since it is not known how to obtain the FRSB
solution in the diluted systems, this prevents us from computing
concrete values of limiting energies for adiabatic simulated annealing
initialized at high temperatures.  Clearly this calls for new
investigations and for new ways to approximate the FRSB solution. We
have suggested one such approximation that gives a sensible solution
even in this region in Sec.~\ref{sec_approx}, but it is still a
computationally costly one. A simpler approach thus need to be
developed. Maybe the direction suggested in
\cite{CavagnaGiardina05,MuellerLeuzzi06}, with marginal states (and
supersymmetry broken cavity method) and the use of the so-called two
groups ansatz could be a way out of this problem. Note also that the
"gap" in the $\Sigma(s)$ (how many states of a given size are present)
in the coloring problem reported in \cite{ZdeborovaKrzakala07} might
be related to the instability observed in the present paper.

The formalism of adiabatic evolution of states in temperature should
extend straightforwardly when other external parameters are changed
adiabatically. It would be interesting to see if one can take as the
adiabatic external parameter the density of constraints (or average
degree of the graph) one study the "connectivity" landscape introduced
in \cite{KrzakalaKurchan07}\footnote{At this point, we want to
  mention, that we tried to do this and obtained equations identical
  to those in Sec.~\ref{sec:WP}. This however predicts a duality
  between the value of $c_d$ and $T_d$ which does not hold. It is yet
  to be understood if this prediction fails due to the FRSB
  instability or due to another effect to be discovered.}. Our work
offers extensions in many other directions as detailed description of
the complex energy landscape is immensely useful in understanding
properties of complex glassy materials. Among possible applications
are the studies of memory and rejuvenation protocols
\cite{DupuisVincent01}, of jammed packings \cite{ParisiZamponi09} and
of the quantum adiabatic algorithm
\cite{KadowakiNishimori98,FarhiGoldstone00}.

\acknowledgments The authors would like to thank S. Franz, J. Kurchan,
M. M\'ezard, G. Semerjian and D. Sherrington for very inspiring and
useful discussions. We thank G. Semerjian in particular for thorough
reading of the manuscript.

\appendix
\section{The large connectivity limit of the cavity equations}
\label{pspin_rem}
The solution of the fully connected $p$-spin model was originally
derived from the replica trick
\cite{Derrida80,GrossMezard84,KirkpatrickThirumalai87a,KirkpatrickThirumalai87b,KirkpatrickThirumalai89}. The
cavity approach was developed later on as an alternative to the replica trick,
and the $p=2$ solution recovered in this way
\cite{MezardParisi85c}. In this appendix, we remind how this computation generalizes and how the large connectivity limit of the cavity
equations yields the RS and 1RSB solutions of the fully connected
$p$-spin model. This should facilitate understanding of our derivation of the
equations for states following in the main text.

\subsection{The replica symmetric solution}
To achieve our goal, it is first suitable to rewrite
Eq.~(\ref{BP_chi}) in terms of cavity fields $h^{i\to a}$ and biases
$u^{b\to i}$ defined as \bea
\chi_s^{i\to a}&=&\frac{e^{\beta h^{i\to a}s}}{2\cosh{\beta h^{i\to a}}}\, ,\\
\psi_s^{b\to i}&=&\frac{e^{\beta u^{b\to i}s}}{2\cosh{\beta u^{b\to
      i}}}\, , \eea it gives \bea
h^{i\to a} &=& \sum_{b\in \partial i \setminus a} u^{b\to i} \, ,\\
\tanh{(\beta u^{b\to i})} &=& \tanh{(\beta J_b)} \prod_{j\in\partial b
  \setminus i} \tanh{(\beta h^{j\to b})} \, . \label{def_u} \eea
At this point, the recursion can thus be written in terms of the local
fields $h^{i\to a}$ as 
\be h^{i\to a} = \frac{1}{\beta}
\sum_{b\in \partial i \setminus a} {\rm arctanh}{\left[ \tanh{(\beta
      J_b)} \prod_{j\in \partial b\setminus i} \tanh{(\beta h^{j\to
        b})} \right]}\, . \label{BP_h} \ee

In the fully connected $p$-spin problem every spin is involved in
${N-1\choose p-1} \sim N^{p-1}/(p-1)!$ interactions, and each of
these interactions $J_b$ is small, of $O(N^{1-p})$. Eq.~(\ref{BP_h})
can thus be rewritten introducing a new message $m^{i\to a}=\tanh{(\beta
  h^{i\to a})}$ as \be m^{i\to a} = \tanh{\left( \beta
    \sum_{b\in\partial i \setminus a} J_b \prod_{j\in \partial
      b\setminus i} m^{j\to b} \right)} = \chi^{i\to a}_1 -\chi^{i\to
  a}_0 \, .
\label{cavity_ap}
\ee 
At this point one realizes that the argument of the $\tanh{}$ is a sum
of many terms. In the replica symmetric approximation these terms are
considered independent and thus according to the central limit theorem
the sum is distributed as a Gaussian variable. We denote $m=\langle
m^{i\to a} \rangle$ the mean (first moment) of $m^{i\to a}$ and
$q=\langle (m^{i\to a})^2 \rangle$ its second moment. Then the mean
(first moment) of the sum in the argument of the $\tanh{}$ is $
\mu=\beta J_0 p \, m^{p-1} $, and its variance $ \sigma= \beta^2 J^2 p
\, q^{p-1} /2 $.  Using the mean and variance we can write $m= \int
e^{-(z-\mu)^2/(2\sigma)} \tanh{z} \, {\rm d}z /\sqrt{2\pi \sigma}$ and
$q= \int e^{-(z-\mu)^2/(2\sigma)} \tanh^2{z} \, {\rm d}z /\sqrt{2\pi
  \sigma}$ which after a substitution gives the usual form of the
replica symmetric equation for the $p$-spin model Eqs.~(\ref{eq_m}-\ref{eq_q}), first obtained in \cite{GrossMezard84}:

\subsection{The free energy calculation}
In the cavity formalism the RS free energy reads
\be
      -\beta F = \sum_i \log{Z^{i+\partial i }} - (p-1)\sum_a \log{Z^a} \, ,
\ee
where the free energy shifts are
\bea
    Z^{i+\partial i } &=&   \sum_{s_i}  \prod_{b\in \partial i} \sum_{s_{j\neq i}}  e^{\beta J_b \prod_{j\in \partial b} s_j} \prod_{j\in\partial b \setminus i} \chi_{s_j}^{j\to b}\, , \\
    Z^a &=& \sum_{s_j}  e^{\beta J_a \prod_{j\in \partial a} s_j} \prod_{j\in\partial a} \chi_{s_j}^{j\to a}\, .
\eea

Let us now take the fully connected limit. The link term is (we write
only the terms with $J_b$ and $J_b^2$ as the rest in negligible) 
\be
Z^a = \sum_{s_j} (1+ \beta J_a \prod_{j\in \partial a} s_j +
\frac{1}{2}\beta^2 J_a^2) \prod_{j\in \partial a} \chi_{s_j}^{j\to a}
= 1 + \frac{1}{2}\beta^2 J_a^2 + \beta J_a \prod_{j\in \partial a}
m^{j\to a}\, . \label{z_link} \ee Hence \be \sum_a \log{Z^a} = \sum_a
\left( \frac{1}{2} \beta^2 J_a^2 + \beta J_a \prod_{j\in \partial a}
  m^{j\to a} - \frac{1}{2} \beta^2 J^2_a \prod_{j\in \partial a}
  [m^{j\to a}]^2 \right) = \frac{1}{4} \beta^2 J^2 + \beta J m^p
-\frac{1}{4} \beta^2 J^2 q^p\, .  \ee

The site term is a bit trickier. It is useful to remind the following relations
\be
           \chi^{i\to a}_s = \frac{1+sm^{i\to a}}{2} = \frac{1+s\tanh{\beta h^{i\to a}}}{2} = \frac{e^{\beta h^{i \to a} s}}{2\cosh{(\beta h^{i\to a})}}\, . \label{useful}
\ee
The site term can then be rewritten as
\be
     Z^{i+\partial i } =   \sum_{s_i}  \prod_{b\in \partial i} \sum_{s_{j\neq i}}  e^{\beta J_b s_i\prod_{j\in \partial b\setminus i} s_j} \frac{e^{\beta \sum_j s_j h^{j\to b}}}{\prod_{j\in\partial b \setminus i}2 \cosh{\beta h^{j\to b}} } \, .
\ee
Using trigonometric relation 
\be
    \sum_{s_{j\neq i}}  e^{\beta J_b s_i\prod_{j\in \partial b\setminus i} s_j}  e^{\beta \sum_j s_j h^{j\to b}} = \cosh{\beta J_b s_i} \prod_{j\in \partial b\setminus i} 2 \cosh{\beta h^{j\to b}} +  \sinh{\beta J_b s_i} \prod_{j\in \partial b\setminus i} 2 \sinh{\beta h^{j\to b}}
\ee
and odd/even properties of the $\sinh$/$\cosh$ functions we have
\be
    Z^{i+\partial i } =   \sum_{s_i}  \prod_{b\in \partial i} \left[  2 \cosh{\beta J_b} \frac{1+s_i \tanh{\beta J_b} \prod_{j\in \partial b\setminus i} \tanh{\beta h^{j\to b}}  }{2}\right]\, .
\ee 
Using relation (\ref{def_u}) and (\ref{useful}) we get useful form of the site term
\be
      Z^{i+\partial i } = 2 \cosh{(\beta \sum_{b\in \partial i} u^{b\to i})} \prod_{{b\in \partial i}} \frac{\cosh{\beta J_b}}{\cosh{\beta u^{b\to i}}}    \, . 
\ee

Only now we start developing the large connectivity limit in which interactions strengths are infinitesimal. Writing the site term in terms of messages $m^{b\to j}$ and expanding hyperbolic functions in the leading order we get
\bea
     Z^{i+\partial i } &=& 2\cosh{(\sum_{b\in \partial i} \beta J_b \prod_{j\in \partial b\setminus i} m^{j\to b})} \prod_{b\in \partial i} \frac{\cosh{\beta J_b}}{\cosh{(\beta J_b \prod_{j\in \partial b\setminus i} m^{j\to a}})}\, ,\\ &=& 2\cosh{(\sum_{b\in \partial i} \beta J_b \prod_{j\in \partial b\setminus i} m^{j\to b})} \prod_{b\in \partial i} \left[1+\frac{\beta^2 J_b^2}{2} - \frac{\beta^2 J_b^2}{2} \prod_{j\in \partial b\setminus i} (m^{j\to b})^2\right]\, . \label{z_site}
\eea
So that the site contribution to the free energy is 
\be
      -\beta f^{i+\partial i } =  \int {\cal D}y \log{2\cosh{\left( \beta J y \sqrt{p \, q^{p-1}/2}+ \beta J_0 p m^{p-1}\right)}} + \frac{\beta^2 J^2 p}{4} (1-q^{p-1}).
\ee
Adding both terms together we get the replica symmetric free energy density of the fully connected $p$-spin model:
\be
   -\beta f = \frac{1}{4}\beta^2 J^2 (p-1) q^p - \beta J_0 (p-1) m^p + \frac{1}{4}\beta^2 J^2 - \frac{1}{4} \beta^2 J^2 p \, q^{p-1} + \int {\cal D}y \log{2\cosh{\left( \beta J y \sqrt{p \, q^{p-1}/2}+ \beta J_0 p m^{p-1}\right)}}.
\ee

\subsection{1RSB solution}
\label{app:1RSB}
To derive the infinite connectivity limit of the 1RSB cavity equations we
define the mean and overlap parameters
\bea
m &=& \int {\cal D}Q(P)  \left[  \frac{\int {\rm d}P(m^{i\to a})  \, Z(\{m^{i\to a}\},\beta)^x \,  m^{i\to a} }{\int {\rm d}P(m^{i\to a})  \, Z(\{m^{i\to a}\},\beta)^x} \right] \equiv \langle \langle m^{i\to a} \rangle_P \rangle_Q \, , \\
q_1 &=& \int {\cal D}Q(P)  \left[  \frac{\int {\rm d}P(m^{i\to a})  \, Z(\{m^{i\to a}\},\beta)^x \,  (m^{i\to a})^2 }{\int {\rm d}P(m^{i\to a})  \, Z(\{m^{i\to a}\},\beta)^x } \right] \equiv \langle \langle (m^{i\to a})^2 \rangle_P \rangle_Q \, ,\\
q_0 &=& \int {\cal D}Q(P) \left[ \frac{\int {\rm d}P(m^{i\to a}) \,
    Z(\{m^{i\to a}\},\beta)^x \, m^{i\to a} }{\int {\rm d}P(m^{i\to
      a}) \, Z(\{m^{i\to a}\},\beta)^x} \right]^2 \equiv \langle
\langle m^{i\to a} \rangle^2_P \rangle_Q\, .  \eea where the average
over $P$ is over the different states and average over $Q$ is over the
different edges in the graph. The message $m^{i\to a}$ is computed
from the incoming messages according to (\ref{cavity_ap}). The
distribution $P(m^{i\to a})$ follows \be P(m^{i\to a}) = \frac{1}{\cal
  Z} \int \prod_j P(m^{j\to b}) Z^{i\to a}(\{m^{i\to a}\},\beta)^x
\delta(m^{i\to a}-{\cal F}(\{m^{j\to b}\},\beta)) \, , \ee where the
equations for $m^{i\to a}=\tanh{(\beta X)}$ as in (\ref{cavity_ap}),
where \be X= \sum_{b\in\partial i \setminus a} J_b
\prod_{j\in \partial b\setminus i} m^{j\to b}. \ee Note that that the
reweighting factor equals the site term (\ref{z_site}), but only the
part is $\cosh{(\beta X)}$ is relevant, as the rest can be written as
\be e^{\sum_{b\in \partial i} \frac{\beta^2 J_b^2}{2}
  [1-\prod_{j\in \partial b \setminus i} (m^{j\to b})^2] } =
e^{\frac{\beta^2 J^2 p}{4}(1-q^{p-1})}\, , \ee which is self-averaging
and does not depend on the integration variables so it always cancels
out as the reweighting appears in both the numerator and denominator.

The argument $X$ is a sum of two Gaussian random variables and can be determined by computing: 
\bea
\mu &\equiv& \langle \langle X   \rangle_P \rangle_Q  = J_0 p m^{p-1} \, ,\\
 \sigma_1 &\equiv&  \langle \langle  X^2   \rangle_P \rangle_Q - \langle \langle X \rangle_P \rangle^2_Q= J^2 \frac{p}{2} q_1^{p-1} \, , \\
\sigma_0 &\equiv&  \langle \langle X \rangle^2_P \rangle_Q- \langle \langle X \rangle_P \rangle^2_Q=  J^2 \frac{p}{2} q_0^{p-1} \, .
\eea

At this point we are able to realize that the average over states $\langle \cdot \rangle_P$ can we written as an Gaussian integral of $\tanh{X}$ where the mean of $X$ over states $\langle X \rangle_P$ and the variance $\sigma = \langle  X^2   \rangle_P- \langle  X \rangle^2_P$. The average over the graph  $\langle \cdot \rangle_Q$ is also an average over a Gaussian variable with mean $\langle \langle X   \rangle_P \rangle_Q =\mu$ and variance $\sigma_0 =  \langle \langle X \rangle^2_P \rangle_Q- \langle \langle X \rangle_P \rangle^2_Q$. Finally $\langle \sigma \rangle_Q = \sigma_1-\sigma_0$. All that gives averaged infinite connectivity 1RSB equations (\ref{1RSB_m}-\ref{1RSB_q0}). The parameter $q_1$ is the average self-overlap and $q_0$ the average overlap between states.

\section{Computing the Franz-Parisi potential in diluted models}

We described the connection between the states following method and the Franz-Parisi potential. For completeness in this appendix we give the equations according to which the Franz-Parisi potential is computed in the diluted (sparse) models.

The Franz-Parisi potential is the free energy $f(q)$ of the system at temperature $\beta_a$ depend on the overlap $q$ with an equilibrium configuration $\{\sigma_i\}$ at temperature $\beta_e$. In order to fix the overlap $q$ we introduce a local uniform field $h$ in the direction of the equilibrium configuration. Our goal is then to compute
\bea
 f(h)   &=& \frac{\sum_{\{\sigma_i\}} \tilde f(h,\{\sigma_i\}) e^{-\beta_e {\cal H}(\{\sigma_i\})} }{\sum_{\{\sigma_i\}} e^{-\beta_e {\cal H}(\{\sigma_i\})} }\, , \\
    e^{-\beta_a \tilde f(h,\{\sigma_i\})} &=& \sum_{\{s_i\}} e^{-\beta_a {\cal H}(\{s\}) + \beta_a h \sum_i s_i \sigma_i}\, .
\eea
The Franz-Parisi potential is then
\be
      f(q) = f(h)  + h q \, , \quad \quad \frac{\partial f(h)}{\partial h} = -q\, . \label{fq}
\ee
To obtain the value of $f(q)$ we need to solve equations similar to Eq.~(\ref{final_m1}) where the field $h$ is taken into account
\bea
 && \overline \psi_\sigma \overline P_\sigma(\psi|\overline \psi) {\cal P}_{\rm RS}(\overline \psi) = \sum_J Q(J)
\sum_{\{l\}} q(\{l\})  \int \prod_{i=1}^{K-1} \prod_{j_i=1}^{l_i} \left[ {\rm d} \overline \psi^{j_i} {\cal P}_{\rm RS}(\overline \psi^{j_i}) \right] \,  \delta\left[\overline \psi - {\cal F}(\{\overline \psi^{j_i}\},\beta_e)\right] \nonumber \\ && \sum_{\{\sigma_i\}}  e^{J\beta_e \sigma \prod_{i} \sigma_i}  \frac{\prod_{i=1}^{K-1} \prod_{{j_i}=1}^{l_i} \overline \psi^{j_i}_{\sigma_i}}{Z(\{\overline \psi^{j_i}\},\beta_e)}  \int \prod_{i=1}^{K-1} \prod_{j_i=1}^{l_i} \left[{\rm d}\psi^{j_i} \overline P_{\sigma_i}^{j_i}(\psi^{j_i}|\overline \psi^{j_i})\right]  \,  \delta\left[\psi - {\cal F}(\{\psi^{j_i}\},\beta_a,\{\sigma_i\},h)\right]\, .  \label{FPFP}
\eea
where
\be
{\cal F}_s(\{\psi^{j_i}\},\beta_a,\{\sigma_i\},h) = \psi_s =  \sum_{\{s_i\}}\frac{ e^{ \beta_a J s \prod_{i} s_i}}{Z(\{\psi^{j_i} \},\beta_a)} e^{\beta_a h \sum_i \sigma_i s_i} \prod_{i=1}^{K-1} \prod_{j_i=1}^{l_i} \psi_{s_i}^{j_i}\, .
\ee
The free energy $f(h)$ is computed from the fixed point of (\ref{FPFP}) as 
\bea
     - \beta_a f(h) &=& \alpha\sum_J Q(J) \sum_{\{l\}} q(\{l\})  \int \prod_{i=1}^{K} \prod_{j_i=1}^{l_i} \left[ {\rm d} \overline \psi^{j_i} {\cal P}_{\rm RS}(\overline \psi^{j_i}) \right] \,  \delta\left[\overline \psi - {\cal F}(\{\overline \psi^{j_i}\},\beta_e)\right]  \\ && \sum_{\{\sigma_i\}}  e^{J\beta_e \prod_{i} \sigma_i}  \frac{\prod_{i=1}^{K} \prod_{{j_i}=1}^{l_i} \overline \psi^{j_i}_{\sigma_i}}{Z^{a+\partial a}(\{\overline \psi^{j_i}\},\beta_e)}  \int\prod_{i=1}^K \prod_{j_i=1}^{l_i} \left[ {\rm d}\psi^{j_i} P_{\sigma_i}^{j_i}(\psi^{j_i}) \right] \log{Z^{a+\partial a}(\{\psi^{j}\}, \beta_a,\{\sigma_i\},h)}\nonumber\\ &-& \sum_{l} {\cal Q}(l) (l-1) \int \prod_{i=1}^l  \left[ {\rm d} \overline \psi^{i} {\cal P}_{\rm RS}(\overline \psi^{i}) \right] \frac{\sum_{\sigma} e^{\beta_e \sigma}\prod_{i=1}^l \overline \psi^i_{\sigma} }{Z^i(\{\overline \psi^i\},\beta_e)} \int \prod_{i=1}^l \left[ {\rm d}\psi^{i} P_{\sigma}^{i}(\psi^{i}) \right] \log{Z^{i}(\{\psi^{i}\},\beta_a,\sigma,h)}\, . \label{free_pl_FP}
\eea
with the $h$-dependent partition function contributions being equal to
\bea
   Z^{a+\partial a}(\{\psi^{j}\},\beta_a,\{\sigma_i\},h) &=&   \sum_{\{s_i\}} e^{\beta_a J \prod_{i} s_i} e^{\beta_a h \sum_i \sigma_i s_i} \prod_{i=1}^{K} \prod_{j_i=1}^{l_i} \psi_{s_i}^{j_i}\, , \\
   Z^{i}(\{\psi^{i}\},\beta_a,\sigma,h) &=& \sum_{s} e^{\beta_a s}  e^{\beta_a h \sigma s}\prod_{i=1}^l  \psi^i_{s} \, .
\eea
The overlap is obtained by derivative with respect to $h$ according to (\ref{fq}). The non-convex parts of $f(q)$ are computed by iteratively choosing a new value of $h$ that gives the expected values of the overlap $q$, in the same manner as total magnetization was fixed in \cite{KrzakalaRicci09,SulcZdeborova09}. 

Note that the states following methods developed in this paper corresponds to the Franz-Parisi potential at $h=0$ initialized in the equilibrium configuration. In other words the states following is looking directly at the minimum of the Franz-Parisi potential that corresponds to the Gibbs state at temperature $\beta_a$.

\bibliography{myentries}

\end{document}